\documentclass[
aps,
prb,
twocolumn,
superscriptaddress,
floatfix,
showpacs,
longbibliography
]{revtex4-1}
\usepackage[T1]{fontenc}
\usepackage[utf8]{inputenc}
\usepackage[english]{babel}

\usepackage{amsmath,amssymb}
\usepackage{bm}
\usepackage{physics}     

\usepackage{graphicx}
\usepackage{dcolumn}

\usepackage[dvipsnames]{xcolor}
\usepackage{comment}

\usepackage[
unicode=true,
pdfusetitle,
bookmarks=true,
colorlinks=true,
linkcolor=blue,
citecolor=blue,
urlcolor=blue
]{hyperref}
\begin{document}
\title{Extended Su-Schrieffer-Heeger  model with competing Rashba spin-orbit interactions, higher order hopping, and domain wall}

\author{Hemant Kumar Sharma}
\email{hemant.sharma@iopb.res.in}
\affiliation{Institute of Physics, Sachivalaya Marg, Bhubaneswar-751005, India}
\author{Arijit Saha}
\email{arijit@iopb.res.in}
\affiliation{Institute of Physics, Sachivalaya Marg, Bhubaneswar-751005, India}
\affiliation{Homi Bhabha National Institute, Training School Complex, Anushakti Nagar, Mumbai 400094, India}
\author{Saptarshi Mandal}
\email{saptarshi@iopb.res.in}
\affiliation{Institute of Physics, Sachivalaya Marg, Bhubaneswar-751005, India}
\affiliation{Homi Bhabha National Institute, Training School Complex, Anushakti Nagar, Mumbai 400094, India}
\date{\today}

\begin{abstract}

We investigate  an extended spin-full Su-Schrieffer-Heeger (SSH) model preserving Chiral symmetry. It contains spin-mixing and non-mixing  Rashba spin-orbit interaction, complex next to next nearest neighbor (NNNN) hopping. Using theory of non-Hermitian matrices, we  construct Hamiltonian function facilitating exact analysis of the system. When the two types of Rashba interactions are present, the spectrum changes from band crossing to band touching having Dirac cone.   The gap-closing condition for the spectrum and its symmetry is derived in detail. The phase diagram shows  rich topological phases with winding number one or two, re-entrant phase transitions and multi-critical points. An extension of the model to $\mathcal{M}$ coupled chains exhibits topological phases with winding number varying from  one to $2 \mathcal{M}$ and it is further characterized by Zak phase.  Remarkably, a domain-wall profile in nearest-neighbor hopping  can tune the location of mid-gap zero energy states. The localization properties of the edge modes  are characterized by inverse participation ratio and compared with earlier works.   Our model and method could bring new perspective to certain classes of topological systems.
\end{abstract}

\maketitle
\section{ Introduction}

Recent studies in condensed matter physics have found a strong interest on various topological aspects \cite{Haldane_NobelLecture,KlitzingIntegerQHalleffect,40yearsHalleffect,LiangKaneTopologicalinsulator,KaneMeleeSpinHallGraphene,BernevigQuantumspinHall,qi2011topological,ReviewAnyonstopologicalQcomputation}. Further understanding of topological band theory \cite{AltlandZirnbauer_tenfold,ReviewTenfoldwayShinseiRyu,NakaharaBook,Cayssol_FuchsIntroductiontotopologybandtheory} and  role of various effects such as 
disorder, dislocations~\cite{GrothPRL2009,JianLiPRL2009,XingYPRB2011,Nag2021,Yamada2022,Salib2023}, 
and effects of interactions \cite{FractionalQHELaughlin,sayan-int-2019,sayan-int-2020,Mandal_2022} continue to draw genuine interests. In these aspects, few paradigmatic tight binding models such as Haldane model~\cite{49-PhysRevLett.61.2015}, one dimensional Kitaev model \cite{Kitaevchain} and Su-Schrieffer-Heeger (SSH) model \cite{SSHoriginalpaper,RevModPhys.60.781} have contributed  far reaching understanding. There have been several  proposals to experimentally realize these models on real or synthetic platform \cite{Jotzu2014,kim-2016,Kim2017,Zhao2024,Dvir2023,Pandey2023,Oreg2010,LutchynPRL2010,SauPRL2010,Lienhard:19,Thatcher_2022,PhysRevResearch.4.013185,10.1063/5.0191076,Bloch_TopologicalPhase,DeLeseleuc_ExpRydbergSSH,Jalochowski2024, Leumer_2020,kvande-2023,Gang-Li-2018,wakatsuki-2014}. These pioneering models have been explored further by considering various extensions \cite{38-PhysRevLett.95.226801,sudarshan_2021,Saha_2023,sayan-saurabh-2023,sayan-saurabh-2022,mondal-basu-2022,arunava-da,20-PhysRevB.96.125418,21-PhysRevA.106.022216,22-PhysRevA.95.061601,23-PhysRevB.89.085111,yonatan-2024}  to investigate new topological aspects with aims of pushing the theoretical knowledge as well as possibility of novel practical  applications~\cite{3-RevModPhys.82.3045,Hou_2020,ma10070814,8-s42005-021-00569-5}. 
Such extensions appeared to be very useful and non-trivial to establish additional topological phases\cite{38-PhysRevLett.95.226801,49-PhysRevLett.61.2015,sayan-saurabh-2023,sayan-saurabh-2022,mondal-basu-2022,sudarshan_2021,Saha_2023,sayan-saurabh-2023,mondal-basu-bilayer-haldane,Wu2022,aldea-2024,30-PhysRevLett.118.076803,PhysRevB.108.104101,PhysRevB.107.045135,PhysRevB.108.245140}. SSH model also went through many extensions and modifications\cite{arunava-da,20-PhysRevB.96.125418,21-PhysRevA.106.022216,22-PhysRevA.95.061601,23-PhysRevB.89.085111,yonatan-2024} including next and next to next nearest neighbour interactions (NNN and NNNN) \cite{tarun-da-2024,23-PhysRevB.89.085111,24-Li-2018-EPL-124-37003,25-Agrawal-2022-J.-Phys.-Condens.-Matter-34-305401,26-s41534-019-0159-6,satyaki-2024}.  Recently the effects of impurities \cite{scollon2020persistence}, long range hopping \cite{dias2022long}, and spin-orbit interactions \cite{kokhanchik2022modulated,ahmadi2020topological} have been studied in the extended SSH model \cite{zhou2023exploring} with an aim either to examine the stability of topological phases or its dependence on extended parameter space. Furthermore the studies on coupled SSH chains\cite{14-Adv-Quantum-Tech-2019-Zurita-2020,15-PhysRevB.106.205111,16-s41598-020-71196-3,17-Hetenyi-2018-J.-Phys.Condens.Matter-30-10LT01,18-PhysRevResearch.2.043191,19-Borja-2022-J.-Phys.-Condens.-Matter-34-205701,20-PhysRevB.96.125418,21-PhysRevA.106.022216,22-PhysRevA.95.061601} have provided valuable insight to understand the emergent topology  in higher dimensions such as  two\cite{30-PhysRevLett.118.076803,PhysRevB.108.104101} and three dimensions\cite{PhysRevB.107.045135,PhysRevB.108.245140}. \\
\indent
In this paper we propose an extended  SSH (e-SSH) model to establish new topological phases and how to manipulate those. The e-SSH model considered here contains i)  Rashba spin-orbit interaction (RSOI) of two kinds  ii) modified inter-cell hopping and all possible NNNN hopping preserving chiral symmetry. The parameters are considered to be complex in general. The main motivation is to investigate the consequence of  above parametric combinations and how it changes the topological phases. For example competition between two different kind of Rashba interactions and effect of imaginary hopping have not been investigated. To our surprise we find that inclusion of two kind of Rashba interaction causes a Dirac cone in conduction band which becomes asymmetric when imaginary hopping is present.   We have shown that for such case, the winding number can be manipulated from $0 \rightarrow 1 \rightarrow 2$ for a single spinful SSH chain and we lay down  the criteria for the same.  This implies that the winding number for each spin-sector can be controlled independently. Next we analysis the case of $\mathcal{M}$ such e-SSH chains coupled in a particular way. It shows that one  can generate topological phases  characterized by arbitrary winding number ranging from $0$ to $2\mathcal{M}$. Finally we extend our study to include a domain wall potential \cite{scollon2020persistence} and interestingly e-SSH model allows us to control  the position of mid gap zero energy (MGZE) state. We calculate inverse participation ratio (IPR) to examine the localization aspect of these zero energy states.
\indent

The structure of the paper is following. In Sec. \ref{sec:level1} we define the  e-SSH model explaining all the parameters in detail, present analytical solution  and  discuss its symmetry in detail. Sec. \ref{sec:topocharac} discusses procedure to determine winding number($W$) and analyses general properties of gap-closing conditions for different parameter regimes. Sec. \ref{sec:bbc} contains plots of spectrum, gap-closing point, phase diagram and IPR for different set of parameters elucidating their role in detail. In Sec. \ref{sec:twoandfourlevel}, we consider $\mathcal{M}$ coupled e-SSH chains show that the winding number can vary from zero to $2\mathcal{M}$. We present study of Zak-phase  in two dimensional limit of this coupled chain system and correlate with winding number. In \ref{sec:edge-spin-texture-ipr}, we study the property of edge states through study of inverse participation ratio and spin textures which shows many interesting features. In   \ref{domain-wall}  we  study the effect of  domain wall on this  extended SSH model, examine how it can tune the localization center of the topological edge mode via different parameters. We compare the localization properties of such topological zero-energy modes via IPR for different parameters as well as with Aubry-Andre potential. We summarize and conclude our paper in Sec. \ref{sec:conclusion} with future directions. In appendix various auxiliary aspects complimentary to main texts are presented.

\begin{figure}[h!]
	\centering
	\includegraphics[width=\linewidth]{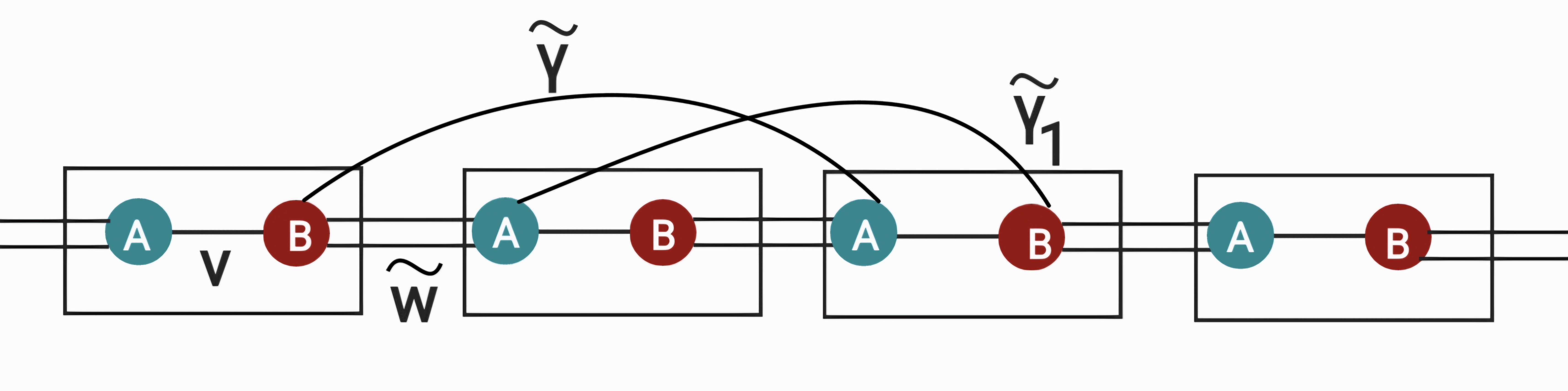}
	\caption{A cartoon of e-SSH model is shown. Here, $v$ and $\tilde{w}$ denote the intra and inter cell NN hopping respectively whereas $\tilde{\gamma}, \tilde{\gamma}_1$ represent NNNN hopping. $\tilde{w}$ and $\tilde{\gamma}, \tilde{\gamma}_1$ are considered  complex in general. In addition the model includes NN Rashba interaction as discussed in the text.}
	\label{fig:1d-cartoon}
\end{figure}

\section{Extended SSH model}
\label{sec:level2}
\subsection{ Model and Methods}
\label{sec:level1}

{\bf Hamiltonian:} The e-SSH model contains  i) usual NN intercell and intracell hopping represented by $v$ and $\tilde{w}$ respectively, ii) NNNN hopping among in-equivalent sub-lattices between NN and NNN unit cells represented by $\tilde{\gamma}$ and $\tilde{\gamma}_1$ respectively, iii) inter-cell and intra-cell NN time reversal (TR) broken Rashba interaction(represented by $\alpha_{R_1} {\rm and}~ \alpha_{R_2}$ respectively) which does not cause spin-flip process, and iv) inter-cell and intra-cell NN TR invariant Rashba interaction(represented by $\alpha_{R_3}~ \rm{and}~ \alpha_{R_4}$ respectively) which causes spin-flip hopping. We consider  $v, \tilde{w}, \tilde{\gamma}, \tilde{\gamma}_1$ to be complex.  For the readers, we refer the expression of full Hamiltonian in position space in  appendix, Eq. \ref{fullHamiltonian}.  In momentum space the Hamiltonian is given as,
\begin{eqnarray}
	\label{firsteq1}
&&	H = \sum_k \sum_{\sigma=\uparrow,\downarrow}\left(  h^k_{\sigma} c^{\dagger}_{k,a\sigma} c_{k,b\sigma}  +  g^{k}_{\sigma} c^{\dagger}_{k,a\sigma} c_{k,b\bar{\sigma}} + h.c \right) .~~~~~
\end{eqnarray}
\indent
In the above first term represents the diagonal blocks of the Hamiltonian for each spin sectors and second term represents the off-diagonal blocks. Here $h^k_{\uparrow}=h^k_{0} + i f^k$,  $h^k_{\downarrow}=h^k_{0} - i f^k$ with  $h^{k}_0= v + \tilde{w}^* e^{-ik} + \tilde{\gamma} e^{ik} + \tilde{\gamma}_1 e^{-2ik}$ and $f^k=\alpha_{R_1} - \alpha_{R_2} e^{-ik}$. $g^k_{\uparrow}= g^k=\alpha_{R_3} - \alpha_{R_4} e^{-ik}$ and $g^k_{\downarrow}=g^{k*}$.  $f^k$ and $g^k$ denote two different types of Rashba interactions  mentioned before. The complex hopping parameter $\tilde{w}, \tilde{\gamma}, \tilde{\gamma}_1$ are written in terms of real and complex part as follows, $\tilde{w}=w + i \xi, \tilde{\gamma}=\gamma + i \frac{\eta}{2}, \tilde{\gamma}_1=\gamma_1+ i \frac{\eta_1}{2}$. Without loss of generality and to reduce the parameter dependency, we mainly considered $\xi=\eta, \eta_1=0$. In some cases we keep both $\eta, \eta_1$ or otherwise $\eta=\eta_1$ as will be mentioned appropriately.  Below we discuss the methods to solve Eq. \ref{firsteq1}.\\
\indent
{\bf \underline{Case of either $\alpha_{R_3}, \alpha_{R_4}=0$ or $\alpha_{R_1}, \alpha_{R_2}=0$ :}}  For the choice  $\alpha_{R_3}, \alpha_{R_4}=0$,  Eq. \ref{firsteq1} is manifestly block diagonal in each spin-sector (with the basis $\Psi^{\dagger}= (c^{\dagger}_{k,a,\uparrow}, c^{\dagger}_{k,b,\uparrow},c^{\dagger}_{k,a,\downarrow}, c^{\dagger}_{k,b,\downarrow})$) with  eigenvalues $E(k)_{\sigma}=\pm| h^k_{\sigma}|$.  We note that in the case $\alpha_{R_1}=\alpha_{R_2}=0$, the Hamiltonian has off-diagonal element in $\Psi$ basis. However, we  can define (see text before  Eq. \ref{hambarbasis} and Eq. \ref{u1u2eq}) an appropriate linear combinations to construct new fermionic operators $d_{k,\bar{a},\bar{\sigma}}, d_{k,\bar{b},\bar{\sigma}}$ where ($\bar{a}, \bar{b}$) and ($\bar{\sigma}=\bar{\uparrow}, \bar{\downarrow}$) denote pseudo sublattice and spin indices respectively. In this representation,  Hamiltonian has the form  $\bar{H} = \sum_{k,\bar{\sigma}}\left(  \bar{h}^k_{\bar{\sigma}} d^{\dagger}_{k,\bar{a}\bar{\sigma}} d_{k,\bar{b}\bar{\sigma}} + h.c \right)$ (see Eq. \ref{hambarbasis} ). The expressions for $\bar{h}^k_{\bar{\sigma}}$ are similar to ${h}^k_{\sigma}$  with $f^k$ replaced by $g^k$. This implies that these two different cases are equivalent upto a basis transformation. \\
\indent
{\bf \underline{Case of $\alpha_{R_1}, \alpha_{R_2}, \alpha_{R_3}, \alpha_{R_4} \ne 0$:}} Here Hamiltonian can not be represented in two separate $2 \times 2$ diagonal blocks.  However it is convenient to define a basis $\Upsilon^{\dagger}=  (c^{\dagger}_{k,a,\uparrow}, c^{\dagger}_{k,a,\downarrow},c^{\dagger}_{k,b,\uparrow}, c^{\dagger}_{k,b,\downarrow})$ where the Hamiltonian is represented by non-zero off-diagonal $2 \times 2$ blocks and vanishing diagonal $2 \times 2$ blocks  as shown in Appendix, Eq. \ref{newphibasisH}. These off-diagonal blocks are non-hermitian.  In Appendix~\ref{exsoltn} we show explicitly that the non-hermitian off-diagonal blocks can be diagonalized  to have complex eigenvalues $\lambda_1, \lambda_2$. Using the orthonormality condition of eigenvectors of both the off-diagonal blocks, one may construct an appropriate fermionic basis such that the  Hamiltonian has only non-zero diagonal blocks where each block has only off-diagonal elements (see Eq. \ref{finalphiHam}). The Hamiltonian looks $\tilde{H}=\sum_{k} ( \lambda^k_1 \eta^{\dagger}_{k,\bar{a},\bar{\uparrow}} \eta_{k,\bar{b},\bar{\uparrow}} + \lambda^k_2 \eta^{\dagger}_{k,\bar{a},\bar{\downarrow}} \eta_{k,\bar{b},\bar{\downarrow}} + h.c)$. Here $\eta_1, \eta_2$ are the new fermionic operators introduced. The expressions of $\lambda^k_i$ and dispersion are obtained as,
\begin{eqnarray}
\lambda^k_{1(2)} = h^{k}_0 \pm i \sqrt{f^{k^2} + g^{k^2}},~ E_i(k)= \pm |\lambda^k_i|, ~i=1,2.
\end{eqnarray}
The above formula is one of the central results of this work and it facilitates to investigate the topology in terms real and imaginary part of $\lambda^k_{1/2}$. This also helps to investigate the gap closing condition and other aspect of topology  of the system in an analytically tractable  way.\\
\indent
 For simplicity in the following we present the case $\alpha_{R_1}, \alpha_{R_2} \neq 0$ only. We note that this case is equivalent to $\alpha_{R_3}, \alpha_{R_4} \neq 0$ as well.  When all $\alpha_{R_i}$s are present, results are changed qualitatively and would be mentioned in due course.\\
 \indent
{\bf Symmetry of the system:}  When all the hopping parameters are real the system preserves time-reversal ($T$), particle-hole ($C$) and chiral symmetry($S$) implying its symmetry class to be BDI\cite{Ahmedi-2020-PhysRevB.101.195117,polina-2023-PhysRevB.107.075422}. The respective operators are obtained as $C= \sigma_0\otimes \sigma_z \kappa$, $T= \sigma_x \kappa \otimes \sigma_0$ and $S= T C= \sigma_x \otimes \sigma_z$. One can easily check that the block Hamiltonian transforms in $\Psi$ basis as $C H(k)C^{-1}=-H(-k)$, $TH(k)T^{-1}=H(-k)$ and $SH(k)S^{-1}=-H(k)$. However,  $\tilde{\gamma}, \tilde{\gamma_1}$ (which are complex) breaks the $T$ and $C$ symmetry making  the system  is in  AIII class. Note that  complex ($v, \tilde{w}$) do not break the $T$ symmetry as this can be taken into account by a gauge transformation and dispersion is related by $E_{\uparrow}(k)=E_{\downarrow}(-k)$. The gap closing condition for both the spin sector occurs at $k_{\uparrow}=-k_{\downarrow}=k$ with $\cos k = \frac{\alpha_{R_1} \alpha_{R_2} - vw} {w^2 + \alpha^2_{R_2}}$ implying same topological phase for both sectors. 
\subsection{\label{sec:topocharac}On determining topology and gap closing point}
As the Hamiltonian can be written into two $2 \times 2$ blocks with only off-diagonal elements $\lambda_i$ where $i=1,2$, the winding number for each sector can be defined as ~\cite{ryu2010topological,chiu2016classification,mondal2023topological},
\begin{equation}
W_i=\Bigg| \pm \frac{i}{2 \pi} \int_{- \pi}^{\pi} dk \hspace{0.1 cm}  \left[\left\{ \lambda^k_i\right\}^{-1} \partial_{k} \lambda^k_i\right] \Bigg|
\label{eq:static_wind}.
\end{equation}
For investigating the gap-closing condition one needs to  solve the equations for real and imaginary part of $\lambda_i$ to be zero separately. Presence of both $f^k$ and $g^k$ makes the analysis little complex though completely tractable as discussed in Appendix (see Sec. \ref{gapcloseallalpha}). For simplicity here we consider the case $f^k=0$ (which is equivalent to $g^k=0$). Inclusion of $f^k$ and $g^k$ only qualitatively changes the result. To understand this, we write the real($h_{1x}$) and imaginary part($h_{1y}$) of $\lambda_1$  as given below.
\begin{eqnarray}
\label{firsthx}
&h^k_{1x} =v+ \mathcal{C}^1_{x} \cos k + \mathcal{D}^1_{x} \sin k + \gamma_1 \cos 2k -\frac{\eta}{2} \sin 2k~,~~~~~~~ & \\
\label{firsthy}
&h^k_{1y} =\alpha_{R_1}+ \mathcal{C}^1_{y} \cos k + \mathcal{D}^1_{y} \sin k + \frac{\eta_1}{2}\cos 2k -\gamma_1 \sin 2k~.~~~&
\end{eqnarray}
In the above $\mathcal{C}^1_x= \gamma + w$, $\mathcal{C}^1_y= \frac{\eta-\xi}{2} - \alpha_{R_2}$, $\mathcal{D}^1_x= - (\frac{\eta+ \xi}{2} + \alpha_{R_2})$ and $\mathcal{D}^1_y= \gamma-w$.
Corresponding equation for sector-2 is obtained by changing the sign of $\alpha_{R_i}$.\\
\indent
A simple inspection of Eq. \ref{firsthx} and Eq. \ref{firsthy} and the corresponding equations for sector-2, tells that in the absence of $\eta,\xi,\eta_1$ i.e when $w,\tilde{\gamma},\tilde{\gamma}_1$ are real, then for a given pair of $\left( k, k^\prime(-k)\right)$, $h^k_{1x}, h^k_{1y}$ can be mapped to $\pm h^{k^{\prime}}_{2x}, \pm h^{k^{\prime}}_{2y}$. This renders the winding number of both the sector to be identical. However in the presence of finite $ \eta, \xi, \eta_1$, no such simple mapping is present and one gets  $h^k_{1x} \neq h^{k^{\prime}}_{2x}$ and $h^k_{1y} \neq h^{k^{\prime}}_{2y}$ for any pair of $(k,k^{\prime})$. In the appendix Sec. \ref{w1w2discussion}, we explain this fact in more details. This opens up the possibility of having different winding number in each sectors for complex $\tilde{\gamma},\tilde{\gamma}_1$. \\
\indent
Now we briefly discuss the parametric equation in $h_x-h_y$ plane which is widely used in investigating SSH model. Normally it is obtained by eliminating  `$k$' dependency from $h_x, h_y$.  From Eq.  \ref{firsthx} and Eq. \ref{firsthy}, it is clear that in the absence of $\sin 2k, \cos 2k$, terms i.e for $\gamma_1,\eta_1=0$, one can easily solve for $\cos k $ and $\sin k$. The trigonometric identity  $\cos^2k + \sin^2k =1$ yields  equation of ellipse in simple cases. In Appendix, Fig. \ref{contour}, we present the various scenarios. It shows that presence of a real $\gamma$ in addition to real $v,w$ makes the contour elliptic. Inclusion of $\alpha_1, \alpha_2$  makes the contour for each spin sector to be different but with identical circumference length ($\mathcal{S}$). Now making $\gamma$ complex  yields different $\mathcal{S}$ for each spin sectors.\\
\indent
 It is clear that in the absence of $\gamma_1, \eta_1$, an easy solution of $\cos k$ and $\sin k$ are obtained from the gap-closing condition $h_{ix}=h_{iy}=0$. In the presence of $\gamma_1, \eta_1$, the  Eq. \ref{firsthx} and Eq. \ref{firsthy} can be used either to eliminate $\cos 2k$ or $\sin 2k$ and obtain a cubic equation in $\sin k$ or $\cos k$ which can be solved numerically as given in Eq. \ref{cubicx123}.  The details of these solutions are given in  Appendix Sec. \ref{gapclosealpha34zero}. In addition to the algebraic method, it can also be solved transcendentally which can be applied even for finite $\alpha_{R_3},\alpha_{R_4} \neq 0$ as discussed in Appendix Sec. \ref{gapcloseallfinite}. It is remarkable to note that our formalism yields a simple  gap-closing condition for real $v,w,\alpha_1,\alpha_2,\alpha_3,\alpha_4$. We define  vectors $\vec{V}= v \hat{x} + \alpha_1 \hat{y} + \alpha_3 \hat{z}$ and $\vec{W}= w \hat{x} - \alpha_2 \hat{y} - \alpha_4 \hat{z}$ and the gap-closing condition reads as, $V^2 + W^2 \cos 2k + 2 \vec{V}\cdot\vec{W} \cos k=0 $.
The above equation resembles the equation $v^2 + w^2 \cos 2k + 2 v w \cos k=0$ in the absence of $\alpha_{R_i}=0$ and manifest the mathematical equivalence between the parameters. Thus our analysis provides a complete scope of understanding the topological aspect of an e-SSH model in the case of unbroken chiral symmetry. 
\begin{figure}[h!]
\centering
\includegraphics[width=1.0\linewidth,height=0.3\textwidth]{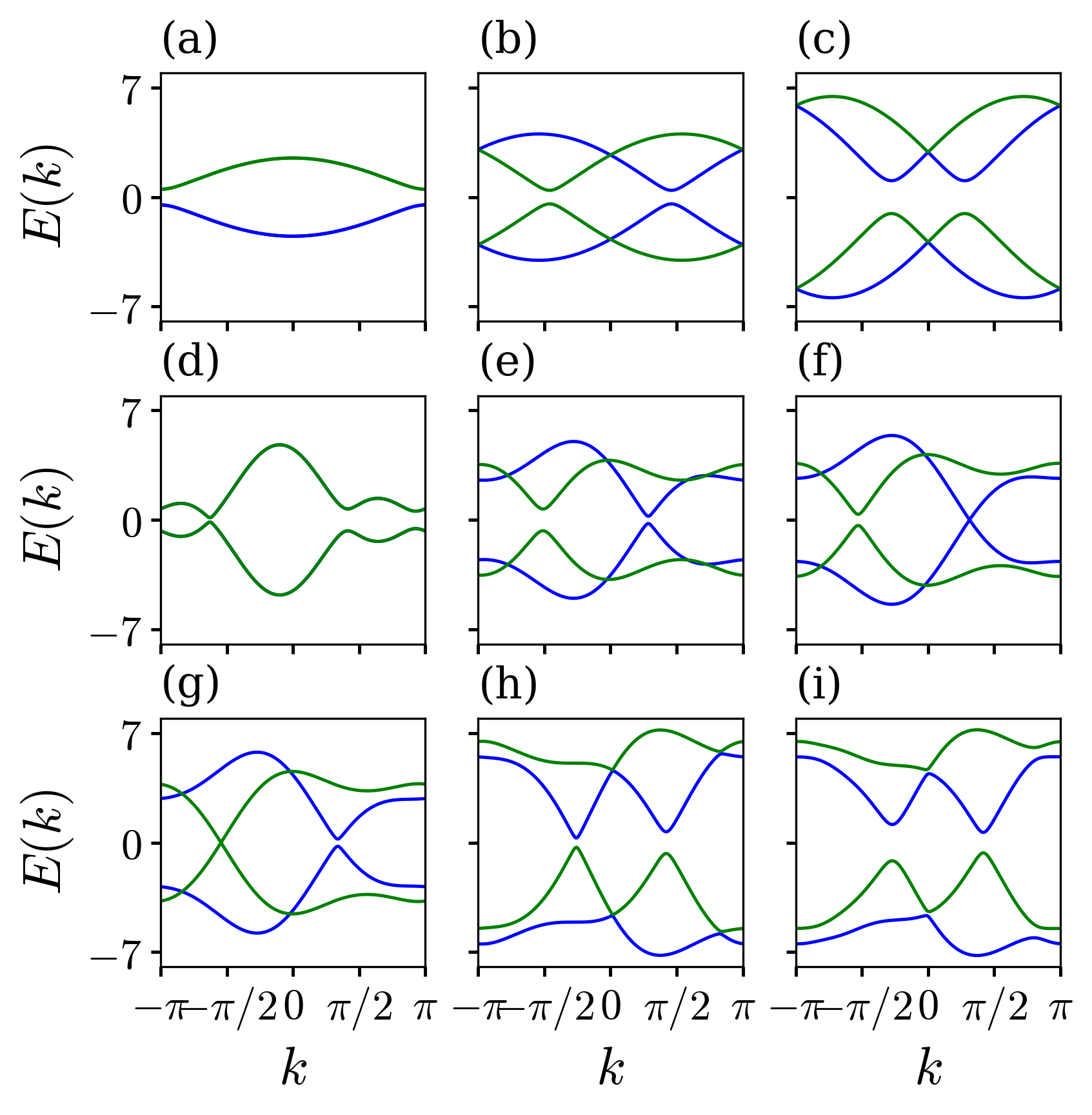}
\caption{Dispersion relation ($E_k$ vs $k$) plots for various sets parameters. In all panels, we set $v = 1$. Panels (a–c) use parameters $\gamma = \gamma_1 = 0$, $\eta = \eta_1 = 0$, while panels (d–i) use $\gamma = \gamma_1 = 0.5$, $\eta = \eta_1 = 1$. Panels (a–e) correspond to $w = 1.5$, while for panel f $w$ is 1.9 and for panels (g,h,i) $w = 2.3$. $\alpha_{R_1} = 2$ and $\alpha_{R_2} = 1$ in panels (b, c, e, f, g, h, i). Additionally, for panels (c, h) and (i), we use $\alpha_{R_3} = 3$ (2) and $\alpha_{R_2} = 2$ (3), respectively.  Panel (c), (h) and (i) manifest the role of finite $\alpha_{R_3}$ and $\alpha_{R_4}$ in producing asymmetric-Dirac nodes. Panel (a) and (d) show that in the absence of $\alpha_{R_i}$ bands are two fold degenerate. While panel (b) and (c) highlight the fact when $\gamma$ is real, bands are time-reversal symmetric.}
\label{EnergySSh}
\end{figure}
\\
\begin{figure}[h!]
\centering
\includegraphics[width=1\linewidth,height=0.45\textwidth]{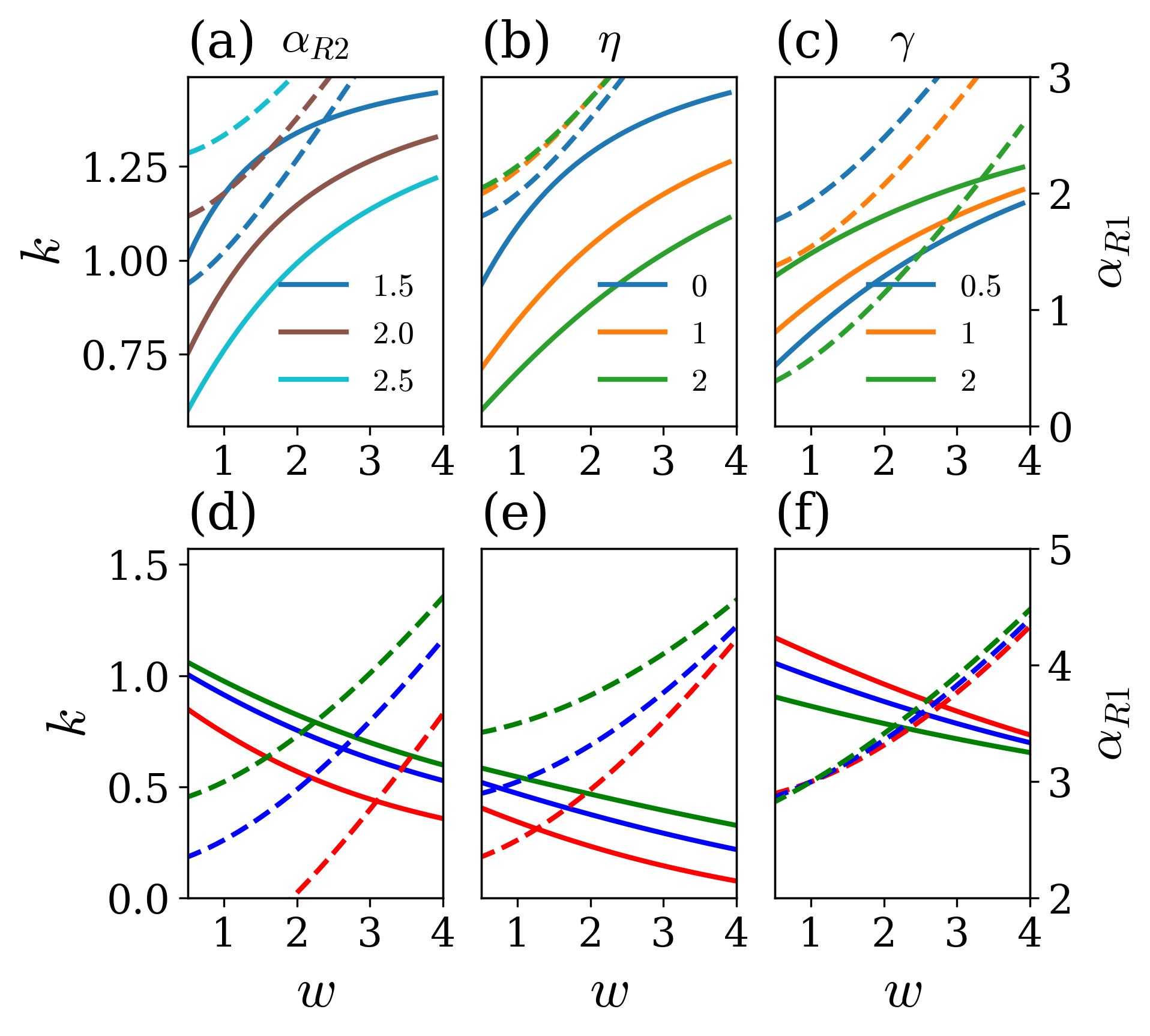}
\caption{Gap closing momentum $k$ is plotted against $w$ solid lines.   $v=1$ is taken for all panels. The dashed line shows $\alpha_{R_1}$ determined self-consistently as labeled by left y-axis. In panel (a) and (d) various solid lines denote various values of $\alpha_{R_2}$.   In panel (b) and (e) various solid lines denote various values of $\eta$.  In panel (c) and (f) various solid lines denote various values of $\gamma$. Further In panel (a) we consider  $\eta=\eta_1=0, \gamma = \gamma_1=0$ while $\alpha_{R_1}$. In the panel (b) parameters  considered are $\alpha_{R_2}=2.0, \gamma=0.5, \eta_1=\gamma_1=0$. In panel (c) we show the dependence on various values of $\gamma$ with   $\eta=2, \eta_1=\gamma_1=0, \alpha_{R_2}=2$. The panel (d) corresponds various values of $\alpha_{R2}$ with $\eta=\eta_1=2, \gamma=\gamma_1=0.5$. Panel (e) depicts the case   with different values of $\eta$ with fixed $\gamma=\gamma_1=0.5, \eta = \eta_1=2$. Finally panel (f) corresponds $\eta=\eta_1=1, \gamma= \gamma_1, \alpha_{R_2}=2$.  }
\label{kpoints}
\end{figure}
\subsection{\label{sec:bbc} Results}
\begin{figure}[h!]
\centering
\includegraphics[width=\linewidth]{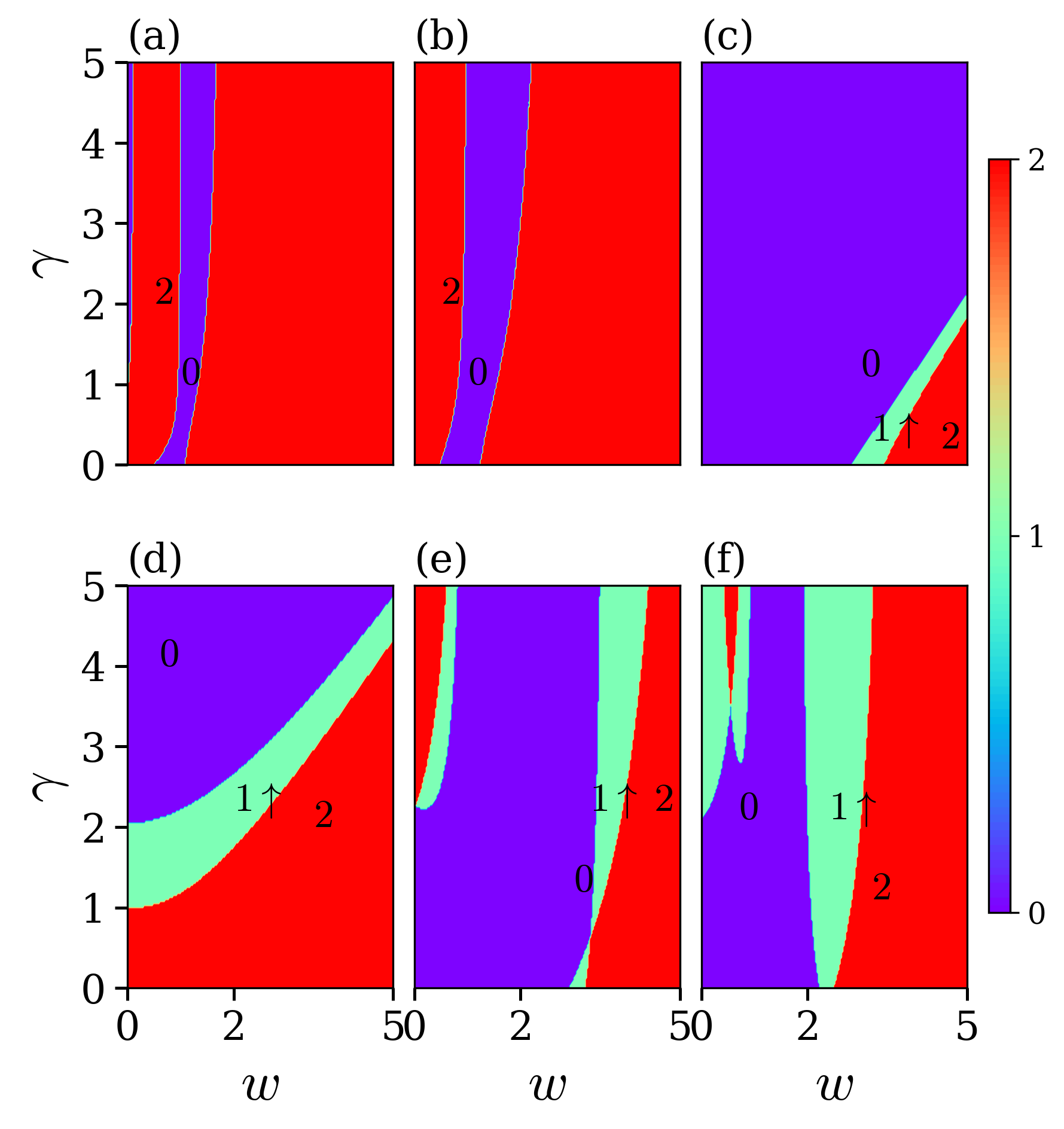}
\caption{Phase diagrams are plotted in $\gamma-w$ plane for different set of parameters.  $v=1$ is kept for all panels.  Panel (a) and (b) shows effect of $\gamma$ and $\eta$ only with $\gamma=\gamma_1, \eta=\eta_1=1$ and $2$ respectively. Panel (c) and (d) consider finite $(\alpha_{R_1}, \alpha_{R_2})=(3,1)$ and (1,3) respectively in addition to $\eta=1, \eta_1=\gamma_1=0$. Panel (e) and (f) depict the case of finite $\gamma_1=\gamma, \eta=\eta_1$ compared to panel (c) with value of $\alpha_{R_2}$ being 1 and 2 respectively.}
\label{phasediag2}
\end{figure}

 We now present graphically some of the aspects discussed above.  We first begin with energy-spectrum and gap closing point. In Fig. \ref{EnergySSh}, we present dispersion for various  sets of parameters.  Panel (a) shows the dispersion for  original SSH model with only $(v,w)=(1,1.5)$. In panel (b), finite $(\alpha_{R_1}, \alpha_{R_2})=(2,1)$ is considered which shows that the dispersion for up and down spins becomes non-degenerate but preserves the symmetry $E_{\uparrow}(k)=E_{\downarrow}(-k) $ with a band crossing at $k=0$ due to time-reversal and chiral symmetry. Panel (c) shows the effect of additional Rashba interaction $(\alpha_{R_3},\alpha_{R_4})=(3,2)$. Interestingly, due to the hybridization between the bands, the band crossing(in panel (b)) is turned to band touching at finite energy at $k=0$. The crossing hosts a linear symmetric Dirac cone. Panel (d) shows the effect of a complex $\gamma=\gamma_1=(0.5 + i )$(with all $\alpha_{R_i}=0$) preserving the spin-degeneracy of the bands. In panel (e) one notices that inclusion of $\alpha_{R_1}, \alpha_{R_2}$ breaks the spin-degeneracy of the bands and also $E_{\uparrow}(k) \neq E_{\downarrow}(-k)$, a consequence of broken time-reversal($T$) and chiral symmetry ($C$). Panel (f) shows the effect of changing $w$ from $1.5$ to $1.9$ to cause the gap closing. Panel (g) shows effect of further increase of $w$ to 2.3 which causes band closing in opposite direction of momenta. Panel (h) and (i) shows the effect of $\alpha_{R_3}$ and $\alpha_{R_4}$ which control the band gap condition in varying degree.  We note that inclusion of all $\alpha_{R_i}$ and $\eta$ makes the band touching at non-zero $k$ value in general and the Dirac cone becomes asymmtric governed by $\eta$. The general expression for the asymmetric Dirac cone is given by $E(\delta k)  = E_0(k_0) + F_1(k_0) |\delta k| +  F_2(k_0) \delta k$.
The expressions for $F_1(k_0), F_2(k_0)$ and detailed derivation is provided in the Appendix \ref{effham}, Eq. \ref{effhameq}. We note that when $\eta=\eta_1=0$, $F_2(k_0=0)=0$.\\
\indent
 In Fig. \ref{kpoints} we show(by solid lines) how the gap-closing point varies for different parameter choices in $k-w$ plane. We  note that each plot contains a free parameter which is determined by the trigonometric identity $\sin^2k + \cos^2k =1$ and we have considered $\alpha_{R_1}$ as the free parameter. Interestingly we notice that in panel (a), (b) and (c) the gap closing condition monotonically increases with $\gamma_1, \eta_1=0$.  Making these parameters finite, gap-closing point monotonically decreases as confirmed by the panels (c), (d) and (e). This shows the presence of $\cos 2k$ and $\sin 2k$ terms in the gap-closing condition. \\
 \indent
\begin{figure}[h!]
	\centering
	\includegraphics[width=0.95\linewidth]{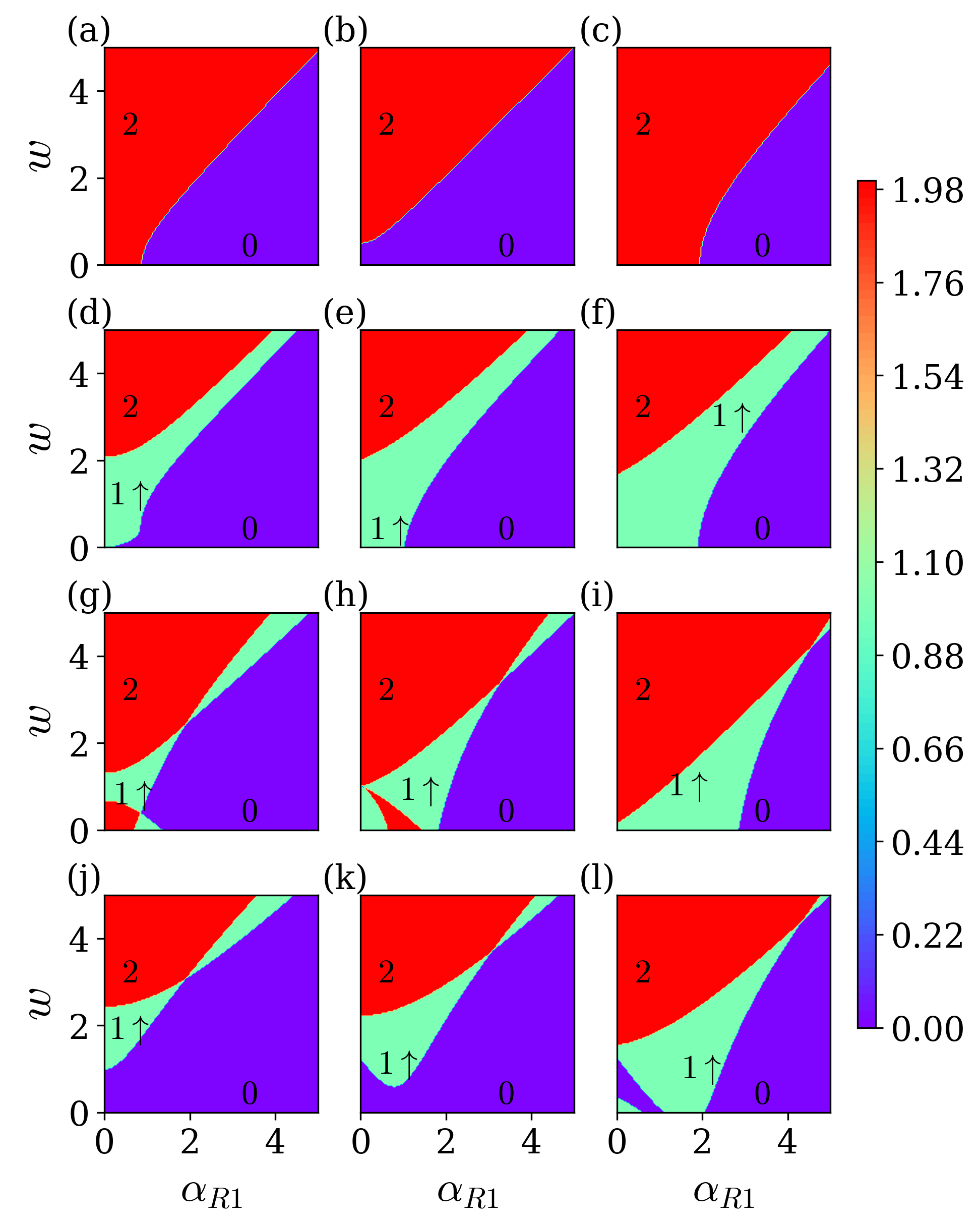}
	\caption{Phase diagram in $\alpha_{R_1}-w$ plane is depicted. Different colored lines separate two different topological phases (which are represented by winding numbers of same color) for a given set of parameters. Values of $\alpha_{R_2}$ are taken as 0, 1, 2 for first, second and third row. For first, second, third and fourth rows  the set of parameters  $\eta, \eta_1, \gamma, \gamma_1$ are taken as $( 0.0, 0.0,0.0,0.0)$, $ (2.0, 0.0, 0.5, 0.0)$, $(1.0, 1.0, 1.5, 1.5) $ and $(2.0, 2.0, 0.5, 0.5)$ respectively. In panel (g), (h), (j) etc we notice re-entrant topological phase transitions. Second row highlights how finite $\eta, \gamma$ cause different topology for different spin sectors. Further, third and fourth row depict how inclusion of $\eta, \eta_1, \gamma, \gamma_1$ cause re-entrant topological phase transition. }	
	\label{phasedgrm-withalpha12}
\end{figure}
  In Fig. \ref{phasedgrm-withalpha12} we show phase diagram in $\alpha_{R_1}-w$ plane.  Yellow, cyan and magenta color shows phases with winding number two, one and zero respectively.  The arrow adjacent to the numeric 1 indicate the spin-sector which is topological.  For the first, second and third column $\alpha_{R_2}$ is taken 0,1 and 2 respectively. In panel (a),(b) and (c) only $\alpha_{R_i}, w, v$ are finite implying winding number to be either zero or two.  Panel (d),(e) and (f) show the effect of $\eta, \gamma$ to break the degeneracy of winding number. In panel (g), (h) and (i)(and (j), (k), (l)) we show that effect of $\gamma_1, \eta_1$ for different values. We note that panel (g) to (i) show re-entrant topological phase.  For the effect of finite $\alpha_{R_3}, \alpha_{R_4}$ we refer Fig. \ref{phasedgrm-withalpha34} in Appendix. In Fig. \ref{windingSSh1}, we plot the phase diagram in $v-w$ plane and show the effect of $(\eta, \gamma)$ and $\alpha_{R_3}, \alpha_{R_4}$. $\alpha_{R_3}, \alpha_{R_4}$ is zero ( and finite with values (0.5,1)) for first (and second row).  Panel (b) and (c) show how the winding number 1 appears in the presence of $\eta_1, \gamma_1$. It also shows  re-entrant topological phases.  Panel (e) and (f) shows how the phase diagram changes in the presence of $\alpha_{R_3}$ and $\alpha_{R_4}$ in correspondence to panel (b) and (c). In Fig. \ref{phasediag2}, we show the phase diagram in $\gamma-w$ plane to depict the interplay of different parameters. For all panels we kept $v=1$. Panel (a) and (b) show what happens of only $\gamma=\gamma_1$ and $\eta=\eta_1$ are present and values of $\eta$ is changed 1 to 2. As there is no $\alpha_{R_i}$ is present, we have only have even winding number. The red region which is topological is separated by a non-topological phase whose width increases as we increase the value of $\eta$.  Panel (c) and (d) shows the effect of $\alpha_{R_1}, \alpha_{R_2} $ which causes winding number 1 to appear. The width of the topological phase also changes as the values of $\alpha_{R_1}, \alpha_{R_2} $ change. Panel (e) and (f) shows the case when $\eta, \gamma$ and $\alpha_{R_1}, \alpha_{R_2}$ are present. This shows how the topological phase with winding number 1 emerges from panel (a) and (b) when $\gamma_1, \eta_1$ is made finite in comparison to panel (c) and (d). \\

\begin{figure}[h!]
	\centering
	\includegraphics[width=\linewidth]{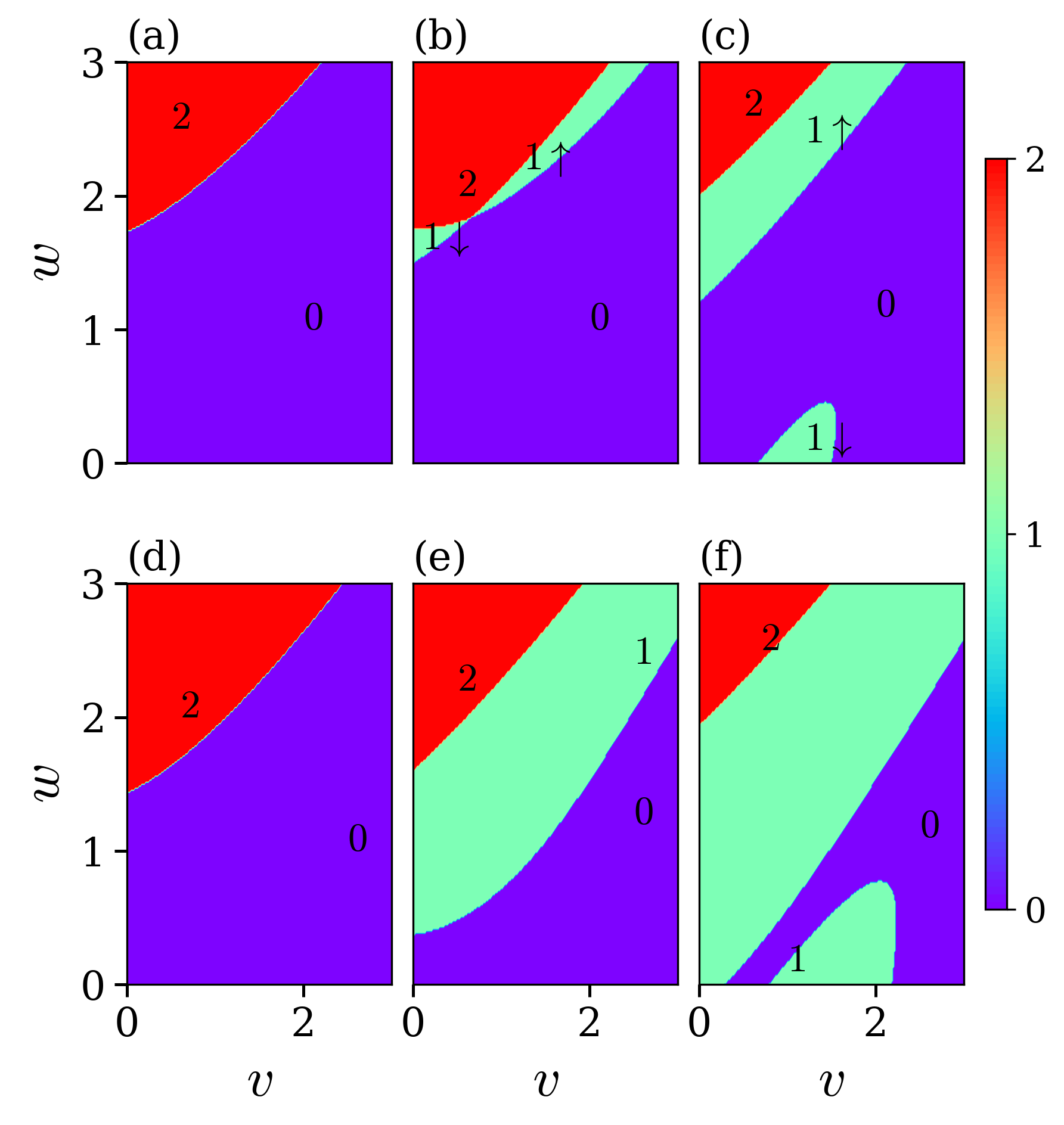}
	\caption{Density plots of winding number are presented in
	$w-v$ plane. The red, cyan and magenta region correspond $W =
		2,1,0$ respectively. In all panel $(\alpha_{R_1},\alpha_{R_2})=(2,1)$ and $\gamma=\gamma_1$ .  Upper panels (lower panels) corresponds ($\alpha_{R_3}, \alpha_{R_4})=(0.0)$ (0.5,1). Value of $(\eta,\gamma)$ for different panel is set as (0,0.5) for panel (a), (1,0.0) for panel (b), (1,1.5) for panel (c), (0,0.5) for panel (d), (2,0.5) for panel (e) and (2,1.5) for panel (f). We notice that effect of finite $\eta$ causes  $W=1$ in all panels except in panel (a) and (d) where $\eta$ is zero.}
	\label{windingSSh1}
\end{figure}
\indent

\section{\label{sec:twoandfourlevel} Coupled chains  and fate of
	SPT in 1D to 2D.}
\subsection{ Topology of finite coupled chains}
\label{sec:coupledchain}
\begin{figure}[h!]
\centering
\includegraphics[width=\linewidth]{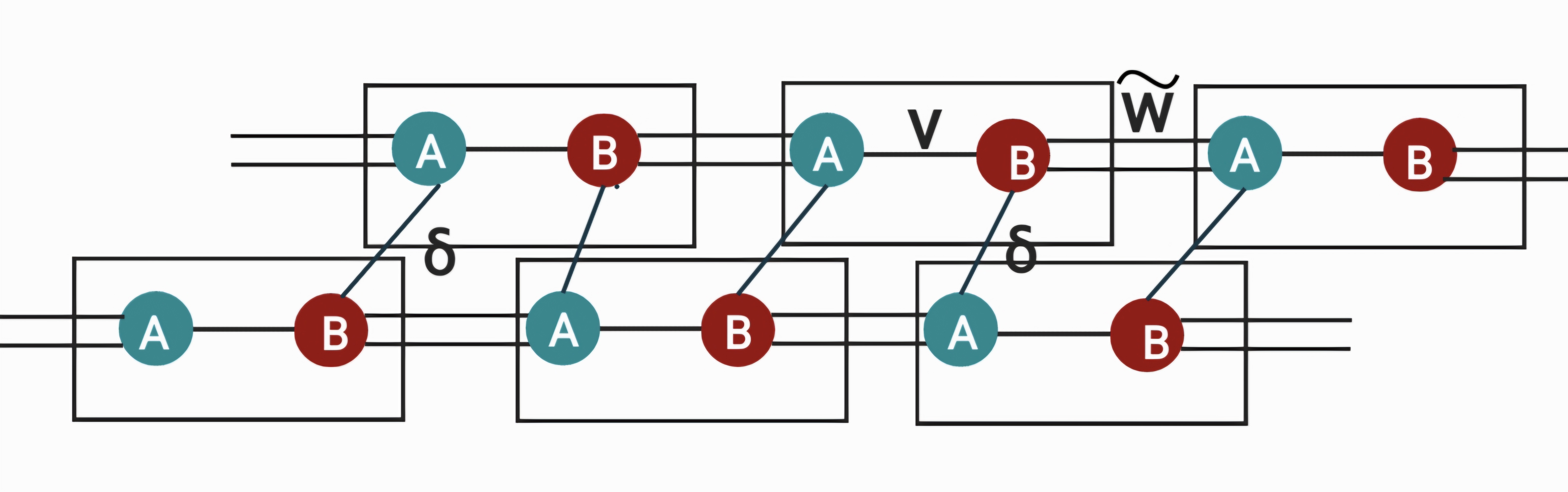}
\caption{A cartoon picture of two coupled e-SSH chains is shown in two different way. In the upper panel pink and magenta  denote the unit cell of each SSH chain where (a,b) (and (c,d)) sites belong to first chain (and second chain). $\delta$ denotes the coupling between the `$b$' and `$c$' (`$a$' and `$d$') sub-lattices.   We note that sites `$c$' and `$d$' of the second chain are equivalent sub-lattices of `$a$' and `$b$' in the first chain.}
\label{SSH_chain2}  
\end{figure}

Having described the various topological properties of single chain e-SSH Hamiltonian elucidating the role of different parameters on how they modify the topology, we now examine few such coupled chains and its consequences. In our scheme, chains are connected by a real coupling $\delta$ which allows only inter sub-lattice hopping between the chains such as to preserve the Chiral symmetry as shown in Fig. \ref{SSH_chain2}. The  coupling between two neighboring e-SSH chains can be represented by the following inter-chain Hamiltonian,
\begin{eqnarray}
	\label{Two_Chain_Hamiltonian1}
H_{c}&& = \sum^{\mathcal{M}}_{m=1} \delta \sum_{n =i}^{N} \left( \psi^{\dagger}_{m,n,d} \psi_{m+1,n,\bar{d}} +  {\rm h.c }  \right) .
\end{eqnarray}
In the above `$m$' and `$m+1$' are indices for chains  and $n$ denotes the $n$-th unit cell in the chain. We note that the interchain coupling only contains intra unit cell hopping for simplicity.  `$d$' and `$\bar{d}$' denote the sublattice indices `$a$' and `$b$' such that $d \ne \bar{d}$ as shown in Fig. \ref{SSH_chain2}. \\
\indent
 To understand the $\mathcal{M}$ couple chain, it is instructive to understand the two chain Hamiltonian in momentum space as given in Appendix Eq. \ref{twochainkham}. This shows that Hamiltonian can be written as a sum of many one dimensional e-SSH models associated with fermions $\xi_i$ whose intra-cell hopping parameter $v$ is modified to $v\pm \delta$ . Various definition of $\xi_i$-fermions are provided in terms of symmetric and antisymmetric combinations of original fermions and we  refer the readers discussion after Eq. \ref{twochainkham}. Similar construction can also be done for four and other even number of  chains. As each $\xi_i$ sectors are evidently chiral symmetric, the winding number for each $\xi_i$ sectors can be independently calculated and an effective analysis be made for such coupled chain system. We also note that the full Hamiltonian can also be written in chiral basis showing the chiral symmetry of the full system. In this case the Hamiltonian can be written  as $2\mathcal{M} \times 1\mathcal{M}$ off-diagonal elements as given in Eq. \ref{two-chain-ham-antd} and Eq. \ref{four-chain-ham-antd} for two and four coupled chains. The procedure can be readily generalized for $2\mathcal{M}$ chains.  \\
 \indent
\begin{figure}[h!]
\centering
\includegraphics[width=\linewidth]{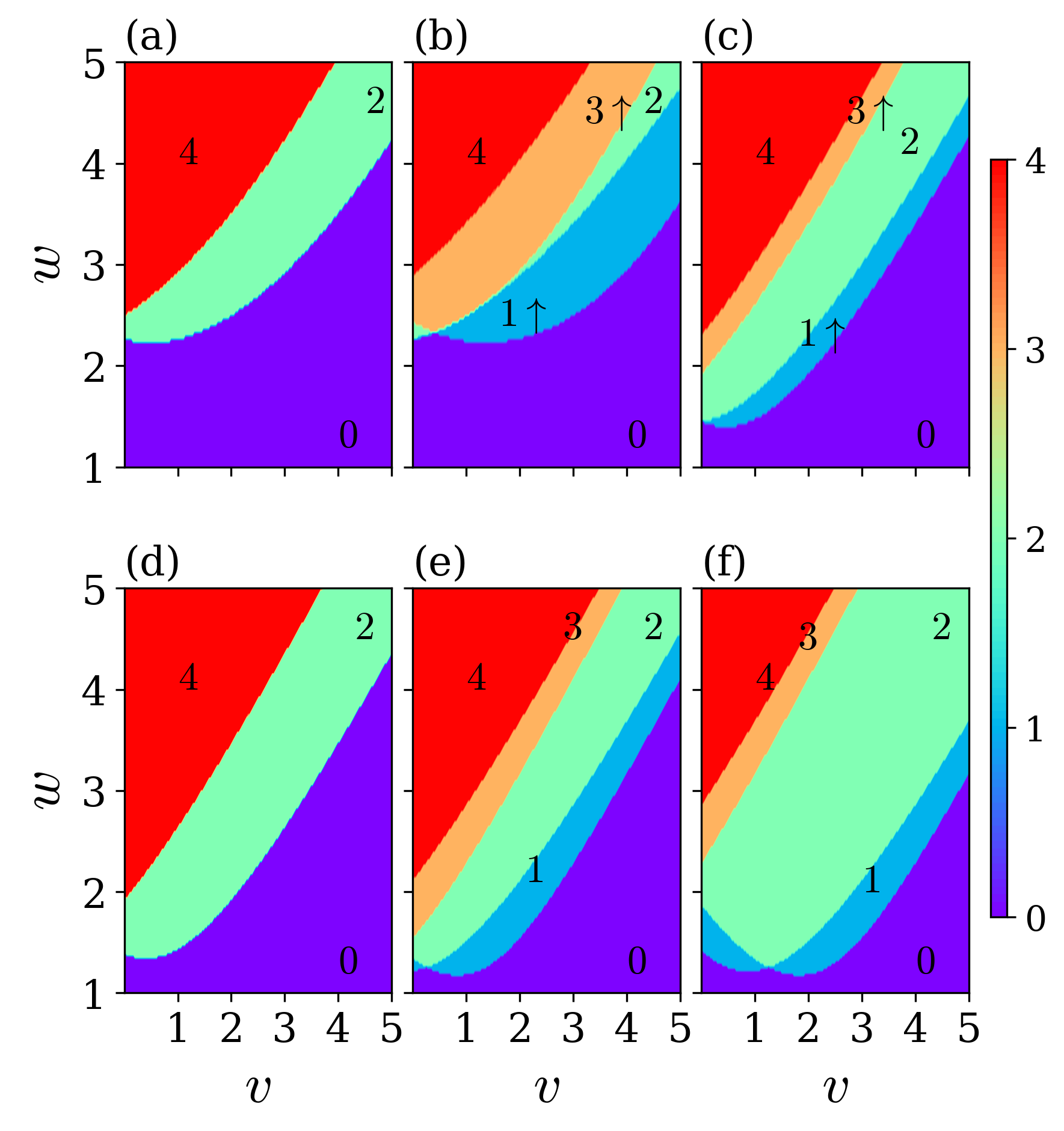}
\caption{Density plot of winding number in $w-v$ plane. Inter-chain hopping parameter $\delta$ is taken 1 for all panel except the panel (f) where $\delta=2$. For all the panels $(\alpha_{R_1}, \alpha_{R_2})=(2,1)$ and $\gamma=0.5$. For first (second) row $(\alpha_{R_3},\alpha_{R_4})=(0.0)$((0.5,1)). $(\gamma_1,\eta_1)=(0,0)$ in panels (a), (b) and (d). For rest of panels it is $(0.5,1)$. $\eta$ is zero for panels (a), (d) and 1 for rest of the panels.}
\label{twochainphased1}
\end{figure}
Fig. \ref{twochainphased1} illustrates the plot for the winding number in $v-w$ plane. Panel (a) depicts the effect of finite $\alpha_{R_1}, \alpha_{R_2}, \gamma$ where winding number takes only even integers. Panel (b) shows the effect of finite $\eta$ in comparison to panel (a). Panel (c) shows the effect  of finite $(\gamma_1, \eta_1)$ in addition to the  parameter considered in panel (b). Panel (d) shows the qualitative changes from panel (a) for  finite $(\alpha_{R_3}, \alpha_{R_4})=(0.5,0.1)$. Panel (e) shows the emergence of odd winding number for inclusion of finite $\eta$ with the parameters considered in panel (d) (though finite $(\gamma_1, \eta_1)=(0.5,1)$ have also been considered). Panel (f) shows the effect of changing the inter-chain coupling $\delta$ from 1 to 2 in comparison to panel (e). Fig. \ref{fourchainphased1} illustrates the plot for the winding number for four coupled chains. Every panels in Fig. \ref{fourchainphased1} have identical parameters considers in same panels in Fig. \ref{twochainphased1}. We note that addition of chains  does not change the odd or evenness of the winding number and also all possible winding number (bounded by 2$\mathcal{M}$)  are observed whenever $\eta$ or $\eta_1$ are finite.  The  result can be easily generalized for $\mathcal{M}$ coupled chain where winding number can vary from $0$ ro $2 \mathcal{M}$. Whenever we obtain a topological phase with certain winding number $p$, the open chain configuration is associated with $2p$  MGZE states localized at the ends as shown in Fig. \ref{mid-gap-all-chain}.  Thus we observe that the zero-energy modes can be controlled easily by manipulating the hopping parameters which could be of practical uses such as  switching devices. \\
\begin{figure}[h!]
\centering
\includegraphics[width=\linewidth]{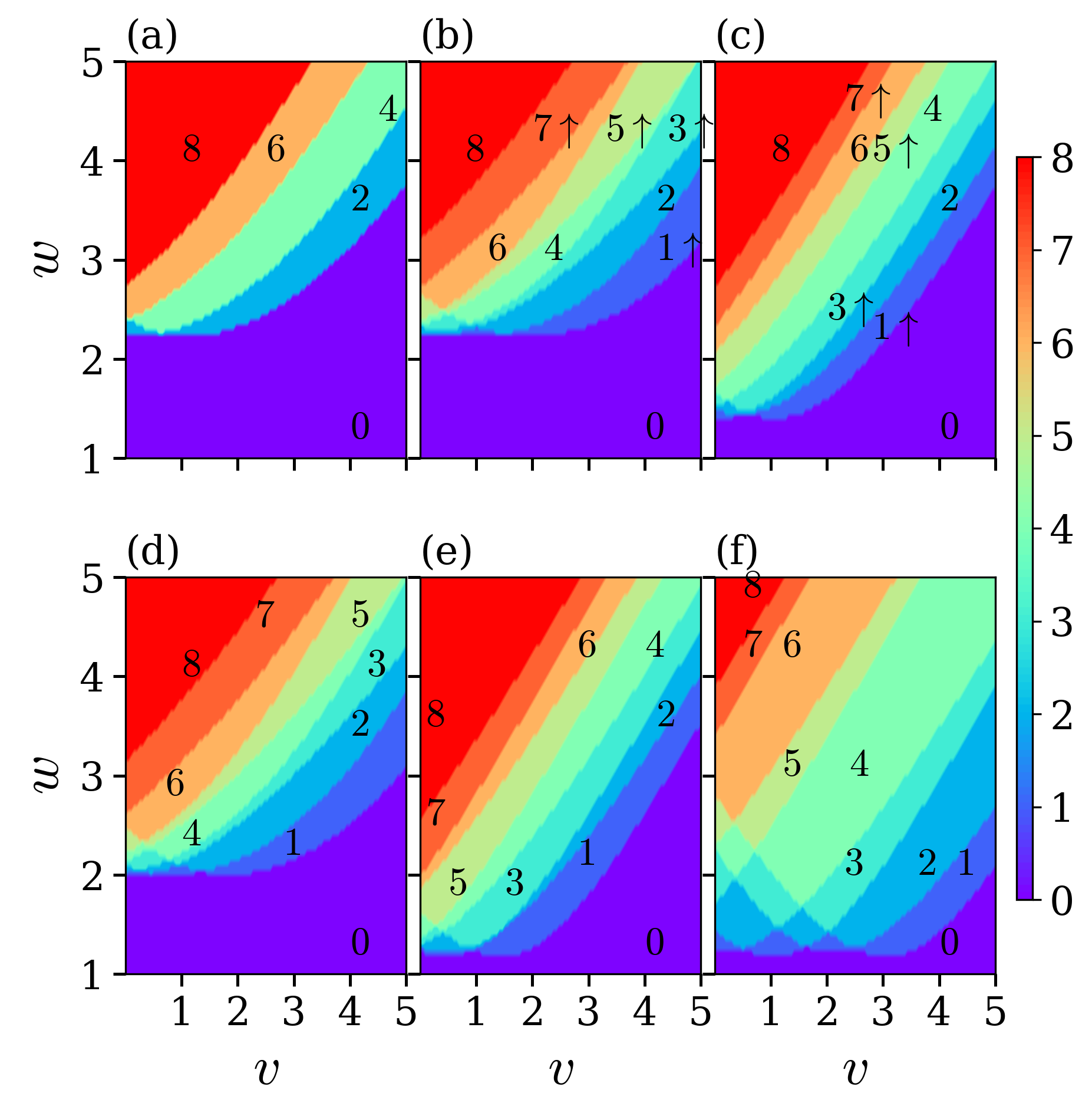}
\caption{Winding number is plotted in $w-v$ plane for four coupled chains. The parameters of each panel considered corresponds to the paraneters of panel in Fig. \ref{twochainphased1}.}
\label{fourchainphased1}
\end{figure}
\begin{figure*}[h!]
\centering
\includegraphics[width=0.8\textwidth]{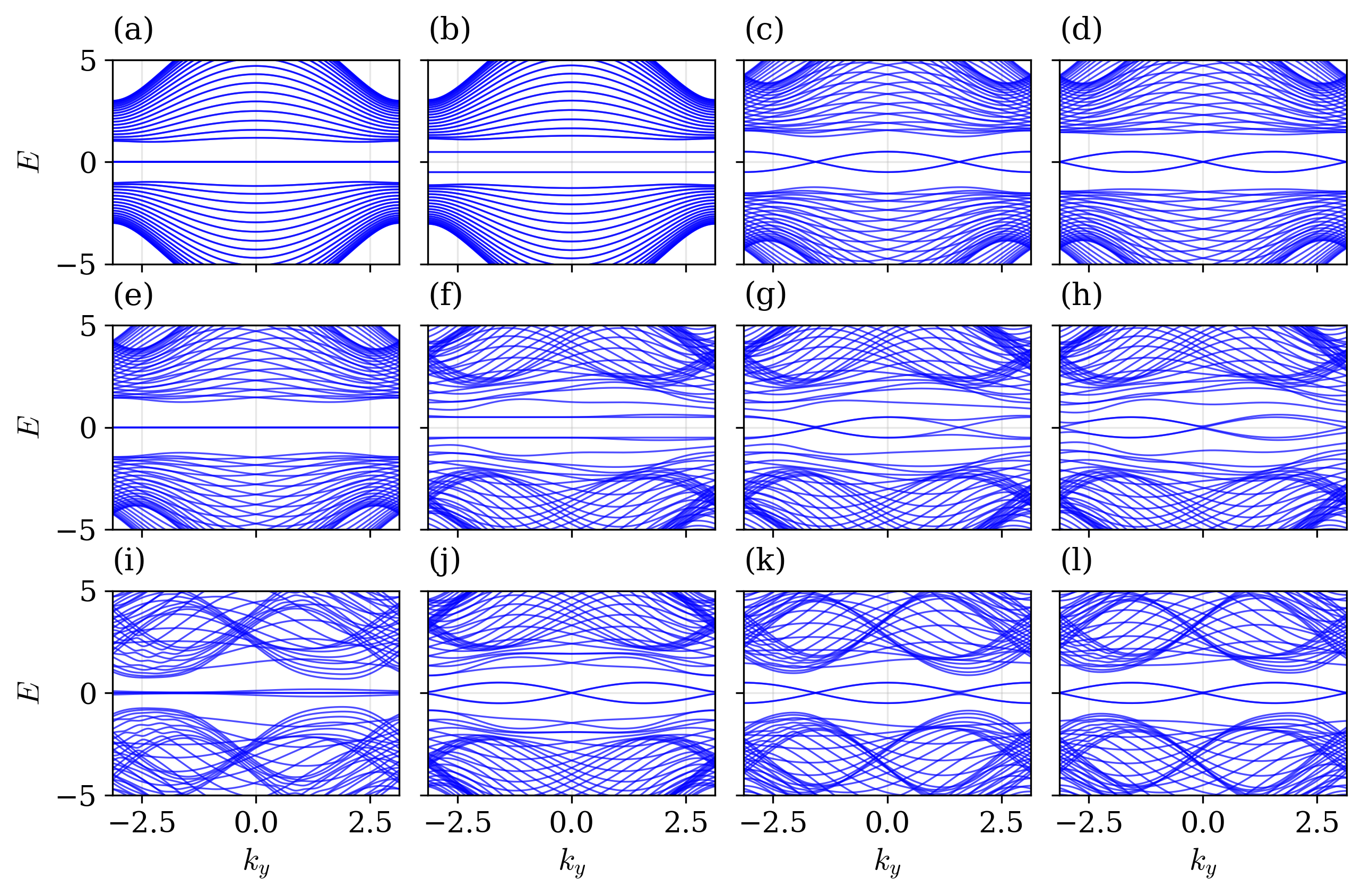}
\caption{ Energy dispersion is plotted for the system open in $x$-direction and closed in $y$-direction. Energies are plotted against a given $k_y$.  For all the panels we consider $(v,w,\delta)=(2,3,1)$. For the first column $\mu=0$ is considered. For second, third and fourth  column $\mu \sigma_z, \mu \cos k_y \sigma_z, \mu \sin k_y\sigma_z$ respectively with $\mu=0.5$. Second row considers effect $(\alpha_{R_1}, \alpha_{R_2}, \gamma, \eta)=(1,2,1,1)$. In third row, we additionally consider $\gamma_1=\eta_1=1.$  From the first column, top to bottom, one notice that the spectrum and edge modes are symmetric around $k_y=0$ in panel (a) and (e) but develop asymmetry in panel (i) due to presence of $\gamma_1, \eta_1$. In comparison to first column, the second column demonstrates how the spectrum and edge modes splits into non-degenerate levels due to inclusion of inversion asymmetric $\mu \sigma_z$ term. The third and fourth column depicts how the nodes of edge modes changes due to inclusion  of $\mu \cos k_y \sigma_z$ and $ \mu \sin k_y\sigma_z$ term respectively. }
\label{2dband}
\end{figure*}
\begin{figure*}[h!]
\centering
\includegraphics[width=0.7\textwidth]{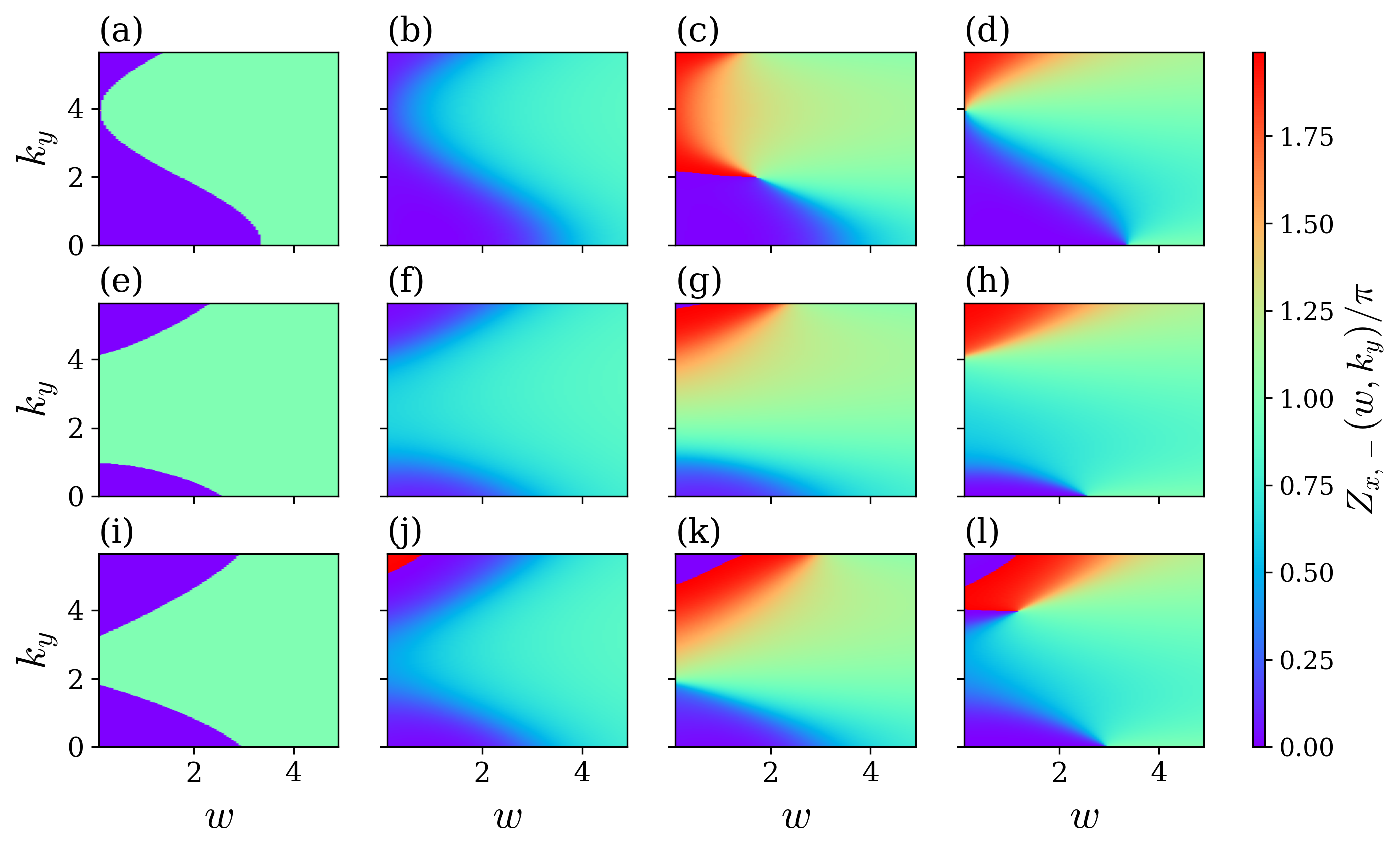}
\caption{Parameters for each panel corresponds to those in Fig. \ref{2dband}. We notice that for the first column when the inversion symmetry is preserved, Zak phase is quantized. Interestingly only for large values of $k_y$, Zak phase becomes two and it occupies more region as we go left to right in each  row. A comparison between second and third row highlights the effect of finite $\gamma, \eta$ (for second row) and finite $\gamma, \eta, \gamma_1, \eta_1$ (for the third row) respectively. In the later case we observe reappearances of Zak phase to be 1 at high $k_y$ and low $w$ values. }
\label{zak}
\end{figure*}
\subsection{ Fate of SPT from 1D to 2D limit}
\label{sec:2dlimit}
We complement our study of coupled SSH chains by briefly describing the two dimensional limit. We begin by considering the band structure. We note that for our system, there are no edge modes exist if it is periodic in $x$-direction and open in $y$ direction, no MGZE edge mode exist. The MGZE exists only for system open along $x$-direction either periodic or open in $y$-direction. For convenience we provide here the band structure periodic in $y$-direction i.e for a cylindrical geometry in consistent with the study of topological band structure. The corresponding figure is Fig. \ref{2dband}.  Following \cite{30-PhysRevLett.118.076803}, we have investigated the  system using Zak phase and how it depends on additional chiral symmetry broken term otherwise Zak phase is always integer. The first column considers no chemical potential i.e $\mu=0$. In the second column we consider $\mu\sigma_z$ which causes zero energy modes of different spin sectors to gain finite energy but of opposite sign. In the 3rd and 4th column a term $\mu \cos k_y\sigma_z$ and $\mu \sin k_y\sigma_z$ have been taken respectively. A comparison of third and fourth rows shows interesting dependencies of edge modes dispersion. Whether panel (f), (g) and (h)  shows little  asymmetries with respect to $k_y=0$, panel (j), (k) and (l) show complete symmetry due to interplay of $\eta$ and $\eta_1$. \\
\indent
To understand the two dimensional limit with $\mathcal{M} \rightarrow \infty$, we investigate the Zak phase, $Z_x(k_y)= \int _{\rm BZ} A_x(k_x,k_y) dk_x$) where $A_x(k_x,k_y)$ is the Berry connection. In Fig. \ref{zak} we plot $Z_x(k_y)$ in $k_y-w$ plane for various combination of chiral symmetry broken term artificially introduced to make  $Z_x(k_y)$  non-quantized. First row depicts how $Z_x(k_y)$ depends in isolated chain case  i.e $\delta=0$ and $\mu=0.5$. For first column all parameter except $v,w$ are zero. Second column considers finite $(\alpha_{R_2}, \alpha_{R_1})=(2,1))$ and in third column additionally $(\eta=\eta_1, \gamma=\gamma_1)=(1,1)$ have been considered. 2nd row  presents clean coupled chain in chiral symmetric limit.  3rd, 4th and 5th row consider finite $\delta$ with $\mu \sigma_z, \mu \cos k_y \sigma_z, {\rm and}  ~\mu \sin k_y \sigma_z $ respectively with $\mu=0.5$. We notice that $\cos k_y$ and $\sin k_y$  produces  different features in Zak phase. One notes that Zak phase has a region with value 2  for 4th and 5th rows which is mostly absent for 2nd and 3rd rows. We also checked the results for $\mu=0.01$ where all the finite $\delta$ case looks identical to 2nd row with red region in 4th and 5th panel reduced significantly. This shows how the coupled chain system can be characterized by different symmetry broken terms. 
$Z_x(k_y)$  shows a finite value mostly when winding number is away from zero or fully saturated value $2 \mathcal{M}$. 
\begin{figure}[h!]
	\centering
	\includegraphics[width=1\linewidth,height=0.65\textwidth]{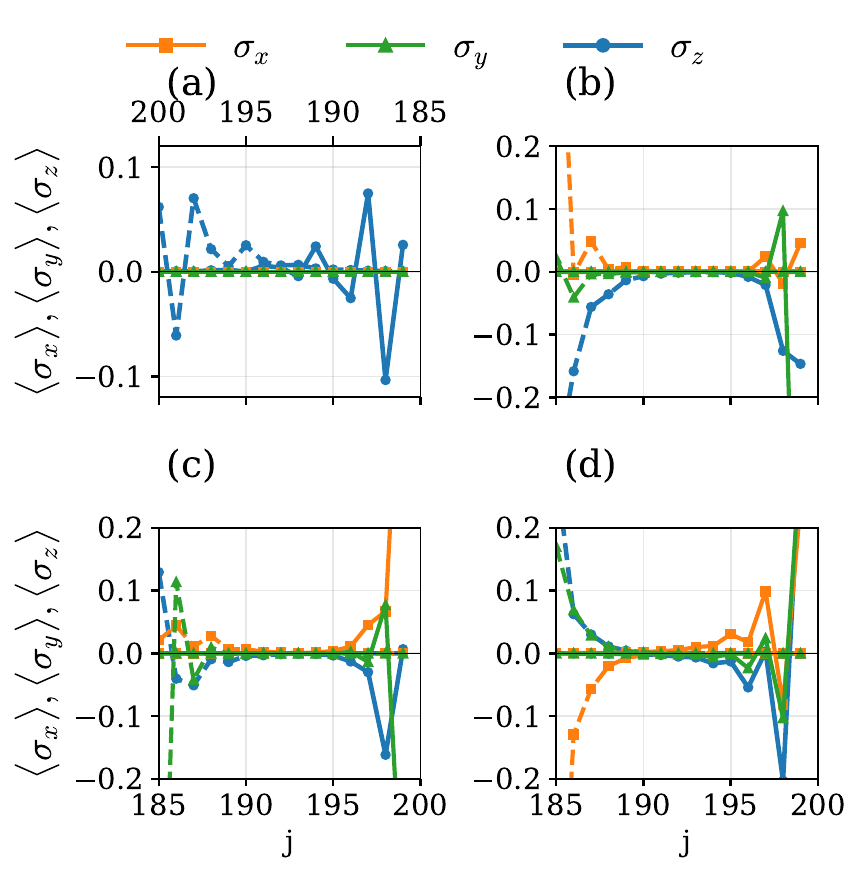}
	\caption{ { Spin-textures of the edge modes are shown. Various colors  represent $\langle \sigma_x\rangle, \langle \sigma_y \rangle, \& \langle \sigma_z \rangle$ as shown at the top. The dashed  plots represent finite $\gamma=\gamma_1=0.5$ and are plotted left to right as denoted by upper $x$-axis labeling. `$\rm j$' denotes the site index for the SSH chain. For solid lines $\gamma=0=\gamma_1$ and are plotted right to left. Panel (a) show the e-SSH model with $v=1,w=3,\gamma=0.5,\alpha_{R_1}=1,\alpha_{R_2}=2, \eta=\eta_1=2$ and vanishing $\alpha_{R_3}$ and $\alpha_{R_4}$.  In panel (b), we consider effect of $(\alpha_{R_3}, \alpha_{R_4})= (0.5,0.5)$. In panel (c) and (d), we consider $(\alpha_{R_3}, \alpha_{R_4})= (1,2)$ and $(2,1)$ respectively. Panel (d) shows the effect of $(\eta,\gamma_1)=(1,0.5)$ in addition to the parameters taken in panel (b).}}
	\label{spin-texture}
\end{figure}
\section{Edge modes: Spin textures and IPR}
\label{sec:edge-spin-texture-ipr}
For one-dimensional system with finite winding number is associated with even number of edge modes as shown in Fig. \ref{mid-gap-all-chain}. The number of mid-gap zero energy states is twice  the total winding number of the one-dimensional system. One of the important consequence of the present study is that depending on the parameters, the edge states show finite spin-component along all the three axis yielding non-zero $\langle \sigma_{\alpha}\rangle$ with $\alpha=x,y,z$. When $\alpha_{R_3}=\alpha_{R_4}=0$, it is obvious that only $\langle \sigma^{z} \rangle$ is only present. Further if $\eta=\eta_1=0$ along with $\alpha_{R_3}$ and $\alpha_{R_4}$, the two spin-sectors are identical and thus $\langle \sigma_{z} \rangle$ is zero. Here, for a single chain,  in Fig. \ref{spin-texture}, we plot few cases to elucidate the spin-textures of the edge modes. In panel (a), the case of $\eta=\eta_1=2, \alpha_{R_3}=\alpha_{R_3}=0$ is shown where only the $\langle \sigma_{z} \rangle$ is present. Effect of finite $\gamma, \gamma_1$ is shown by dashed plots as described in the panel. In panel (b), (c) and (d) values of $(\alpha_{R_3}, \alpha_{R_4})$ are taken (0.5,0.5), (1,2) and (2,1) respectively. The above analysis, though representative, shows promising future usage in spin-dependent  transport for practical usage.

\begin{figure}[h!]
	\centering
	\includegraphics[width=\linewidth]{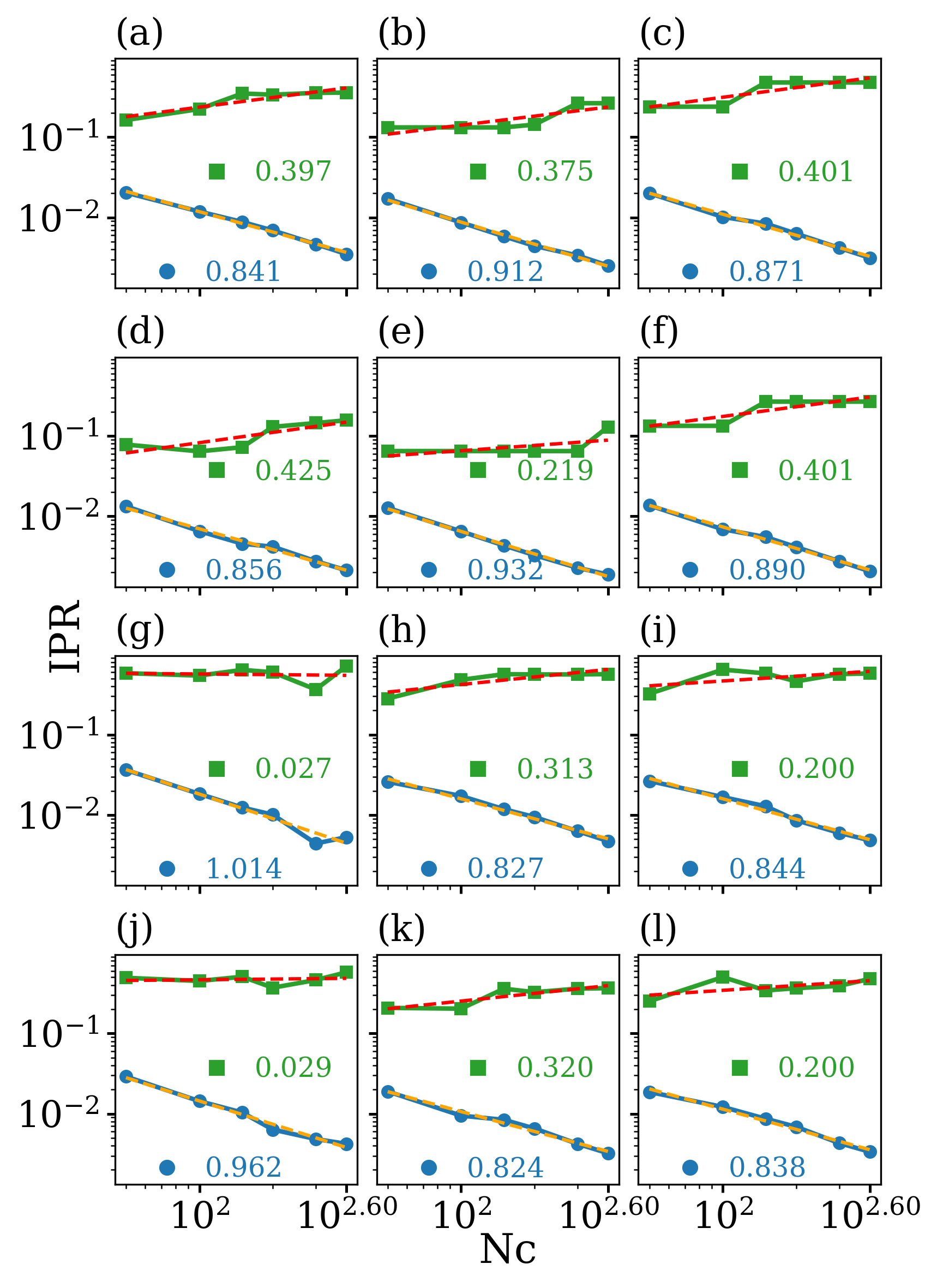}
	\caption{ { IPR is plotted in Log scale with  various system sizes denoted by $\rm N_C$. For the first(last) two rows we consider $(v,w)=(1,0.5) ((1,3))$. The solid green and blue line denotes the bulk and edge modes IPR as described in the text. The dashed lines is a best fit to the actual plot. The numeric in respective color denotes the value of $\alpha$ obtained. For all panels we have considered $(\alpha_{R_1}, \alpha_{R_2})=(1,2)$. All panels except those in left column considers $(\gamma,\eta)=(0.5,1)$. The second and fourth rows also includes $(\alpha_{R_3}, \alpha_{R_4})=(1,0.5)$. For the right column, we consider $(\gamma_1,\eta_1)=(0.5,1)$.}}
	\label{ipr_1}
\end{figure}
We now   examine the localization profile of the edge states through inverse participation ratio (IPR) which serves as a good indicator for localization to delocalization transition  \cite{Li-Li-Sarma-2017,Li-Sarma-2020,Mandal_2022}. IPR for any normalized eigenstate is defined as  $ {\rm IPR}= \sum_i |a_i|^4$ where $a_{i}$ denotes the probability amplitude  for site `$i$'. For an extended state IPR typically varies as $N^{-1}$ and for a localized state it goes as $N^0$, i.e does not depend on system size `$N$'. For critical states and states having fractal dimension the IPR varies as $N^{-\alpha}$ where $0<\alpha<1$ \cite{Li-Li-Sarma-2017,Li-Sarma-2020}, where $\alpha$ specify the fractal dimension denoting a rough estimation of the wave-function's extension. Hence we plot the IPR in Log scale with respect to $N$ such that the slope describe the edge state localization. For a topological edge states value of $\alpha$ often lies close to zero and depends on system parameters.\\
\indent
In Fig. \ref{ipr_1}, we plot the IPR for  edge states for different set of parameters for a comparison \cite{gabriel-PhysRevResearch.4.013185,Munoz2018,Bahmani-24}. The green line denotes that average IPR of the two edge modes at the two edges. The blue solid line denotes the average IPR of the half of the bulk modes chosen from the middle of the spectrum, for a comparison.  A comparison between the 2nd and 3rd row indicates that inclusion of $\gamma_1, \eta_1$ increases the magnitude of $\alpha$ for edge modes. However for the bulk modes, we see for $w>v$ (the 3rd and 4th  row) it decreases $\alpha$ otherwise if $v>w$, it increases $\alpha$. Similarly a comparison between first and second column indicates an increase in magnitude of $\alpha$ for edge modes only for $w>v$ (compare panels (g), (h), (j), (k)). On the other hand a comparison of panels (a,b,d,e) shows that absence of $\gamma,\eta$ decreases magnitude of $\alpha$. The bulk modes also show considerable dependencies on various parameters.
\section{Effect of domain wall}\label{domain-wall}
\indent
\begin{figure}[h!]
\centering
\includegraphics[width=.45\textwidth, height=.1\textwidth]{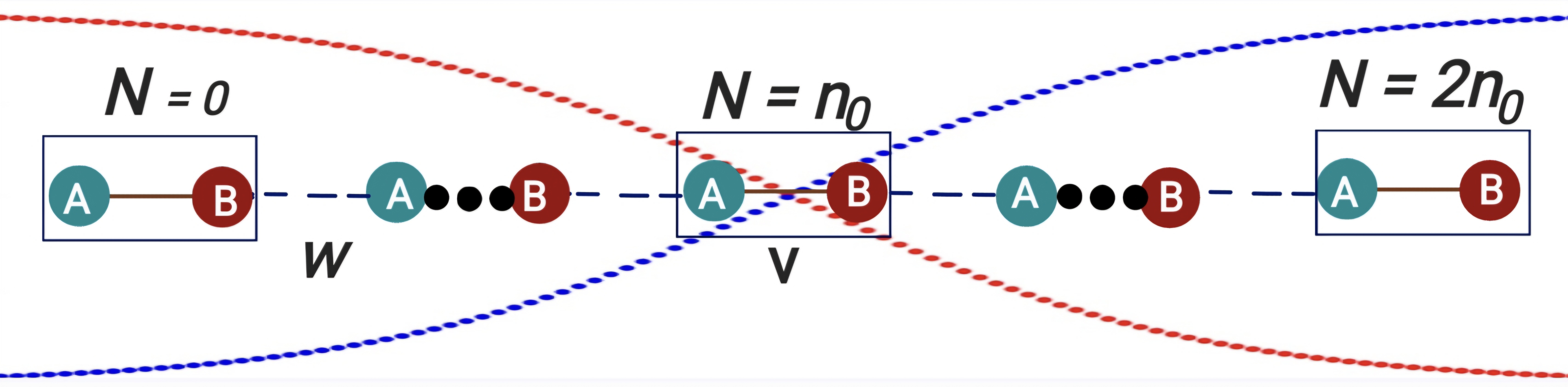}
\caption{ How the  domain-wall profile in the hopping parameter varies is shown pictorially.  The red(blue) line shows the variation of  amplitude of even (odd) bonds  as a distance from left end. $2n_0$ denotes total sites and $N$ denotes the total unit cells in conventional SSH model without domain wall.}
\label{domain_1}
\end{figure}
 We now examine how a domain wall (DW) profile \cite{scollon2020persistence} in the hopping parameter affects the  system. Following~\cite{scollon2020persistence}, we make the parameters $v$ and $w$ as site dependent with the  distribution $t_n= t_0( 1 + (-1)^n u)$ with $u= u_0 \tanh \left[ \frac{n-n_0}{ \xi/a}\right]$. Here the index $n$ denotes the location of $n$-th site. Thus `$n$' even and odd denote inter and intra  sub-lattice hopping respectively. In Fig. \ref{domain_1}, we show how the domain wall varies with increasing distance a given bond from the left end. It is  known that DW affects the system in two non-trivial ways. Firstly, it can cause a topological phase transition from a trivial to a non-trivial phase or vice-versa.  Secondly within a given phase it can effect the eigen-spectrum and edge-states in a non-trivial way.  We investigate the IPR and NPR for the edge mode, the location of  mid-gap energy states and eigen-spectrum for various choices for different such phases. Briefly, we also consider how a Aubry-Andre(AA) potential ( $V_{AA,i}= \sum_i \lambda_i n_i \cos[ 2 \pi \beta i]$ with $\lambda_i=(-1)^i$ and $\beta$ being $(\sqrt{5}-1)/2$ and $n_i=c^{\dagger}_i c_i$.) effect\cite{tapan-2021-PhysRevLett.126.106803,Miranda-PhysRevB.109.195427,Xu2024} the system when DW is present.\\

\subsection{Shifting of localization of edge mode}

\begin{figure}[h!]
	\centering
	\includegraphics[width=\linewidth,height=0.5\textwidth]{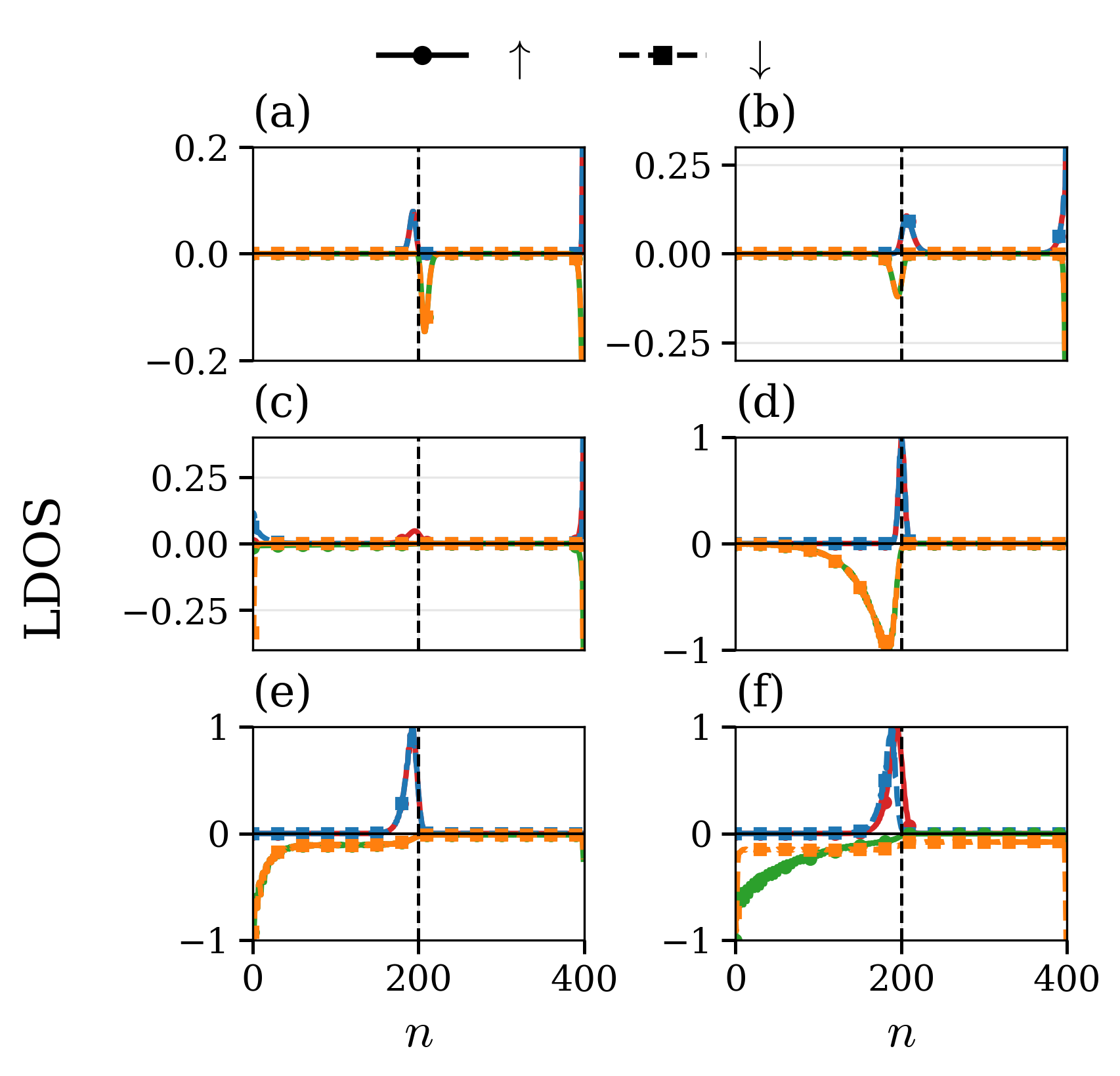}
	\caption{  The probability amplitude $|\Psi(x)|^2$ has been plotted for mid-gap zero enegry states for various parameter choices in the presence of domain wall. The blue (squared)  and red (circled)lines denote MGZE states for up and down spin respectively for a given set of parameters. The lower case also denotes the same for a different set of parameters.  Panel (a) upper (down) plot shows MGZE probability amplitude for only $\gamma=0.5$ ($\gamma_1$=0.5) in the presence of Domain wall. Panel (b) upper (and down) plot shows the effect of $\eta=1$ ($\eta_1=1$) in addition to $\gamma$ ($\gamma_1$). Panel (c) shows effect of $(\alpha_{R_1}, \alpha_{R_2})=(1,2)$ with  $(\gamma,\eta,\eta_1)=0.5,1,0.0$($(\gamma,\eta,\eta_1)=0.5,1,1$) for upper (lower) case. Panel (d) upper case shows the effect of AA potential with DW and lower case shows additional $(\alpha_{R_1},\alpha_{R_2})=(1,2)$ only. Panel (e)   upper (lower ) case shows the effect of AA for the parameter choice $(\alpha_{R_1}, \alpha_{R_2}, \gamma,\gamma_1=1,2,0.5,0.0)(1,2,0.5,0.5)$. Panel (f) upper (lower) case shows the effect of $(\eta,\eta_1=(1,0))(1.0,1.0)$ to the upper (lower) case of panel (e) respectively.  }
	\label{LDOS_1}
\end{figure}

\begin{figure}[h!]
	\centering
\includegraphics[width=\linewidth]{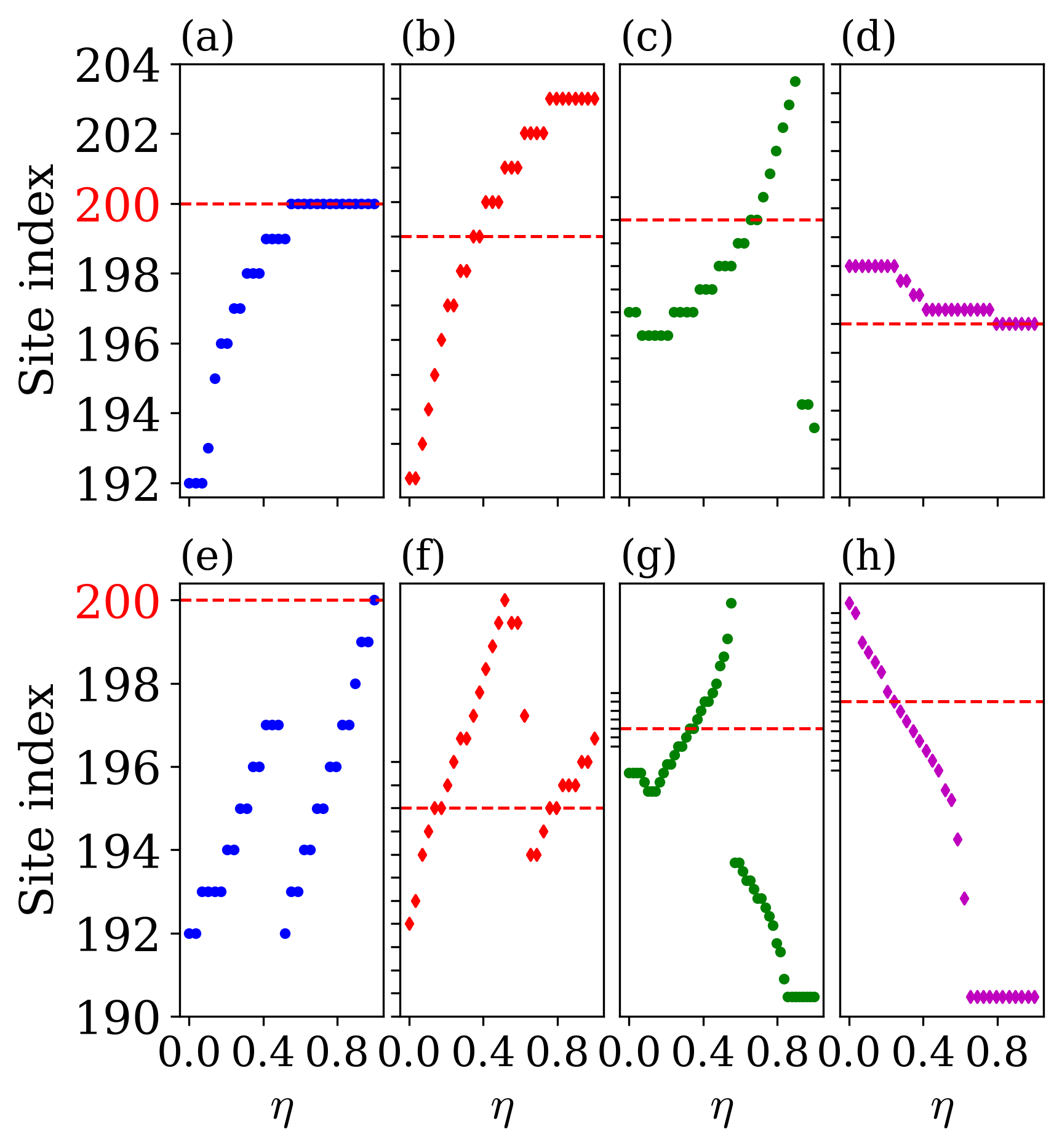}
	\caption{  Here y-axis represents the site index. The orange line in every plot represents the chain centre with site index $200$. Thus the actual site indexing is proportionally adjusted in comparison to panel (a) and (e) for upper and lower row respectively. Panels (a)–(d) correspond to  single chain with parameters: $\alpha_{R_2} = 1$, $\alpha_{R_1} = 0.6$ . $\alpha_{R_3} = 0.5$ and $\alpha_{R_4} = 1$ are present in (d) only, and $\gamma = \gamma_1$. $\gamma = 0$ in (a), and $\gamma = 1$ in (b)–(d). $\eta_1 = 0$ in (a), (b), and $\eta_1 = \eta$ in (c), (d). Panels (e)–(h) show the corresponding two-chain case with interchain hopping $\delta = 0.5$, using the same parameters as (a)–(d), respectively.
	}
	\label{domain-wall-position}
\end{figure}
As described, one of the key effect of a domain wall potential is to cause the shifting of the localization center of MGZE states. This happens due to the specific modulations of the hopping amplitude due to the domain wall. We note that, given the form of the domain wall potential as shown in Fig. \ref{domain_1}, the hopping parameter at an odd bond with index $2n-1$ is greater in magnitude than the even bond next to it with index $2n$ until $n<n_0$ which correspondences the domain center. For $n>n_0$ the above nature of relative magnitude becomes opposite. Thus, the recurrence relation are obtained such that  amplitude of the left edge mode  are localized only at odd sites and it is gradually increasing and becomes maximum at site $2n_0+1$. In appendix \ref{midcenter} we have explained this in more detail. Now, interestingly, when other parameters are included then the recurrence relations among the amplitudes of the wave-functions at different sites are modified in a nontrivial way and the localization centre could depend on them. In appendix \ref{midcenter}, we have explained the case of finite $\gamma$ changing the localization center of the MGZE modes away from center.\\
\indent
\indent
The probability amplitude $|\Psi(x)|^2$ has been plotted for mid-gap zero energy states for various parameter choices in Fig. \ref{LDOS_1}. Panel (a) upper (down) plot shows MGZE probability amplitude for only $\gamma=0.5$ ($\gamma_1$=0.5) in the presence of domain wall. It clearly demonstrates that in the presence of finite $\gamma$, the localization center shifts left to the middle. On the other hand a finite $\gamma_1$ shifts the localization to the right of middle as shown by the lower case of panel (a). Other panels also describe the effect of other parameters in changing the localization center as explained.   Panel (b) upper (and down) plot shows the effect of $\eta=1$ ($\eta_1=1$) in addition to $\gamma$ ($\gamma_1$). Panel (c) shows effect of $(\alpha_{R_1}, \alpha_{R_2})=(1,2)$ with  $(\gamma,\eta,\eta_1)=0.5,1,0.0$($(\gamma,\eta,\eta_1)=0.5,1,1$) for upper (lower) case. Panel (d) upper case shows the effect of AA potential with DW and lower case shows additional $(\alpha_{R_1},\alpha_{R_2})=(1,2)$ only. Panel (e)   upper (lower ) case shows the effect of AA for the parameter choice $(\alpha_{R_1}, \alpha_{R_2}, \gamma,\gamma_1=1,2,0.5,0.0)(1,2,0.5,0.5)$. Panel (f) upper (lower) case shows the effect of $(\eta,\eta_1=(1,0))(1.0,1.0)$ to the upper (lower) case of panel (e) respectively. \\
\indent
 In Fig. ~\ref{domain-wall-position}, we show how the localization center  changes for  representative parameter choices. Panels (a)-(d) ((e)-(h)) is plotted for single(double) chain. Panel (a) shows that when only $v,w,\alpha_{R_1},\alpha_{R_2}$ are present, the localization center gradually moves near the center (from the left) and at some critical $\eta$, it stops moving after reaching the center. In panel (b), we show how a finite $\gamma$ can further move the localization center from left to right side. Panel (c) represents the case $\eta=\eta_1$ in addition to the parameters of panel (b). It cause a discontinuous jump of localization center from left to right side. Panel (d) shows that finite $\alpha_{R_3}$ and $\alpha_{R_4}$ induce a qualitatively different behavior, it moves from right to left instead of left to right in comparison to all other panels mentioned before. Panels (e)-(g) show that when the inter-chain coupling is present, there is a discontinuous movements of the localization center. Interestingly for panel (h) when $\alpha_{R_3}, \alpha_{R_4}$ are present, this disconitnuos jump is absent.\\
\subsection{IPR and NPR in the presence of domain wall}
\begin{figure}[h!]
	\centering
	\includegraphics[width=\linewidth]{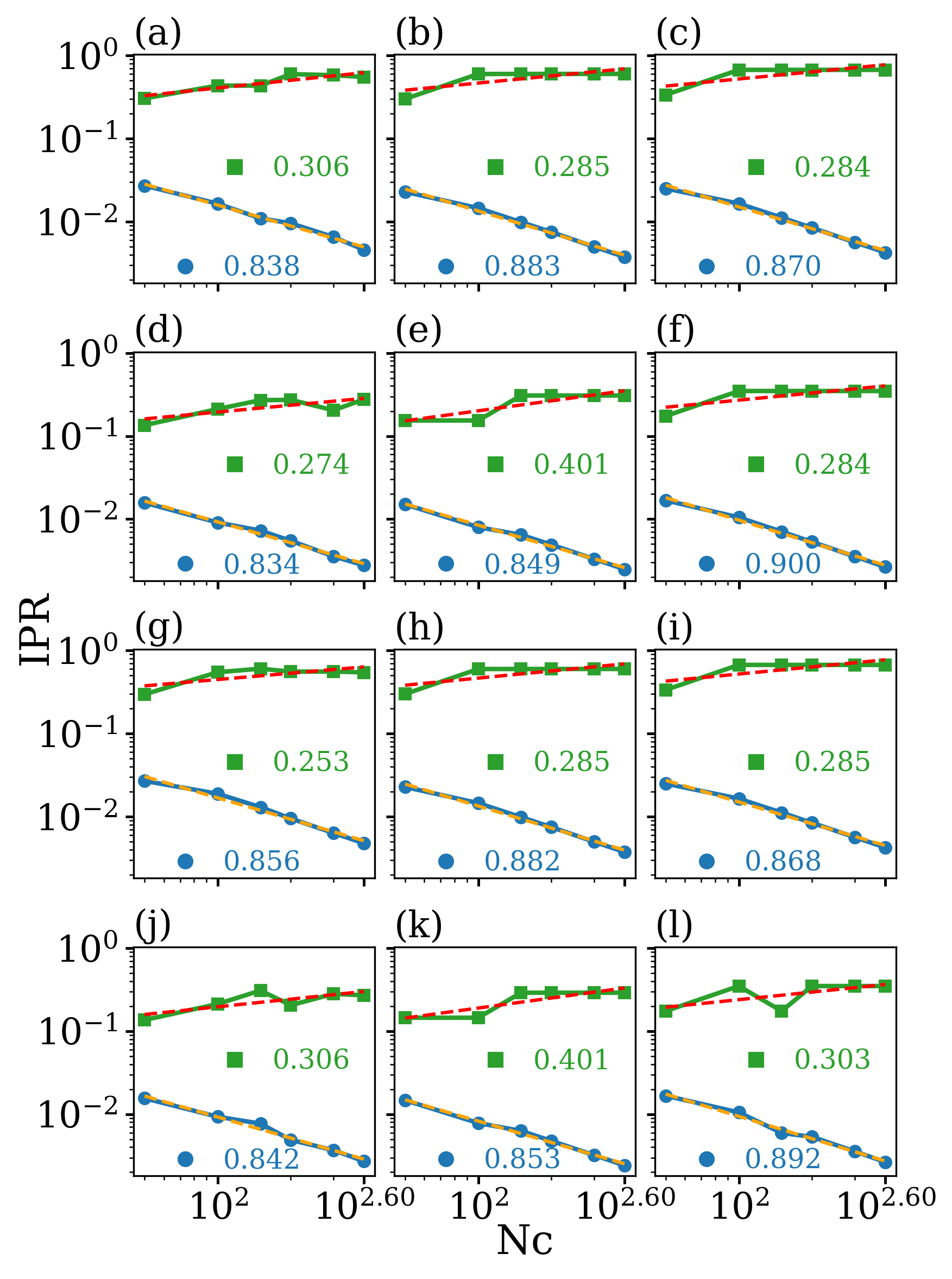}
	\caption{{IPR of single chain in the presence  domain wall potential is shown. All the parameter choices of panels of Fig. \ref{ipr_1} are repeated, except $v,w$ which has been replaced by the domain-wall distribution. For the odd (even) rows $(v,w)$ is replaced by $t_n= t_0( 1 + (-1)^n u)$  (  $t_n= t_0( 1 - (-1)^n u)$.} } 
	\label{ipr-domain-wall-one-chain}
\end{figure}

\begin{figure}[h!]
	\centering
	\includegraphics[width=0.5\textwidth]{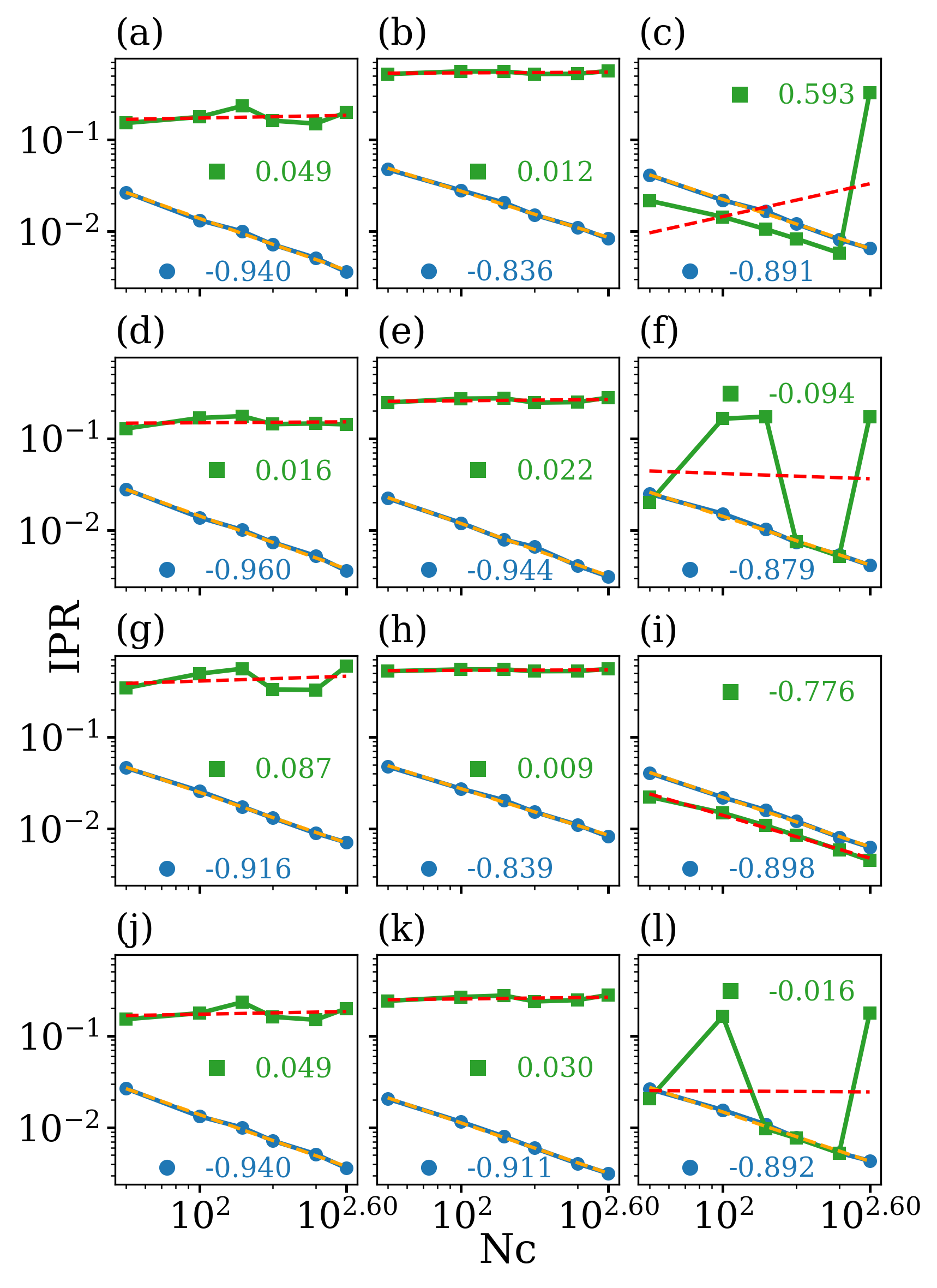}
	\caption{{The IPR is plotted in Log scale with the presence of Domain wall and AA potential. The parameter choices of each panel remain same to the corresponding panels in Fig. \ref{ipr-domain-wall-one-chain}. It is interesting to note that the right column which considers effect of finite $(\gamma_1, \eta_1)$ shows unusual behavior.}} 
	\label{ipr-domain-wall-AA}
\end{figure}

In Fig \ref{ipr-domain-wall-one-chain} we  present the IPR for a single chain in the presence of domain wall potential.  Further, Fig. \ref{ipr-domain-wall-AA} shows the results in simultaneous presence of domain wall and AA potential. All the panels in Figs. \ref{ipr-domain-wall-one-chain} and \ref{ipr-domain-wall-AA} have identical parameter values of the corresponding panels in Fig. \ref{ipr_1}. A comparison with first row of Fig. \ref{ipr_1} with the first row of Fig. \ref{ipr-domain-wall-one-chain} and Fig. \ref{ipr-domain-wall-AA} shows that in general domain wall does not change the IPR values much though AA potential    seems to increase the magnitude of $\alpha$. However the middle panel i.e panel (b) (in Figs. \ref{ipr-domain-wall-one-chain}, \ref{ipr-domain-wall-AA} ) shows different behaviors. We note that panel (b) has $\gamma,\eta$ finite but panel (c) has both $\gamma, \gamma_1$ and $\eta, \eta_1$ are finite. Thus turning on only $\gamma, \eta$ yields anomalous dependency. While comparing the second row of Fig. \ref{ipr_1}  with that of Fig. \ref{ipr-domain-wall-one-chain} and \ref{ipr-domain-wall-AA} shows that  while domain wall mostly decreases the value of $\alpha$, AA potential with domain wall increases the magnitude of it except the panel(f), again highlighting the effect of both $\gamma, \gamma_1, \eta, \eta_1$ finite.

In a similar way a comparison  among the  third rows indicate that it   yields a different trend in comparison to second row. We note that the difference between second and third row is that in the former $v>w$ and in the later $w>v$ in Fig. \ref{ipr_1}. While in Fig. \ref{ipr-domain-wall-one-chain} and Fig. \ref{ipr-domain-wall-AA} such a notion of $v,w$ are less relevant. Very interestingly, the fourth row of Fig. \ref{ipr-domain-wall-one-chain} and Fig. \ref{ipr-domain-wall-AA} , shows that AA potential in addition to domain wall potential increases the value of $\alpha$.  In regard to the bulk band, we see a remarkably different behavior when AA potential is finite and also $\gamma,\gamma_1, \eta, \eta_1$ is present. The fourth column of Fig. \ref{ipr-domain-wall-AA} shows that the IPR of edge modes do not follow a consistent linear scaling with respect to $N$, Though the IPR of the bulk modes consistently follow a linear scaling. It is interesting to note that generally for multi-fractal\cite{SKINNER1990333,PhysRevB.42.8121,PhysRevLett.84.3690} states the IPR of the bulk modes(not the edge modes)  show such non-uniform scalling though such non uniform scalling for edge or surface modes has also been discussed for certain context\cite{RevModPhys.80.1355}. Further we see that there are wide range of parameters for which the bulk IPR for domain wall and AA (Fig. \ref{ipr-domain-wall-AA}), yields very low values of $\alpha$\\
\indent
\indent 
\begin{figure}[h!]
	\centering
	\includegraphics[width=0.45\textwidth]{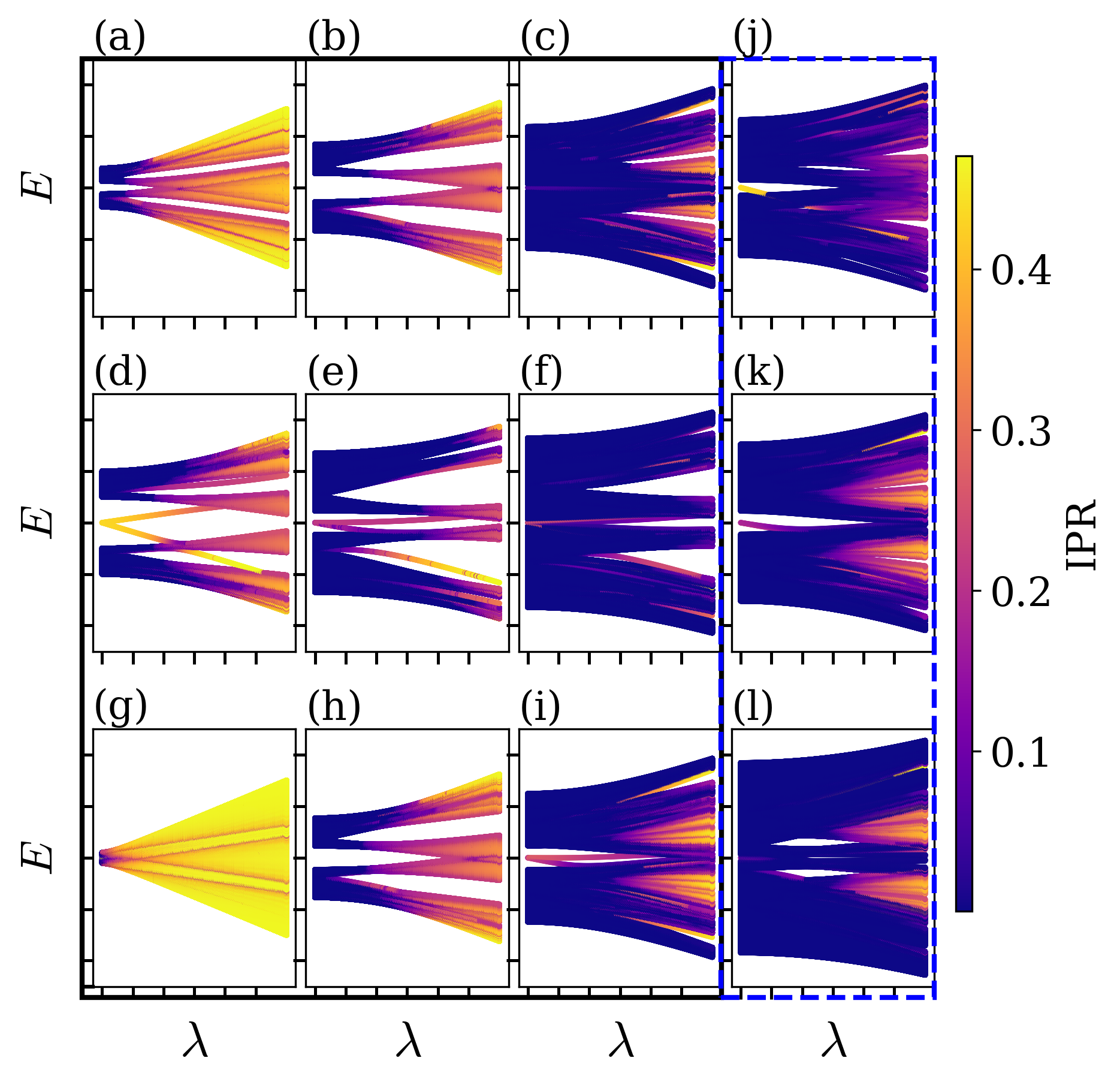}
	\caption{ Comparison of AA-vs-DW profile. Panel (a)- (c)  correspond $v> w$ and panel (d)-(i)correspond  $w >v$. For first column only $v, w$ are finite. Parameters in panel (g) are same as in panel (d) but contains domain wall. Except these three panels, all others have $\alpha_{R_1}=2,\alpha_{R_2}=1, \gamma=0.5, \eta=1$. $\gamma_1, \eta_1$ are also kept zero for (b),(e) and (h). Panels in second row contain no DW. Panels in third and fourth row are same as second row but with domain wall. Only panels in fourth row are with finite  $\alpha_{R_3}$ and $\alpha_{R_4}$. For panel (j), (k) and (l) values of $(\alpha_{R_3}, \alpha_{R_4})$ are takes as $( 1,2),~(\pm 2,\pm 1),~(\pm2,\mp1)$ in addition to  $\gamma_1=\gamma=0.5, \eta=\eta_1=1$.  }
	\label{ipr-density-plot}
\end{figure}
In Fig. \ref{ipr-density-plot} IPR of all eigenstates for different parameters are compared for AA and DW potential. Panels (a)((d)) corresponds pure SSH model for $v>w$($w>v$)  with AA profile. Panel (g) shows effect of  DW  included in (d) which completely closes the gap between the bands and causes localization of eigenstates. All panels (except those in first rows) have the common parameters  $(\alpha_{R_1}, \alpha_{R_2}, \gamma, \eta)= (2,1,0.5,1)$ with   $\gamma=\gamma_1, \eta=\eta_1$. Panels in second column shows effect of finite parameters other than $v,w$ in the presence of AA potential. We find significant broadening of bands and causing delocalization of eigenstates indicating role of other parameters than $v,w$. Panels in third row depicts what happens to panels in second column in the presence of DW. It is clear that additional states appear and low values of IPR suggests significant delocalization. Interestingly panel(i) shows appearances of a mid gap states due to DW. Panels in fourth row shows effect of different signs but same magnitude of $\alpha_{R_3}$ and $\alpha_{R_4}$. One notices that the gap structure depends on their relative signs.
\\
\indent
\section{\label{sec:conclusion} Conclusion}
In summary, this paper investigates an extended SSH model which considers all possible Rashba spin-orbit interaction ( represented by $\alpha_{R_i}, i =1,4$), a longer range NNNN hopping $\tilde{\gamma}, \tilde{\gamma}_1$ obeying  chiral symmetry.  We assume all hopping  parameters to be complex in general.  When $\tilde{\gamma}$ and $\tilde{\gamma}_1$  are real, the system has all the three symmetries $T, C, \& S$. When $\tilde{\gamma}, \tilde{\gamma}_1$ are made complex, the system has only the $S$ symmetry. We provide an exact method using the theory of non-hermitian eigenstates to  examine all aspect of the system  analytically.  Inclusion of NNNN hopping introduces two distinct  features in the system. It brings a $\cos 2k$ function in the gap-closing equation and also makes the winding number of both the sector to be different. We have provided exact analytical method to find the solution for gap-closing equations and discuss its various parameter dependencies and symmetries.   There has been previous studies \cite{23-PhysRevB.89.085111, 24-Li-2018-EPL-124-37003,yonatan-2024} which explored NNNN hopping preserving $S$ \cite{24-Li-2018-EPL-124-37003} or not \cite{23-PhysRevB.89.085111,Beatriz-PhysRevB.99.035146}. However  in \cite{24-Li-2018-EPL-124-37003} only $\tilde{\gamma}$ term has been considered. Ref \onlinecite{yonatan-2024} includes  real $\tilde{\gamma}_1$ yielding  $W= \pm 1$ only. Our works yields a more richer phase diagram such as re-entrant topological phases with possible values of $W$ ranging from 0 to 2 and present exact analytical solution of the Hamiltonian as well as gap-closing equation. \\
\indent
\indent
 Further we study $2 \mathcal{M}$ coupled  e-SSH chains where they are connected by a simplest $S$ preserving hopping. For  two and four coupled chains, we explicitly show  that $W$ can take any integer from zero to 4 and 8 and show existence of multi-critical points\cite{song-2017}. Our finding is true for arbitrary  $S$ preserving inter-chain coupling though exact parametric dependency may change.  We show that the two dimensional limit of such coupled chains can be characterized by appropriate Zak phase $Z_x(k_y)$. High(low )values of winding number corresponds to high (low) values of $Z$. Also it has different dependencies on various  $S$-symmetry broken terms.   Next we consider a domain-wall profile for the hopping parameter $v, w$. It shows  that the mid-gap zero energy states which are now localized around  middle of  the chain, changes its position as $\eta$ and other parameters are varied. We calculate IPR and NPR to characterize the properties of this mid-gap zero energy states and it shows when domain wall is present, the parameter dependencies are widely varied. A comparison of AA potential and DW shows that AA potential dominates over DW as per as IPR is considered. DW also causes re-entrant topological phase transition observed before in the presence of only AA  potential\cite{tapan-2021-PhysRevLett.126.106803}. But inclusion of AA potential ceases this phase transition.\\
\indent
There has been some progress in realizing various aspects of SSH model recently \cite{Lienhard:19,Thatcher_2022,PhysRevResearch.4.013185,10.1063/5.0191076,Bloch_TopologicalPhase,DeLeseleuc_ExpRydbergSSH}. Bosonic version of SSH model is realized using Rydberg atoms trapped in one dimensional array of optical tweezers\cite{Lienhard:19,DeLeseleuc_ExpRydbergSSH}. A mechanical set up using metallic masses connected by string \cite{Thatcher_2022} is proposed to observe the topological phase transitions found in 1d-SSH model. On the other hand  a lattice of coupled waveguides\cite{PhysRevResearch.4.013185} with staggered hopping is reported to realize the edge modes similar in SSH model. Interestingly mean-chiral displacement is used as an order parameter to establish finite winding number in an extended SSH model \cite{26-s41534-019-0159-6}. Recently study of disorder in SSH model is proposed in simulated topo-electric circuits \cite{julio-2024}. All these studies are particularly promising to realize the extended SSH model proposed in this article. The NNN hopping connecting same sub-lattices can easily be adopted in one-dimensional mechanical set up where the one-dimensional chain forms a zig-zag pattern so that the `A' and `B' sub-lattices reside in two different parallel straight line\cite{tarun-da-2024}. This geometry will help connecting springs between same sub-lattice easily and without making contact with the springs that connect nearest neighbor masses. For the NNNN coupling discussed in this article, either a suitable curved spring be used or  an additional heavy mass could be used as a connecting pathways between the two masses in question such that  the motion of the additional mass is negligible. The spin-degree of freedom can be simulated by incorporating additional SSH-chain and with suitable coupling with the previous one. We note that quantum spin hall insulator assisted with spin-orbit interaction has already been achieved in mechanical system\cite{Zhou_2018,CRPHYS-2018}.  In cold atomic  set up or  electric circuits, similar indigenous mechanism can be utilized to realize such extended SSH model. Recently an interesting work \cite{english2026topologicallyprotectedspatiallylocalized}  proposes the extended-SSH model containing the NNNN hopping with chiral symmetry in an electric-circuit network. Thus we hope many of the results discussed in this article could be verified. One interesting aspect of such extended SSH model, for example, the case of chiral symmetry broken SSH model\cite{tarun-da-2024}, if realized in a real system,  would results in anomalous differential conductivity at finite energy\cite{Klaassen2025}, alternatively  scanning tunnelling microscopy in real space be used to detect these finite energy edge (at finite voltage) to establish bulk-boundary correspondence\cite{Yin2021,PhysRevB.94.041102,HAN2023417}. Further proximity induced long range hopping or Floquet engineering can also be used to realize certain aspect of the extended SSH model which we leave for future study. \\
\indent
 All of these study shows a possibility to realize the present extended SSH model in some setting. However we leave it as a future study to examine how and what are the necessary conditions to be satisfied for such practical realizations. However our exact analytical formula in terms of functions $\lambda_i$ would be very useful in analyzing any experimental realization of certain extended SSH model. On the other hand, possible theoretical extension incorporating non-hermitian effect or Floquet dynamics could reveal further topological structure which we leave for future scope. In a recent study \cite{Zurita2023fastquantumtransfer}, a protocol for quantum state transfer network has been implemented in multi-domain SSH model. {In a multi-domain SSH chain with $N_d$ numbers of domains, numbers of localized zero energy topological modes are $N_d+1$. Further these modes are spatially more nearer in comparison to single domain case where two  localized topological modes are well separated\cite{Munoz2018}. This increased overlap between the localized modes positively contributes to faster and efficient transport of the topological modes. The adiabatic transport  begins with the initial condition of all odd bond strength to be zero and then follows an increment and subsequent variation following an optimized protocol.  For the present model it remains to see the role of different parameters in controlling the optimization of such state transfer.  Interestingly when `$v$' becomes close to `$w$', the quantum state transfer seems to be more faster though its topological protection becomes weaker given the fact the topological gap depends on $|v-w|$. However given the role of different parameters in the model, in particular the role of $\eta$(which causes NNNN hoping complex) which makes the gap asymmetric could be of interesting practical usage. The effect of complex  NNNN hopping and other parameters on such quantum state transfer protocol and robustness of zero energy  modes\cite{CHEN2020126168} are also left for future study. Further, the coupled chain system considered in the present study offers a collection of many one-dimensional extended-SSH chains whose spectral gaps are modified and depend on $k_y$. This might also be of practical use for the manipulation and state transfer. Thus we think the present study will generate further interest along these lines.}

\bibliography{AA_bib}
\appendix
\label{appendix}
\begin{center}
{\bf Appendix}
\end{center}
\section{ Hamiltonian in real space and details of exact solution}
\label{exsoltn}
The Hamiltonian in position space reads as follows, \\
\begin{eqnarray}
\label{fullHamiltonian}
&&H= \sum \left( v c^{\dagger}_{a,n} c_{b,n} + \tilde{w} c^{\dagger}_{b,n} c_{a,n+1}  + h.c \right) \nonumber \\
&& ~~~~+ \sum \left( \gamma c^{\dagger}_{a,n} c_{b,n+1} + \gamma_1 c^{\dagger}_{b,n} c_{a,n+2} + h.c\right) \nonumber \\
&& ~~~~ + \sum \left( i\alpha_{R_1} c^{\dagger}_{a,n}\sigma_zc_{b,n} + i\alpha_{R_2}c^{\dagger}_{b,n}\sigma_zc_{a,n+1} + h.c \right)\nonumber \\
&&~~~~ + \sum \left( i\alpha_{R_3} c^{\dagger}_{a,n}\sigma_yc_{b,n} + i\alpha_{R_4}c^{\dagger}_{b,n}\sigma_yc_{a,n+1} + h.c \right). ~~~~~~~~~
\end{eqnarray}

In the above $n$ denotes the cell index and $a$ and $b$ are sub-lattice indices. First line denotes pure SSH model. The second line corresponds NNNN interactions. The third (fourth) line  denotes  Rashba interaction which does not (does) mix spin sectors. Spin-mixing Rashba interaction follows from the representation $i \psi^{\dagger}_i \hat{n}_z \cdot ( \vec{\sigma} \times \vec{d}_{ij})\psi_{j}$ with  $\vec{d}_{ij}= \hat{n}_x$. Here  $\psi^{\dagger}_i= (c^{\dagger}_{i,\uparrow}, c^{\dagger}_{i,\downarrow})$.  After simplification and a straightforward Fourier transformation, we obtain the Hamiltonian in momentum space as,
\begin{eqnarray}
\label{hupdnfgk}
H &&= \sum_k\left(  h^k_{\uparrow} c^{\dagger}_{k,a\uparrow} c_{k,b\uparrow}  + h^k_{\downarrow} c^{\dagger}_{k,a\downarrow} c_{k,b\downarrow} + h.c \right) \nonumber \\
&& + \left( f^k c^{\dagger}_{k,a\uparrow} c_{k,b\downarrow} + g^k c^{\dagger}_{k,a\downarrow} c_{k,b\uparrow} + h.c \right) .
\end{eqnarray}
In the above first line represents the diagonal blocks of the Hamiltonian for each spin sectors and second line represents the off-diagonal blocks. For future analysis, we define $h^k_{\sigma}= h^{k}_{\sigma,x} + i h^k_{\sigma,y}$. $h^{k}_{\sigma, x(y)}$, $f^k$ and $g^k$ are given by($w_{\pm}=\gamma\pm w, \eta_{\pm}= \frac{\eta}{2} \pm \frac{\xi}{2} \pm \alpha_{R_2}$),
\begin{eqnarray}
\label{hxhyupdn1ap}
&&h^k_{\uparrow x} =  v+ w_+ \cos{k} -\eta_+ \sin{k}\nonumber \\
&& ~~~~~~~~ +  ~\gamma_1 \cos{2k}  - \frac{\eta_1}{2} \sin{2k},~ \\
\label{hxhyupdn2ap}
&&h^k_{\uparrow y} =  \alpha_{R_1} + w_- \sin{k} + w_- \cos k  \nonumber \\
&&~~~~~~~~-\gamma_1 \sin{2k} - \frac{\eta_1}{2} \cos{2k},~ \\
&&g^k=  - \alpha_{R_3} + \alpha_{R_4} e^{-i k}, ~~~~g^k= -f^{k*}. ~~~~
\label{hxhyupdn3ap}
\end{eqnarray}
We note that expression for $h^k_{\downarrow x/y}$ can be obtained from $h^{k}_{\uparrow x/y}$ by changing the signs of $\alpha_{R_1}$ and $\alpha_{R_2}$.
We note that in the absence of $\alpha_{R_3}, \alpha_{R_4}$, the Hamiltonian is block diagonal in $\Psi^{\dagger}_k = (c^{\dagger}_{k,a\uparrow}, c^{\dagger}_{k,b\uparrow}, c^{\dagger}_{k,a\downarrow},c^{\dagger}_{k,b\downarrow})$ basis.
On the other-hand, in the absence of $\alpha_{R_1}, \alpha_{R_2}$, one  can do a basis  transformation be defining new fermionic operator $\Phi= \mathcal{U}^{\dagger}_1 \Psi$ with $\Phi^{\dagger}= (d^{\dagger}_{k,a1},d^{\dagger}_{k,b1}, d^{\dagger}_{k,a2},d^{\dagger}_{k,b2})$. The expression for $\mathcal{U}^{\dagger}_1= \frac{1}{\sqrt{2}} \left( c (\sigma_x - \sigma_z) + c^*( \sigma_y +  I) \right) \otimes \sigma_x$ with $c= \frac{1}{2}(1+i)$.
 Then the Hamiltonian is again  block  diagonal  with the expression as below.\\
 \begin{eqnarray}
 	\label{hambarbasis}
 &&H=\sum_k \left(\tilde{h}^k_{1} d^{\dagger}_{k,a1} d_{k,b1} +  \tilde{h}^k_{2} d^{\dagger}_{k,a2} d_{k,b2} + h.c \right)~. ~~~ 
\end{eqnarray}
The expressions for $\tilde{h}^k_{\alpha,\sigma}$ is similar to $h^k_{\alpha,\sigma}$ as in Eq \ref{hxhyupdn1ap} and in Eq. \ref{hxhyupdn2ap} with $(\alpha_{R_1}, \alpha_{R_2})$ being replaced by $(\alpha_{R_3}, \alpha_{R_4})$. 
Now we proceed to discuss the method to obtain the Hamiltonian in an off  block diagonal form  in an intermediate basis which greatly helps in analyzing the model.  Firstly, we define a  basis  $\Upsilon^{\dagger}_k=( c^{\dagger}_{k,a\uparrow}, c^{\dagger}_{k,a\downarrow}, c^{\dagger}_{k,b\uparrow}, c^{\dagger}_{k,b\downarrow})= \mathcal{U}_2\Psi^{\dagger}$. The expression of $\mathcal{U}_1, \mathcal{U}_2$ are give below.
\begin{eqnarray}
	\label{u1u2eq}
\mathcal{U}_1=\frac{1}{\sqrt{2}}\begin{pmatrix}
	0 & i & 0 & -i \\
	i & 0 & -i & 0 \\
	0 & 1 & 0 & 1 \\
	1 & 0 & 1 &0 
\end{pmatrix},~~~\mathcal{U}_2=\begin{pmatrix}
		1 & 0 & 0 & 0 \\
		0 & 0 & 1 & 0 \\
		0 & 1 & 0 & 0 \\
		0 & 0 & 0 &1 
	\end{pmatrix}.
	\end{eqnarray} 
The explicit  expression of Hamiltonian in the $\Upsilon$ basis is given below,
\begin{eqnarray}
\label{newphibasisH}	
	\mathcal{H}=\begin{pmatrix}
		0_{2\times2} & H  \\
		H^* & 0_{2\times2} 		 
	\end{pmatrix},~~ H= \begin{pmatrix}
	h^k_{1} & y^k  \\
	-y^k& h^k_{2} 		 
\end{pmatrix} .
\end{eqnarray}

$H^*$ represents conjugate transpose of $H$. We have $y^k=g^k, h^k_1= h^k_{\uparrow}, h^k_2=h^k_{\downarrow}$.  We note that $H$ is in general a non-Hermitian matrix . The eigenvalues of $H$ and $H^*$ are complex conjugate to each other. We represent the eigenvalues and eigenvectors by $\lambda_i, |\lambda_i \rangle$ and $\tilde{\lambda}_i, |\tilde{\lambda}_i \rangle$ respectively. The expressions for them are given below. For brevity, in the rest,  we ommit explicite $k$-dependence. 
\begin{eqnarray}
\label{lambda1234}
&&\lambda_{1}= \frac{ \left(x_0 - x_1  \right)}{2},~\lambda_2= \frac{ \left(x_0 + x_1  \right)}{2},~\tilde{\lambda}_i= \lambda^*_i,~\\
 && |\lambda_{i}\rangle= A_{i} \begin{pmatrix}- \frac{x\mp x_1}{2y} \\1  \end{pmatrix},~  |\tilde{\lambda}_{i}\rangle= \tilde{A}_{i} \begin{pmatrix} \frac{x^*\mp x^*_1}{2y^*} \\1  \end{pmatrix} .
\end{eqnarray}
In the above $i=1$ and 2 are asigned with $-$ and $+$ respectively. Also $x_1= \sqrt{x^2-4y^2}$, $x_0=h_1 + h_2, x=h_1-h_2$. $A_i$ and $\tilde{A}_i$ denotes normalization constants needed.  One can easily check that $\langle \tilde{\lambda}_i | \lambda_{j} \rangle =0, ~{\rm for}~ i \ne j, \langle \tilde{\lambda}_{i} | \lambda_{i} \rangle \ne 0, ~~ \langle \lambda_{i} | \lambda_{j}\rangle \ne 0 $.
 To proceed we denote $U$ and $V$ are $2\times2$ matrices which are made of right  eigenvectors of $H$ and $H^*$. We note that in general $U^{\dagger} U$ and $V^{\dagger} V$ is not unit matrix. However we can use the fact that $U^{\dagger}V$ and $V U^{\dagger}$ may be made unit matrix to construct a new complex fermionic basis. In the below we outline this in detail. The Hamiltonian $\tilde{\mathcal{H}}= \Upsilon^{\dagger} \mathcal{H} \Upsilon$ can be written as $\tilde{\mathcal{H}}= \Upsilon^{\dagger} \mathcal{K} \mathcal{H} \tilde{\mathcal{K}}  \Upsilon$.
We need to prove explicitly that,
\begin{eqnarray}
 &&\mathcal{K}=\begin{pmatrix} 0 & U \\ V & 0 \end{pmatrix} \begin{pmatrix} 0 & U^{\dagger} \\ V^{\dagger} & 0 \end{pmatrix}= \begin{pmatrix} UV^{\dagger}&0 \\
0&VU^{\dagger} 
\end{pmatrix}=1~. ~~~~~~
 \end{eqnarray}
 Similar expression for $\tilde{\mathcal{K}}$ can be obtained by interchanging $U$ and $V$.  For convenience we define $\mathcal{X}= UV^{\dagger}, ~~~\mathcal{Y}= VU^{\dagger} $. Below we explicitly show that by choosing the normalization constant $A_i, \tilde{A}_i$ one is able to satisfy the above conditions. Our calculation shows that $\mathcal{X}_{12}= -\mathcal{X}_{21}=$  and $\mathcal{X}_{12}=0$ yields the following condition $\frac{A^*_1 \tilde{A}_1  }{A^*_2 \tilde{A}_2 }= \frac{\sqrt{x^*-4y^{*2}} + x^*}{\sqrt{x^*-4y^{*2}} - x^* }$.
To solve for $A_{i}, \tilde{A}_i$, we choose a symmetric solution ,  $A_i=\tilde{A^*}_i$ as given below, $A^*_1= \tilde{A}_1= \sqrt{z} \left(\sqrt{x^*-4y^{*2}} + x^*  \right)^{1/2} $ and $ A^*_2= \tilde{A}_2= \sqrt{z} \left(\sqrt{x^*-4y^{*2}} - x^*  \right)^{1/2}$.
Various elements of $\mathcal{X}$ and $\mathcal{Y}$ are obtained as $\mathcal{X}_{11}= \mathcal{X}_{22}= 2 z \sqrt{x^{*2} - 4 y^{*2}}$ and $ \mathcal{X}_{11}=\mathcal{Y}^*_{11},~\mathcal{X}_{22}=\mathcal{Y}^*_{22},~ \mathcal{Y}_{21}= \mathcal{X}^*_{12},~\mathcal{Y}_{12}= \mathcal{X}^*_{21}$.
One finds that $\mathcal{K}$ and $\tilde{K}$ becomes unity with $z=\frac{1}{2 \sqrt{x^{*2} - 4y^{*2}}}$.
 Now we construct a new basis defined by $\tilde{\Upsilon}$ as  below,
\begin{eqnarray}
	\label{upsilon0}
\tilde{\Upsilon}= \begin{pmatrix} \alpha \\
	\beta \end{pmatrix}=  \begin{pmatrix} 0 & U^{\dagger} \\ V^{\dagger} & 0 \end{pmatrix} \Upsilon, \alpha= \begin{pmatrix} \alpha_1\\ \alpha_2	\end{pmatrix},    \beta= \begin{pmatrix} \beta_1\\ \beta_2	\end{pmatrix} ~~
\end{eqnarray}
$\alpha_i, \beta_j$ are mutually anti-commuting but as  $U$ and $V$ are not unitary, $N_{\alpha}=\alpha^{\dagger}_i \alpha_i + \alpha_i \alpha^{\dagger}_i \neq 1.$ Same goes with $N_{\beta}$. Explicit calculation yields, 
\begin{eqnarray}
&&N_{\alpha} =\langle \tilde{\lambda}_i|\tilde{\lambda}_i\rangle=r^{-1}_i ~,~~N_{\beta}=\langle {\lambda}_i|{\lambda}_i\rangle= s^{-1}_i~.
\end{eqnarray} 
The above finding suggests that to satisfy the ferminoc normalization, we must  redefine Eq. \ref{upsilon0} as follows,
\begin{eqnarray}
	\label{upsilon1}
 \begin{pmatrix} \alpha \\
	 \beta \end{pmatrix}= W \begin{pmatrix} 0 & U^{\dagger} \\ V^{\dagger} & 0 \end{pmatrix} \Upsilon,~~W=\begin{pmatrix}\mathcal{R}&0_{2\times 2}\\0_{2 \times2}&\mathcal{S}\end{pmatrix}~.~~~
\end{eqnarray}
The diagonal matrix $\mathcal{R}(\mathcal{S})$ has elements $r_1, r_2(s_1,s_2)$.
This suggests one needs to redefine $\mathcal{K}$ and $\tilde{\mathcal{K}}$ as,
\begin{eqnarray}
	&&\mathcal{K}=\begin{pmatrix} 0 & U \\ V & 0 \end{pmatrix} W^{-1}W \begin{pmatrix} 0 & U^{\dagger} \\ V^{\dagger} & 0 \end{pmatrix} .
\end{eqnarray}
$ \tilde{\mathcal{K}}$ can be obtained by replacing $U, V, W$ by $V,U, W^{-1}$. This renders the final Hamiltonian $\tilde{\mathcal{H }}$   in $\tilde{\Upsilon}$ basis to be as follows, 
\begin{eqnarray}
\tilde{\mathcal{H}}&&= W^{-1}\begin{pmatrix} 0 & U^{\dagger} \\ V^{\dagger} & 0 \end{pmatrix} \mathcal{H}\begin{pmatrix} 0 & V \\ U & 0 \end{pmatrix}W^{-1}\\
&&= \begin{pmatrix}
	0_{2\times2} & H^{\prime}  \\
	H^{\prime*} & 0_{2\times2} 		 
\end{pmatrix},~~ H^{\prime}= \begin{pmatrix}
	\frac{\lambda^*_1}{r_1s_1} & 0  \\
	0& \frac{\lambda^*_2}{r_2s_2}		 
\end{pmatrix} . 
\label{tildeH}
\end{eqnarray}
With  $\mathcal{L}_{\pm}= \sqrt{x\pm x_1}$.  $r_i, s_i$ are obtained as $s^{-1}_i=|z| |\mathcal{L}_+ - \mathcal{L}_-|^2,~r^{-1}_i=|z| |\mathcal{L}_+ + \mathcal{L}_-|^2$ with $i=1,2$.
Interestingly one finds  $r_i s_i=1$. Thus  after a basis transformation (using the matrix $\mathcal{U}_2$) the Hamiltonian in Eq. \ref{tildeH} reduces ( $k$-dependencies being explicit),
\begin{eqnarray}
\label{finalphiHam}
\mathcal{H}^{\prime}= \sum_k\left(  \lambda^k_1 \beta^{\dagger}_{1,k} \alpha_{1,k} + \lambda^k_2 \beta^{\dagger}_{2,k} \alpha_{2,k} + {\rm h.c} \right) .
\end{eqnarray}
It is interesting to note that the scaling factor of mode $\alpha_i$ and $\beta_i$ is inversely proportional to each other and their eigenvalues are related to complex conjugate to each other.
Before moving further, we note that the Hamiltonian in Eq. \ref{newphibasisH} becomes  normal 
when  $HH^*=H^*H$. This yields the condition $fg^*=gf^*$ implying $\alpha_1 \alpha_4 = \alpha_2 \alpha_3$ This means either $\frac{\alpha_1}{\alpha_2} = \frac{\alpha_3}{\alpha_4} = n_1 {\rm or}   \frac{\alpha_1}{\alpha_3} = \frac{\alpha_2}{\alpha_4} = n_2 $. In both the case one obtains $f=ng$ where $n$ is a real constant.

\subsection{Gap closing condition  and  its symmetry}
\label{gapcloseallfinite}
Here we derive the gap closing condition for the system when all $\alpha_i$s are finite and discuss its symmetry in some simple cases. In case of $f=0, g \neq 0$ or $f \neq 0, g=0$ or $f = n g$, the analysis of  gap closing condition are equivalent mathematical problem. In general case when $f \neq g$, the eigenvalues are expressed as $\lambda_{1/2}= h_0 + i \sqrt{f^2 + g^2}$. Here onward for simplicity we omit the $k$-indices in $h, f, g$.   $f^2 + g^2$ can be simplified as $(a + i b)$ and with the formula $\sqrt{a+ i b}= x + i y$ with $a , b, x, y$ given below (with $\alpha^{34}_{12}=(\alpha_1 \alpha_2 + \alpha_3 \alpha_4)$),
\begin{eqnarray}
&& a = \alpha_1^2 + \alpha_3^2 + (\alpha_2^2 + \alpha_4^2) \cos 2k - 2 \alpha^{34}_{12} \cos k , ~~~~~~~~~~~~~~~~~\\
&& b  = 2 \alpha^{34}_{12} \sin k - (\alpha^2_2 + \alpha^2_4) \sin 2k,~z = \sqrt{a^2 + b^2} , \\
&& x =\pm \frac{1}{\sqrt{2}}\sqrt{|z| + a},~~ y = \frac{1}{\sqrt{2}}\frac{b}{|b|} \sqrt{|z|-a} .
\end{eqnarray}
We note that `$a$'  and `$b$' are even and odd function of  `$k$' respectively.
Thus the eigenvalues $\lambda_{1,2}$ reduce to be, $\lambda_1=  h_0 - y + i x ,~~\lambda_2= h_0 + y - ix$.

\subsubsection{{\bf Case of real $v,w$, $\gamma, \gamma_1=0$ and all $\alpha_i$ finite.}}
When the gap is closed $\lambda_i=0$ which yields, 
\begin{eqnarray}
&& v+ w \cos k = \frac{b}{|b|} \frac{\sqrt{|z|-a}}{\sqrt{2}},~ w \sin k =  \frac{\sqrt{|z|+a}}{\sqrt{2}}~.~~~~~~~
\end{eqnarray}
Above condition is re-written as,
\begin{eqnarray}
&&(v^2 +  \alpha^2_1 + \alpha^2_3) + (w^2 + \alpha^2_2 + \alpha^2_4) \cos 2k \nonumber \\
&& + ~~2( vw -\alpha_1 \alpha_2 - \alpha_3 \alpha_4)\cos k=0 .
\end{eqnarray}
By defining a vector $\vec{V}= v \hat{x} + \alpha_1 \hat{y} + \alpha_3 \hat{z}$ and $\vec{W}= w \hat{x} - \alpha_2 \hat{y} - \alpha_4 \hat{z}$, one  re-writes the above gap-closing condition to be,
\begin{eqnarray}
	\label{gapclosingeq00}
V^2 + W^2 \cos 2k + 2 \vec{V}\cdot\vec{W} \cos k=0 .
\end{eqnarray}
Above equation shows how the Rashba and other hopping parameters are related algebraically in equal footings in  the gap closing condition. The solution for `$k$' can be easily found by solving for $\cos k$.
\subsubsection{\bf Gap closing condition for general case }
\label{gapcloseallalpha}
 From Eq. \ref{hxhyupdn1ap} and \ref{hxhyupdn2ap}, we write $h_0=h_{\uparrow}= h_{\uparrow, x} + i h_{\uparrow, y}$. With $h_{\uparrow, x}=h_{0x}, h_{\uparrow, y}=h_{0y} $, the gap-closing condition is obtained as, 
\begin{eqnarray}
\label{gapcloseeq1}
&&h_{0x}=  \pm \frac{b}{|b|} \frac{ \sqrt{|z|-a}}{\sqrt{2}},~~h_{0y} =\mp~  \frac{\sqrt{|z|+a}}{\sqrt{2}} ~. ~~~~
\end{eqnarray}

In the above the upper(lower) sign corresponds to sector 1(2). Below, various scenarios are discussed.\\
\indent
{\bf Case of $\gamma=\gamma_1=0$ and real $w$:} In this case the gap closing condition is provided by Eq. \ref{gapclosingeq00}. Both the sectors have identical equations due to the time reversal symmetry. A solution `$k$' indicates another  at `$-k$' also.\\
\indent
{\bf For real $w, \gamma, \gamma_1 ~i.e~ \xi, \eta, \eta_1=0$:} In this case we see that the set of equations  \ref{gapcloseeq1}  are not identical for both spin-sectors. They are differed by a $-$ sign at r.h.s. However it can be checked that  given a solution for a point `$k$' in sector-1, it admits a solution for sector-2 for `$-k$'.\\
\indent
{\bf For complex $w, \gamma, \gamma_1 ~i.e~ \xi, \eta, \eta_1 \neq 0$:} In this case the gap closing solution at `$k$' for sector-1 is not related to any solution of sector-2. However  for a given $\xi, \eta, \eta_1$ if sector-1, admits a solution, the sector-2 admits a solution for $-k$ with $-\xi,-\eta,-\eta_1$. This indicates the broken time-reversal symmetry of the system.\\
\indent
{\bf Constraints to have solution at $k=0,\pi$:} It is interesting to note that if $k=0, \pi$ are to be a gap closing solution, then the  parameters needs to satisfy the constraints, $v + (\gamma + w) + \gamma_1=0 $ and $ \frac{1}{2}(\eta - \xi) - \frac{\eta_1}{2}= \pm \sqrt{(\alpha_1\pm \alpha_2)^2 + (\alpha_3 \pm \alpha_4)^2}$. This is obtained from Eq. \ref{gapcloseeq1} with $k=0,\pi$. Thus we see that if $\alpha_1= \mp \alpha_2$ and $\alpha_3 = \mp \alpha_4$, then both the sectors admit identical solutions if the real and imaginary part of the system parameters satisfy the constraint. However interestingly, second constraint implies only one of the spin-sector can admit the solutions. It also signifies that by changing the imaginary part of the hopping parameter, the other sector can also be made topological while the former one is no longer topological.
\begin{center}
{\bf General method  when  all parameters are finite}	
\end{center}
For general parameters, the set of equations as given by Eq. \ref{gapcloseeq1} are in general very complex to solve algebraically. An alternative way to solve is graphically. If one denotes the l.h.s and r.h.s of Eq. \ref{gapcloseeq1}  by $A_1$ and $B_1$ ($C_1$ and $D_1$ for  sector-2), then one can graphically solve for $y=A_1(k)$ and $y=B_1$. Similarly for  $y=C_1(k)$ and $y=D_1$. A possible solution thus implies all the four graph meeting at some `$k$', though $A_1, B_1$ would have different values than $C_1, D_1$.
\section{Gap closing point for $\alpha_3=\alpha_4=0$} 
\label{gapclosealpha34zero}
Here we consider the up-spin sector only whose gap closing conditions are given by  Eqs. \ref{hxhyupdn1ap}, and \ref{hxhyupdn2ap}. The solution for down-spin sector can be found analogously. The condition, $h^{k}_{\uparrow, y}=0$ yields  for $\sin k$ as,
\begin{eqnarray}
	\label{apsink}
	\sin k= \frac{\alpha_{R1} - \alpha_{R2} \cos k- \eta \cos 2k}{w- \gamma + 2 \gamma \cos k} .
\end{eqnarray}
The above equation is meaningful when r.h.s is less than 1 and $w- \gamma + 2 \gamma \cos k \ne 0$. Substituting for $\sin k$ in Eqn. \ref{hxhyupdn1ap} and using $\cos 2k = 1- 2 \cos^2 k$, one obtains a cubic equation for $\cos k$ as (with $x= \cos k$)  given below. 
\begin{eqnarray}
a \cdot x^3 + b \cdot x^2 + c \cdot x + d = 0 .
\end{eqnarray}

where the coefficients are defined as (with $\alpha_{Ri}=\alpha_i$):
\begin{eqnarray}
	\label{cubicx123}
&&	a = \eta^2 + 4\gamma^2, ~~~b = 2\alpha_{2} \eta + \eta^2 + 4\gamma w, \\
&&	c = \alpha_{2}(\alpha_{2} + \eta) - \alpha_1 \eta  - \frac{1}{2}\eta^2 - 3\gamma^2 + 2\gamma v + w^2, \\
&&	d = -\alpha_1( \alpha_2 +  \eta) - \frac{\eta}{2}(\alpha_2 + \eta) + \gamma(\gamma  -  v -  w) + v w.~~~~~~~~
\end{eqnarray}
The solutions for $ x $ are given by:
\begin{eqnarray}
x_1&&=-\frac{b}{3a}- \frac{2^{1/3} \bar{z}_0}{3az_2} + \frac{z_2}{3\cdot 2^{1/3} a}~,~~~~~~\\
x_2&&=-\frac{b}{3a}- \frac{(1+ i \sqrt{3}) \bar{z}_0}{3az_2} - \frac{(1-i \sqrt{3})z_2}{3\cdot 2^{1/3} a} , \\
x_3&&=x^*_2,~~~~~ z_2=(\bar{z}_1 + \sqrt{4 \bar{z}^3_0 + \bar{z}^2_1} )^{1/3} .
\end{eqnarray}
Here $\bar{z}_0= -b^2+ 3ac$ and $\bar{z}_1= 2b^3-9abc+27a^2d$.
In the present case only the real value of the root provided by $x_1$ is used for analysis. 

\section{Discussions on possible topological phases in different cases}
\label{w1w2discussion}
 We now discuss the possibility of having different or identical topological phases in various parameter choices. \\
\indent
{\bf Case-I. $w, \gamma, \tilde{\gamma}$ are real and all $\alpha_i=0$:} One notice that in this case $h^k_{1x}=h^k_{2x},~~h^k_{1y}=h^k_{2y}$. This means that the real and imaginary part of the both eigenvalues are identical. Thus this choice leads to identical topology for both the sectors with $W_1=W_2$.\\
\indent
{\bf Case-II- $w,\gamma,\tilde{\gamma}$ complex and all $\alpha_i$ zero: }One  finds $h^k_{1x}=h^k_{2x}$ and $h^k_{1y}=h^k_{2y}$ implying $W_1=W_2$. \\
\indent
 {\bf Case-III- $w,\gamma,\tilde{\gamma}$ are real but $\alpha_i \neq 0$:}  Here all the $\sin$-terms  in $h^k_{0x}$ are zero but `$y$'  changes sign in $h^k_{1x}$ and $h^k_{2x}$. This can be taken care by $k \rightarrow - k$, namely $h^k_{1x} = h^{-k}_{2x}$. In $h^k_{0y}$ all $\cos$  terms are zero and we obtain $h^{-k}_{1y}=- h^k_{0y} + x^k= -(h^k_{0y} - x^k)= - h^k_{2y}$. This relations implies  $W_1=W_2$\\
 \indent
{\bf Case-IV: $w,\gamma,\tilde{\gamma}$ are complex and $\alpha_i \neq 0$:} In this case one finds $h^k_{1x}= v + (\gamma + w) \cos k + \frac{1}{2} (\eta + \xi) \sin k + \gamma_1 \cos 2k + \frac{\eta_1}{2} \sin 2k + y^k$. This is not related with $h^{k^{\prime}}_{2x}$ and thus $h^k_{1x}$ is not related to $h^{k^{\prime}}_{2x}$ for any pair of $(k, k^{\prime})$. Thus $W_1, W_2$ could be different.
\vspace{-0.7cm}
\section{Additional details}
\label{addetails}
\begin{figure}[h!]
\centering
\includegraphics[width=0.8\linewidth]{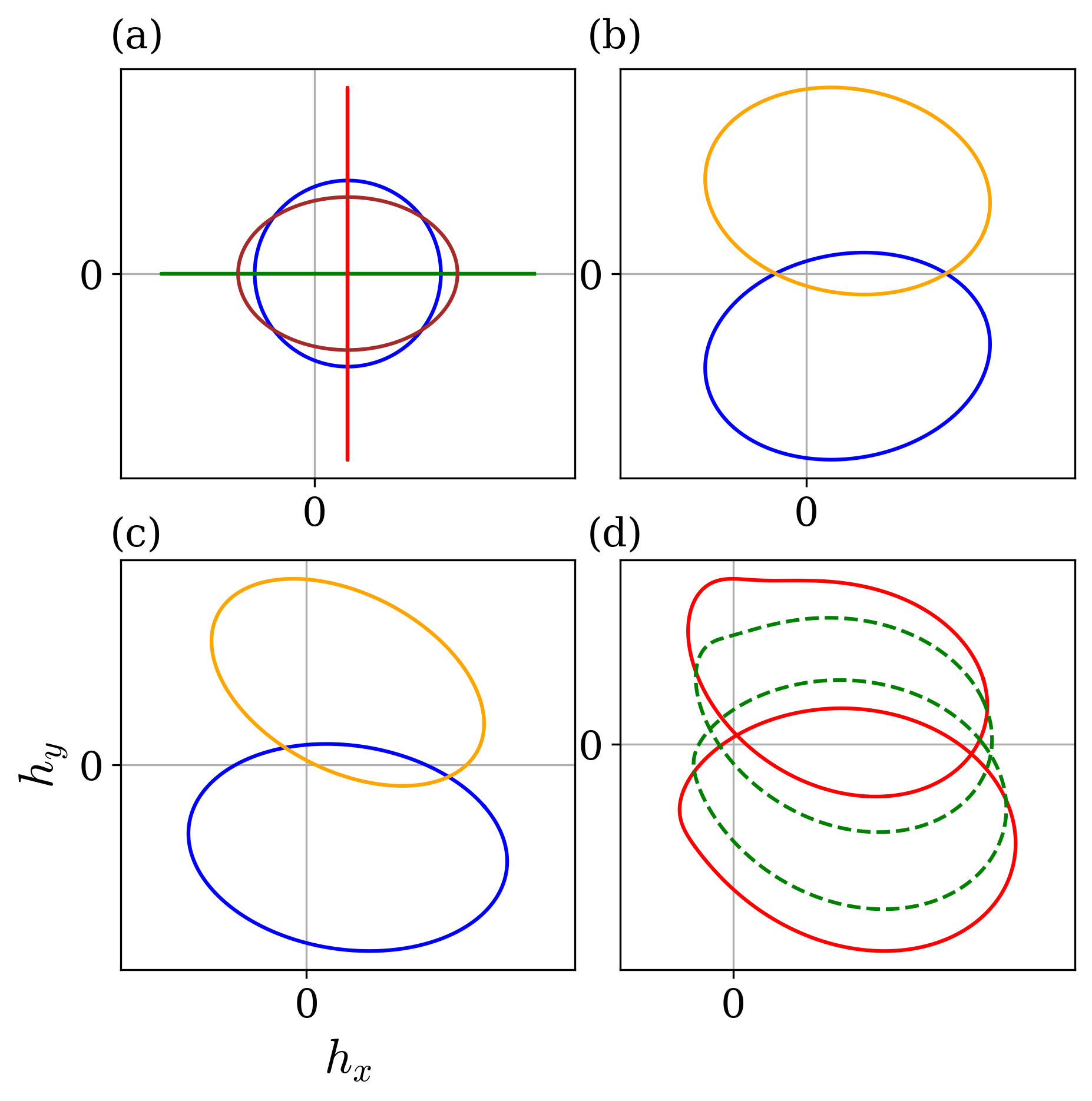}
\caption{ In panel (a) blue  circle denotes the case for only finite $(v,w)=(2,2.8)$, a finite $\gamma=0.5$ makes the contour ellipse as shown by green contour. Making $\gamma=\pm w$ makes the ellipse to be a straight line as represented by brown line along x and y-axis respectively.  In panel (b), effect of finite  $(\alpha_{R_1}, \alpha_{R_2})=(2,1)$ is shown. Here spin up and down sectors produce different contours, blue and yellow  denote spin up and down respectively.  Panel (c) depicts effect of finite $\eta$ which causes difference circumference length for different spins. In panel (d), we show effect of $\gamma_1$ by solid lines.  The dashed line shows the effect of $\alpha_{R3}= 1$ and $\alpha_{R4}= 0.5$.}
\label{contour}
\end{figure}

\begin{figure}[h!]
\centering
\includegraphics[width=0.8\linewidth]{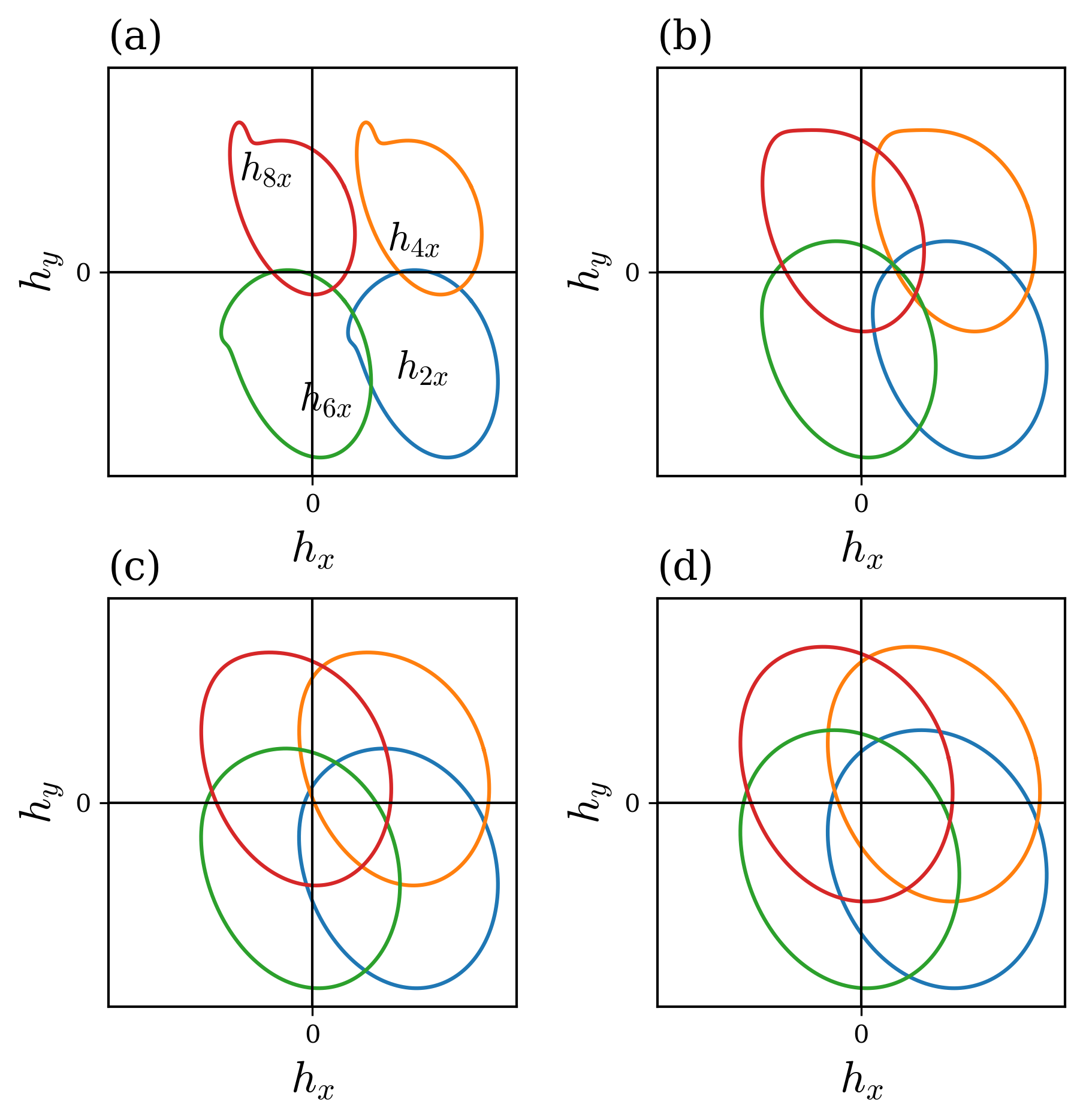}
\caption{ Contours for coupled chain system are plotted.  The four  contours correspond different $\xi_i$ sectors as represented by $\rm h_{i,k}$ in panel (a).  The four panel (a), (b), (c) and (d) correspond two, four , six and eight mid gap states respectively. Here for all panels $v=0.2, \alpha_{R1}=0.3, \alpha_{R2}=0.1, \gamma= \gamma_1=0.05, \eta=0.2, \delta=0.4$. For panel (a), (b), (c) ad (d) value of $w$ is 0.3, 0.5, 0.7, 0.9 respectively.}
	\label{TWo_Chain_ZERo}  
\end{figure}

In Fig. \ref{contour} (and \ref{TWo_Chain_ZERo}) we present contours obtained from hamiltonian function $h$ as given in \ref{hxhyupdn1ap}, \ref{hxhyupdn2ap}(and \ref{twochainkham}). $h_x$ and $h_y$ are generically used to denote the real and imaginary part of $h$. The detail specification regarding spin component etc are described in the caption.

\begin{figure}[h!]
	\centering
	\includegraphics[width=0.5\textwidth]{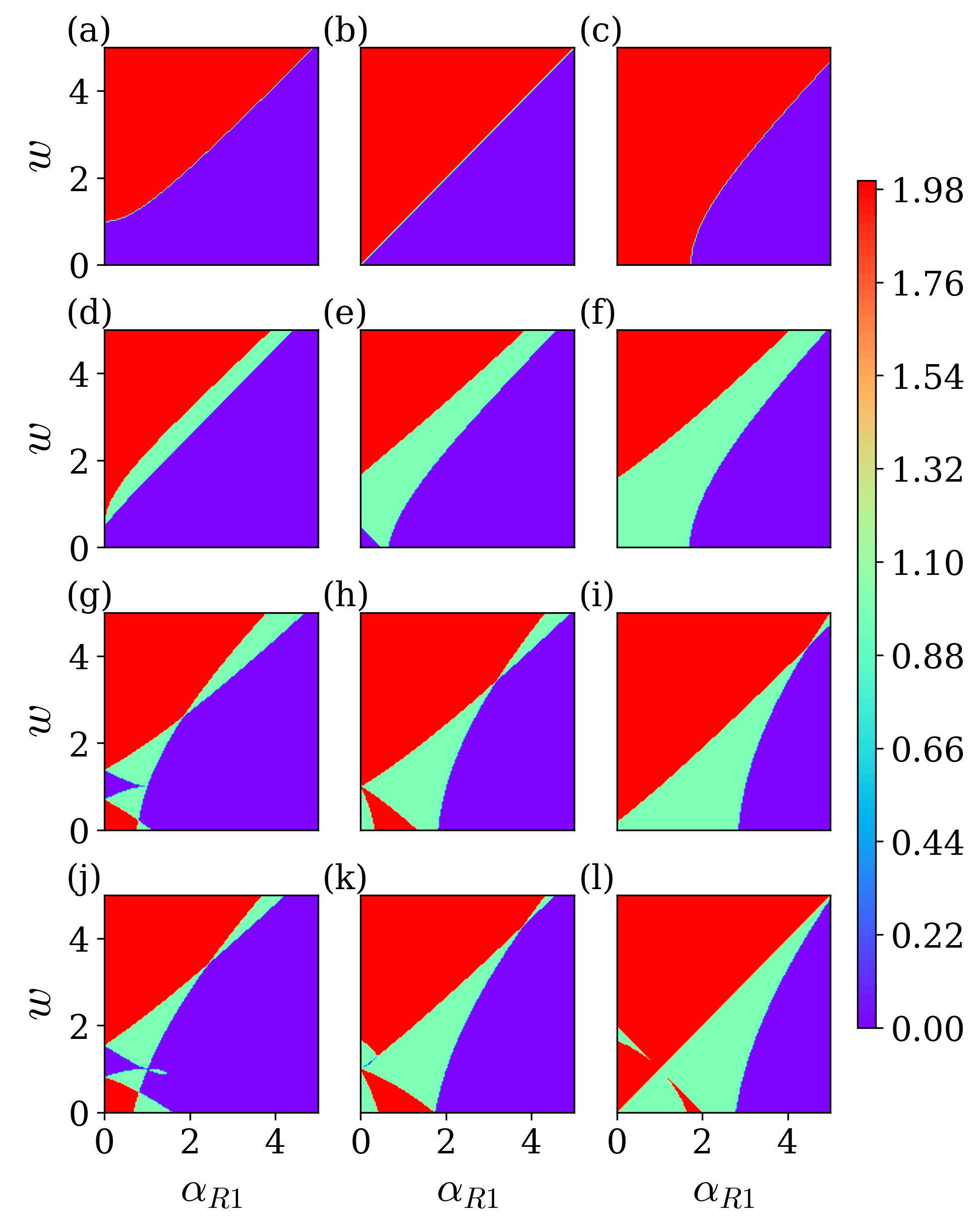}
	\caption{Here we present phase diagram in $\alpha_{R_1}-w$ plane in the presence of $\alpha_{R_3}, \alpha_{R_4}$ to compare with the phase diagram presented in Fig. \ref{phasedgrm-withalpha12}.  The parameters in panel (a) and panel (b) are same as that of Fig. \ref{phasedgrm-withalpha12} and here $(\alpha_{R_3}, \alpha_{R_4})=(0.5,1)$.  Panel (c) and panel (d) have the same values of $( \eta,\eta_1,\gamma,\gamma_1)$ as that of panel (d) in Fig. \ref{phasedgrm-withalpha12}. For panel (c) and (d) values of $(\alpha_{R_3}, (\alpha_{R_4}))$ are  set to $(0.5,1)$ and $(2,1)$ respectively. We observe that inclusion of $(\alpha_{R_3}, \alpha_{R_4})$ can  modify the phase boundary substantially.}
	\label{phasedgrm-withalpha34}
\end{figure}

\begin{figure}[h!]
\centering
\includegraphics[width=0.4\textwidth]{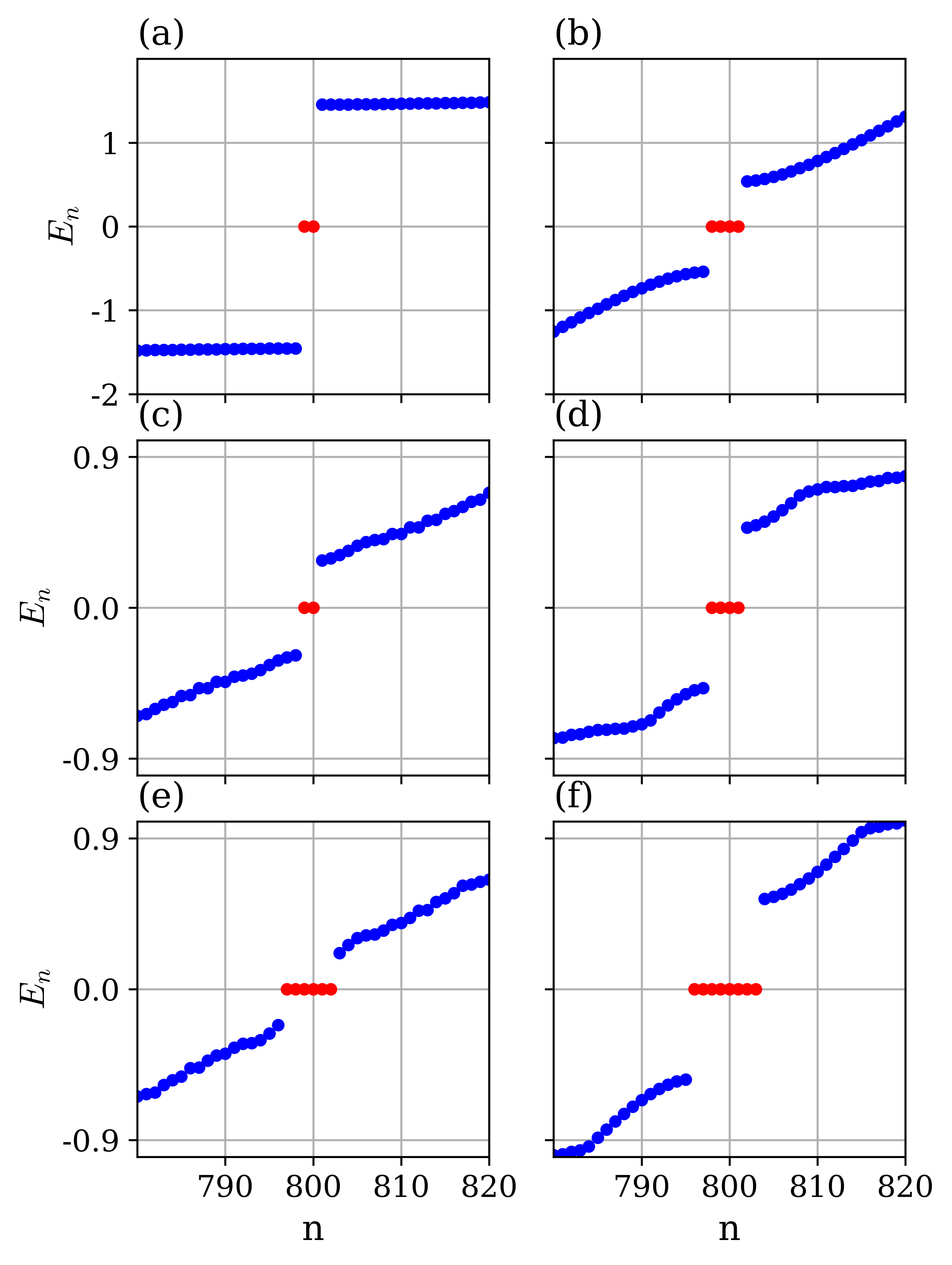}
\caption{ Here we present mid-gap zero energy (MGZE) states for various sets of parameters. x-axis represents index of eigenvalues and y-axis the eigenvalues.  Here we conside an open chain with $N=400$ unit cell. The blue and red points denote bulk and edge modes respectively.  For all panel $ v,\gamma(=\gamma_1), \eta(=\eta_1), \alpha_{R_1}, \alpha_{R_2}$ are taken (1,0.5,1,2,1). Panel (a) and (b) depicts MGZE states for single chain for $w=3$ and $5$ respectively with corresponding winding number $W=1$ and $2$. Panel (c), (d), (e) and (f) depicts MGZE states for coupled chain with $\delta=1$ and $w=0.5,1.5, 2.5$ and $4$ respectively. The corresponding winding numbers are 1, 2, 3, and 4 respectively. }
\label{mid-gap-all-chain}
\end{figure}

\section{Details of coupled chain Hamiltonian}
Considering the single chain Hamiltonian given in Eq. \ref{fullHamiltonian} and the coupling between them in Eq. \ref{Two_Chain_Hamiltonian1}, the coupled chain Hamiltonian becomes (after some algebra ),

\begin{eqnarray}
	\label{twochainkham}
&&	H_{2c} = \sum_k \sum^4_{i=1} \left(\mathfrak{H}^k_{i} \xi^{\dagger}_{i,a,k}\xi_{i,b,k} + \rm{h.c} \right), \\
&&	~ \xi_{i,s}=\left( \psi_{1,s,\downarrow} + t \psi_{2s\downarrow} \right) .\nonumber ~~~
\end{eqnarray}
In the above $s=a,b,~~ i=1,2,3,4,~~ t=\pm $. Here $\sigma=\uparrow (\downarrow)$ for $i$ even (odd). $t=+(-)$ for $i=1,2(3,4)$. $\mathfrak{H}^k_{1}= h^k_{\uparrow} + \delta, \mathfrak{H}^k_{2}=h^k_{\downarrow} - \delta, \mathfrak{H}^k_{3}= h^{k}_{\uparrow} - \delta, \mathfrak{H}^k_{4} =h^k_{\downarrow} + \delta$.  In Eq. \ref{twochainkham} $\xi_{i,\alpha,k}$'s are redefined fermionic operators in terms of the original fermions $\psi_{1,a,\sigma}, \psi_{1,b,\sigma},\psi_{2,a,\sigma}, \psi_{2,b,\sigma} $
where $1$ and $2$ indicate the chain indices, $a$ and $b$ sub-lattice indices and $\sigma$ spin indices.    A comparison of $\mathfrak{H}^k_{1}$ and $\mathfrak{H}^k_{2}$ in Eq. \ref{twochainkham} with the expression of $h^k_{\uparrow}$ and $h^k_{\downarrow}$, we note that $\delta$ re-normalizes the intra-sublattice hopping with values $v \pm \delta$. This changes the criteria of topological phases for each $\xi$ sectors and their dependencies on  $\eta$, $\delta$ and $w$.  In the table \ref{demo-table} we enumerates in details the relative signs of $\eta, \delta$ and value of $w$ for which various fermionic sectors can be made topological.  This understanding allows us to manipulate the topology of a given sector by controlling the sign of $\delta, \eta$ and magnitude of $w$ and could pave way of realizing quantum gates. In Eq. \ref{twochainkham}, each $\xi$ sector is manifestly Chiral invariant. In the original basis, the Chiral operator can be obtained by writing the Hamiltonian in  $\psi$ basis be defining 
($\Psi^{\dagger}_{12}=(\Psi^{\dagger}_1 \Psi^{\dagger}_2) $, 
\begin{eqnarray}
&&	\mathcal{H}_{12}(k)  =[v+ w_+\cos k- \eta \sin k]\Gamma_5+ w_-\sin k \Gamma_6  \nonumber \\ 
	&& + [\alpha_{R1}-\alpha_{R2}\cos k] \Gamma_7-\alpha_{R2} \sin k \Gamma_8 +\delta \Gamma_9 ,
	\label{eq:twochain_Hamiltonian}
\end{eqnarray} 
where  $ \Gamma $ is a $ 8\times 8 $ matrix given as $ \Gamma_5 = \sigma_0 \otimes\sigma_x \otimes \sigma_0 $ ,$ \Gamma_6 = \sigma_0 \otimes \sigma_y \otimes \sigma_0 $, $ \Gamma_7 = \sigma_0 \otimes \sigma_y \otimes \sigma_x $, $ \Gamma_8 = \sigma_0 \otimes \sigma_x \otimes \sigma_z $  and $ \Gamma_9 = \sigma_x \otimes \sigma_x \otimes \sigma_0 $.  The Chiral operator is  defined as,  $ S = \sigma_0 \otimes \sigma_z \otimes\sigma_0 $  yielding $ S^{-1}\mathcal{H}_{12}(k)S = \mathcal{H}_{12}(-k) $. One can compute $	\tilde{H}_{AD}(k) ={U^{\dagger}}_{S}\mathcal{H}_{12}(k){U_{S} }$ such that it contains off-diagonal blocks given below,
\begin{eqnarray}
	\tilde{H}_{AD}(k)&&=
	\begin{pmatrix}
		0&H_{AD}^{+}(k)\\
		H_{AD}^{-}(k)&0
	\end{pmatrix}
	\label{two-chain-ham-antd} .
\end{eqnarray}
Using the above block-diagonal form one can calculate the winding number of the full system. The connection between $\xi$-basis and $\Psi$ basis can be obtained by constructing the an unitary matrix $U_S$ which is obtained by diagonalizing $S^{\prime}$. 
\begin{center}
	\begin{table}[h!]
		\begin{tabular}{ | c | c | c |c |}
			\hline
			W, $\xi$'s& $\eta$ & $\delta$ & $w$ \\ \hline
			$W=1, ~~\xi_3$ & + & + & $w_1$ \\ \hline
			$W=1, ~~\xi_2$ & + & - & $w_1$ \\ \hline
			$W=1, ~~\xi_1$ & - & + & $w_1$ \\ \hline
			$W=1, ~~\xi_4$ & - & - & $w_1$ \\ \hline
			
			$W=2,  ~~\xi_2, \xi_3$ & + & $\pm$ & $w_2 > w_1$ \\ \hline
			$W=2, ~~\xi_1, \xi_4$ & - & $\pm$ & $w_2>w_1$ \\ \hline
			
			$W=3, ~~\xi_2,\xi_3, \xi_4$ & + & $+$ & $w_3 > w_2 > w_1$ \\ \hline
			$W=3, ~~\xi_1, \xi_2, \xi_3$ & + & $-$ & $w_3 >w_2>w_1$ \\ \hline
			$W=3, ~~\xi_1,\xi_2, \xi_4$ & + & $+$ & $w_3 >w_2 > w_1$ \\ \hline
			$W=3, ~~\xi_1, \xi_3, \xi_4$ & + & $-$ & $w_3 >w_2>w_1$ \\ \hline
			
			$W=4, ~~\xi_1,\xi_2,\xi_3, \xi_4$ & $\pm$ & $\pm$ & $w_4 > w_3 > w_2 > w_1$ \\ \hline
			\hline
		\end{tabular}
		\caption{The table depicts how the winding number depends on the signs of $\eta$, $\delta$ and magnitude of $w$. Also  the first column depicts the normalized fermionic sectors which has finite winding number.}
		\label{demo-table}
	\end{table}
\end{center}
\indent
\begin{figure}[h!]
	\centering
	\includegraphics[width=\linewidth]{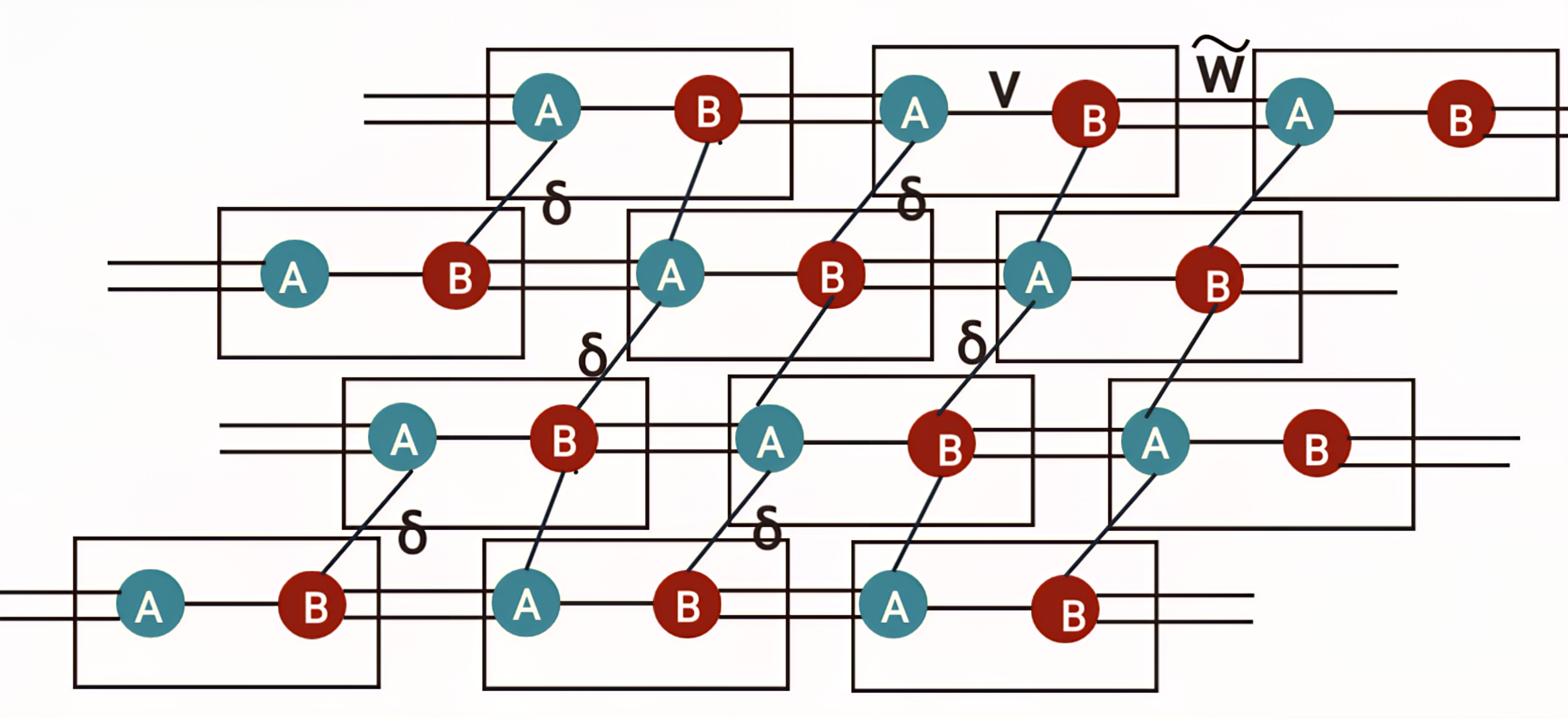}
	\caption{A cartoon picture of four coupled SSH chains is shown in two different equivalent representation. In the upper panel, the pink, light magenta,yellow, and deep magenta denotes the unit cells of each individual SSH chain with sub-lattices $(a,b), (c,d), (e,f)$ and $(g,h)$ respectively. For each chain the first and second alphabets denote the equivalent sub-lattices. The inter-chain coupling are always between two in-equivalent sub-lattices between neighboring chains.}
	\label{fig:SSHchain_4}
\end{figure}
The Hamiltonian $ H_{4c}$ for the four chain model is given in Eq. \ref{Four_Chain_Hamiltonian}. Here $ H_{ab}$, $H_{cd}$, $H_{ef}$ and $H_{gh}$  are the Hamiltonians for each single chain  discussed before and $(\mu\nu)$ in $H_{\mu\nu}$ denotes the equivalent sub-lattices in each chain. 
\begin{gather}
	H_{4c}  = H_{ab}+ H_{cd}+H_{ef}+ H_{gh}+  H_{a\rightarrow h} .
	\label{Four_Chain_Hamiltonian}
\end{gather}

In the above $ H_{a\rightarrow h} $  denotes the coupling terms between two neighboring chains and we are considering periodic boundary condition. Explicitly, we
have

\begin{eqnarray}
	H_{a\rightarrow h} &&=  \delta \sum_{n =i}^{N} \left[ \psi^{\dagger}_{n,\mu} \psi_{n+1,\nu^{\prime}}+ \psi^{\dagger}_{n,\nu} \psi_{n+1,\mu^{\prime}}+h.c\right],
	\label{Four_Chain_Hamiltonian1}~~~
\end{eqnarray} 

where `$n$' denotes the n-th chain  and $(\mu,\nu)$ and $\mu^{\prime}, \nu^{\prime}$ denote the pair of equivalent sublattices of neighboring chains. We perform Fourier transform to obtain for $H_{4c}$, \\
\begin{eqnarray}
&	H_{4c} = [v+w_+\cos k- \eta \sin k]\Gamma_{10}+ w_-\sin k \Gamma_{11} &\nonumber \\
& + ~[\alpha_{R_1}-\alpha_{R_2}\cos k] \Gamma_{12}-\alpha_{R_2} \sin k \Gamma_{13} +\delta \Gamma_{14} ,&~~
\label{eq:static_Hamiltonian}
\end{eqnarray} 
where $ \Gamma $s are  $ 16\otimes16 $ matrices given as $ \Gamma_{10} = \sigma_0 \otimes\sigma_0 \otimes\sigma_x \otimes \sigma_0 $ ,$ \Gamma_{11} = \sigma_0 \otimes\sigma_0 \otimes \sigma_y \otimes \sigma_0 $, $ \Gamma_{12} = \sigma_0 \otimes\sigma_0 \otimes \sigma_y \otimes \sigma_x $, $ \Gamma_{13} = \sigma_0 \otimes\sigma_0 \otimes \sigma_x \otimes \sigma_z $,$ \Gamma_{14} = \sigma_0 \otimes\sigma_x \otimes \sigma_x \otimes \sigma_0+ (\sigma_x \otimes \sigma_x \sigma_x \otimes\sigma_0+\sigma_y \otimes \sigma_y \otimes \sigma_x\otimes\sigma_0)/2 $. The Hamiltonian Eq. \eqref{eq:static_Hamiltonian} preserves the chiral symmetry with the following symmetry operator $ {S''} = \sigma_0 \otimes \sigma_z \otimes\sigma_0 \otimes\sigma_0 $ : $ S''^{-1}H(k)_{4c}S'' = -H_{4c}(k) $. To study the topological properties of four coupled chains system, we first  re-cast our Hamiltonian $H_{4c}$ in a manifestly off-diagonal block form with the help of operator $U_{S''}$  as follows,
\begin{eqnarray}
	\tilde{H}_{AH}(k) ={U}_{S''} ^{\dagger}H_{4c}(k){U_{S''} } = \begin{pmatrix}
		0&H_{AH}^{+}(k)\\
		H_{AH}^{-}(k)&0
	\end{pmatrix}.~~
	\label{four-chain-ham-antd},
\end{eqnarray}
Here ${U}_{S''}$ is the unitary matrix in which $ {S''} $ is diagonal. Explicit form of $U_{S^{''}}$ is obtained from the eigenvectors of $S^{''}$.

\section{Effective Hamiltonian at the Dirac cone}
\label{effham}
Here we ouline a derivation of effective Hamiltonian near the Dirac point. Eigenvalues of the system can be obtained by exact diagonalization of $\mathcal{H}$ in  \ref{newphibasisH}. General expressions of the eigenvalues are given below.
\begin{eqnarray}
&&E(k)= \pm \sqrt{\mathcal{A}^k \pm \sqrt{(\mathcal{B}^{k}_1)^2 + (\mathcal{B}^{k}_2)^2 + (\mathcal{B}^{k}_3)^2}} .
\end{eqnarray}
In the above, $\mathcal{A}^k= |h^k_{\uparrow}|^2 + |h^k_{\downarrow}|^2 + 2 |g^k|^2  $, $\mathcal{B}^k_1= |h^k_{\uparrow}|^2 - |h^k_{\downarrow}|^2 $, $\mathcal{B}^k_2=2 (h^k_{\uparrow x} g^k_x+h^k_{\uparrow y} g^k_y-h^k_{\downarrow x} g^k_x+h^k_{\downarrow y} g_y)  $ and $\mathcal{B}^k_3=2 (h^k_{\uparrow y} g^k_x - h^k_{\uparrow x} g^k_y+ h^k_{\downarrow y} g^k_x- h^k_{\downarrow x} g_y) $.
The expressions for $h^k_{\sigma x}, h^k_{\sigma x}, g^k_x, g^k_y$ can be found in Eqs. \ref{hxhyupdn1ap}, \ref{hxhyupdn2ap} and \ref{hxhyupdn3ap}. Here $g^k_x$ and $g^k_y$ are the real and imaginary parts of $g^k$. We note that in the absence of $\eta, \eta_1, \xi$, the band touching happens at $k=0$ otherwise it occurs at finite $k$. At the band touching point the two conduction (or valence) band eigenvalues touch each other which yields the condition $\mathcal{B}^k_i=0$  for each `$i$'. Using this fact one can show that an expansion of  $(\mathcal{B}^k_i)^2$ around $k_0$ yields $\left( \frac{\partial \mathcal{B}^k_i}{\partial k} \right)_{k_0}^2 (\delta k)^2$. Thus $ \sqrt{(\mathcal{B}^{k}_1)^2 + (\mathcal{B}^{k}_2)^2 + (\mathcal{B}^{k}_3)^2}$ can be approximated as $\mathcal{B}^{k_0}_0 |\delta k|$ with $\mathcal{B}^{k_0}_0= \sqrt{\sum_i \left( \frac{\partial \mathcal{B}^k_i}{\partial k} \right)^2_{k_0}}$. On the other hand $\mathcal{A}^k$ can be approximated as $\mathcal{A}^{k_0}_0 + \mathcal{A}^{k_0}_1 \delta k$. To proceed further we employ $\sqrt{1+ x} \approx 1 + \frac{x}{2}$ for very small $x$. This yields the following expansion around Dirac point,
\begin{eqnarray}
\label{effhameq}
\tilde{E}_k = \left( \mathcal{A}^{k_0}_0 \right)^{1/2} + \left(\mathcal{A}^{k_0}_0 \right)^{-1/2} \left(\bar{\mathcal{A}}^{k_0} \delta k +  \mathcal{B}^{k_0} |\delta k|  \right) .
\end{eqnarray}
We note that the second term has both $|\delta k|$ and $\delta k$ which yields symmetric and asymmetric contribution to Dirac cone respectively. In the below we provide expressions for $\mathcal{A}^{k_0}$ and $\left(\frac{\partial \mathcal{B}^k_i}{\partial k}\right)_{k_0}$.  $\mathcal{A}^{k_0}_0=\sum^3_ {j=0}\left( \mathcal{A}_j \cos k_0j + \bar{\mathcal{A}}_j \sin k_0j \right)$. Where $\mathcal{A}^{k_0}_0=2(v^2 + w^2 + \alpha_1^2 + \alpha_2^2 + \alpha_3^2 + \alpha_4^2 + \gamma^2 + \gamma_1^2) + \eta^2 + \frac{1}{2} \eta_1^2$, $\mathcal{A}_1=-4 \alpha_1 \alpha_2 - 4 \alpha_3 \alpha_4 + 4v(w + \gamma) + 4w\gamma_1 + \eta \eta_1$, $\mathcal{A}_2=4w\gamma + 4v\gamma_1 - \eta^2$, $\mathcal{A}_3=4\gamma \gamma_1 - \eta \eta_1$, $\bar{\mathcal{A}}_1=-2(2v\eta - \gamma_1 \eta + w \eta_1)$, $\bar{\mathcal{A}}_2=-2((w + \gamma)\eta + v \eta_1)$ and $\bar{\mathcal{A}}_3=-2(\gamma_1 \eta + \gamma \eta_1)$.  $\bar{\mathcal{A}}^{k_0}=\sum^3_{j=1} j \bar{\mathcal{A}}_j \cos k_0 j$. $\frac{\partial \mathcal{B}_1}{\partial k}= \sum^2_{i=1}\left( \mathcal{B}^{\prime}_j \cos kj + \mathcal{C}^{\prime}_j \sin kj \right)$, $ \mathcal{B}^{\prime}_1=4 \left( \alpha_2 (\gamma_1 - v) + \alpha_1 (\gamma - w) \right)$, $ \mathcal{B}^{\prime}_2
- 8 \left( \alpha_2 \gamma + \alpha_1 \gamma_1 \right)$, $\mathcal{C}^{\prime}_1=  - 2 \alpha_2 \eta_1 $, $\mathcal{C}^{\prime}_2=  4 \left( \alpha_2 \eta + \alpha_1 \eta_1 \right).$ The expression for $\frac{\partial \mathcal{B}_2}{\partial k} $ is obtained by replacing $(\alpha_1, \alpha_2)$ by $(\alpha_3, \alpha_4)$. Finally $\frac{\partial \mathcal{B}_3}{\partial k}= 4 \left( \alpha_2 \alpha_3 - \alpha_1 \alpha_4 \right) \cos k$. We note that $\mathcal{A}_i, \bar{\mathcal{A}}_i$ involve $\eta$ which causes asymmetry in the dispersion around the Dirac point. When $\eta=\eta_1=0$, all $\bar{\mathcal{A}}_i$ is zero and $\mathcal{A}^{k_0}_0$ $= 2 (v + w + \gamma_1 +\gamma)^2 +2 (\alpha_1-\alpha_2)^2 + (\alpha_3-\alpha_4)^2$. 

\section{Additional details for IPR and NPR}
\begin{figure}[h!]
	\centering
	\includegraphics[width=1\linewidth]{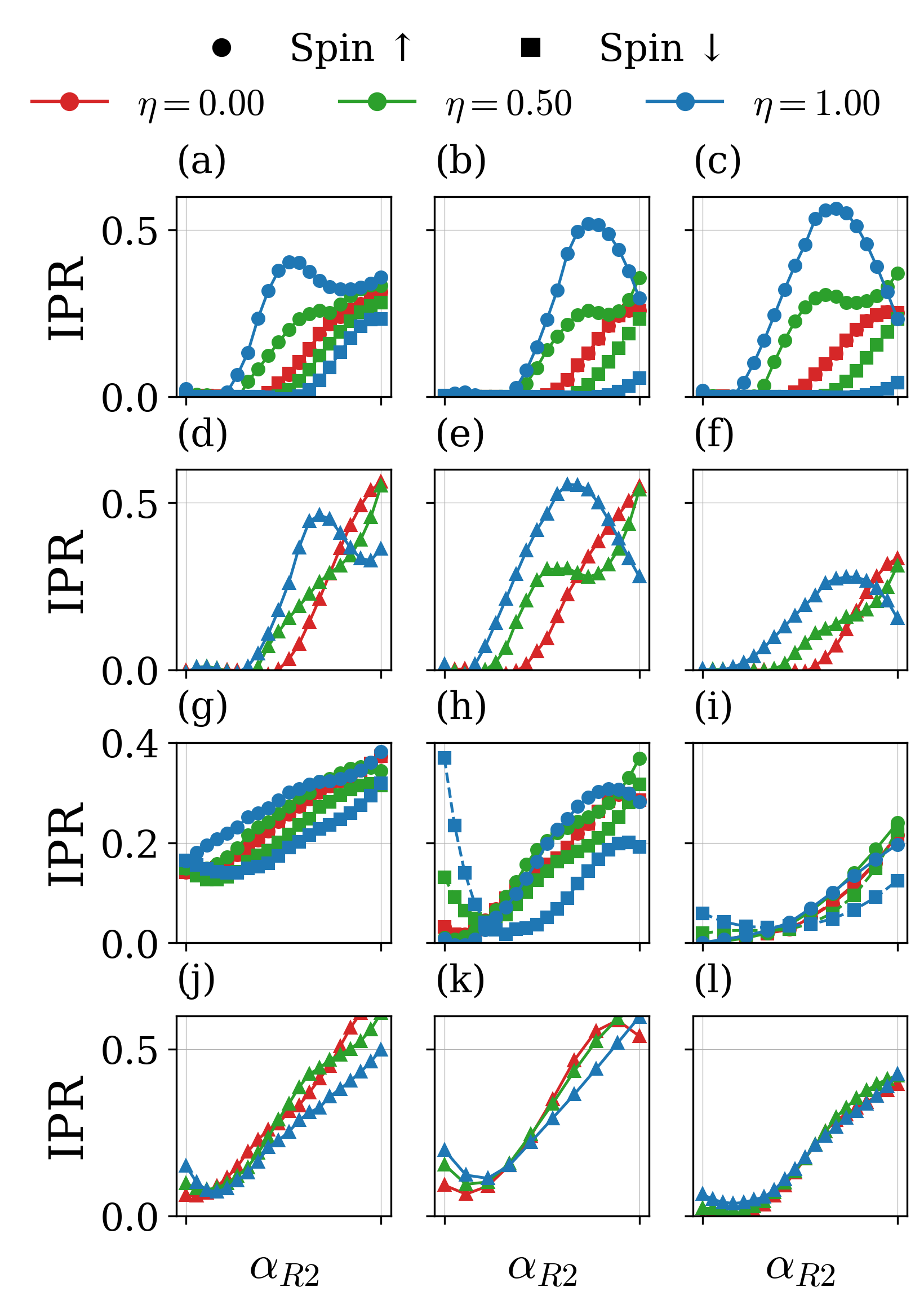}
	\caption{IPR is plotted vs $\alpha_{R_2}$. Panel (a)-(f) is with ($v,w$=(1,0.5)) and panel (g)-(h) is with ($(v,w)=(1,3)$). $\alpha_{R_1}=1, \eta=\eta_1$ for all panels.  Panels in 1st $\&$ 3rd row (2nd $\&$ 4th) rows correspond zero (non-zero) values of $\alpha_{R_3}, \alpha_{R_4}$. $(\alpha_{R_3}, \alpha_{R_4})=(0.5,1)$ for panels (d),(j),(e),(k) and (2,1) for panels (f), (l).  In panels (a) and (g), $\gamma$ is taken zero and for all panel it is set at 0.5. $(\gamma_1,\eta)$ is taken (0,0) ((0.5,1)) for  1st,3rd (2nd, 4th) rows. }
	\label{fig:ipr-decay-ap}  
\end{figure}

\begin{figure}[h!]
	\centering
	\includegraphics[width=0.45\textwidth]{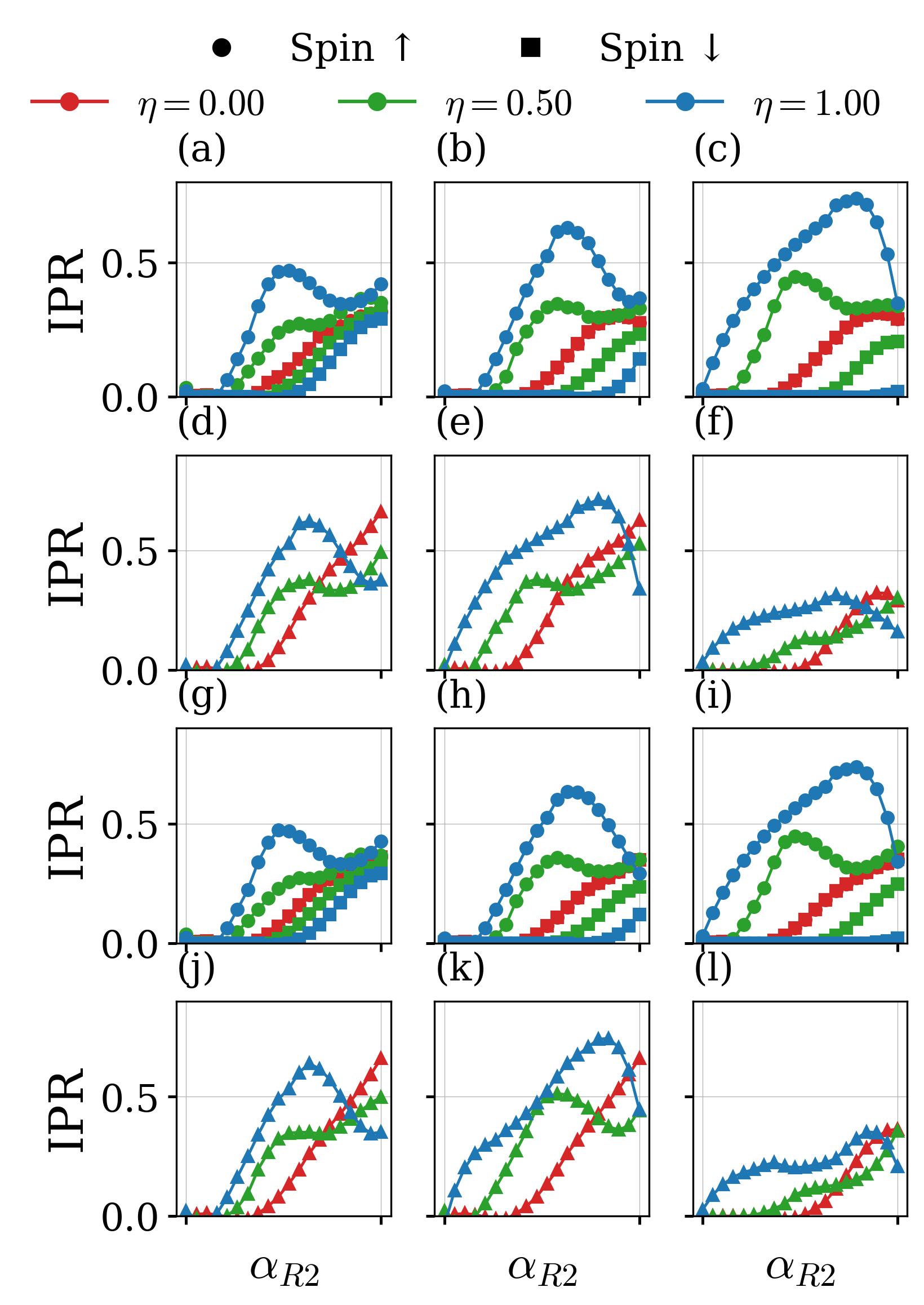}
	\caption{IPR of single chain with domain wall. All parameters are same for single chain except the domain-wall potential taken as $ w = 1+ G$ and $v=1-G $ where $G = u_0 \tanh\frac{n-n_0}{\xi/a}$  $t_0 = 0.2,u_0 = 1,n_0 = 200,\xi = 10,a = 1,N= 400$.  }
	\label{ipr-domain-wall-one-chain-ap}
\end{figure}

\begin{figure}[h!]
	\centering
	\includegraphics[width=0.5\textwidth]{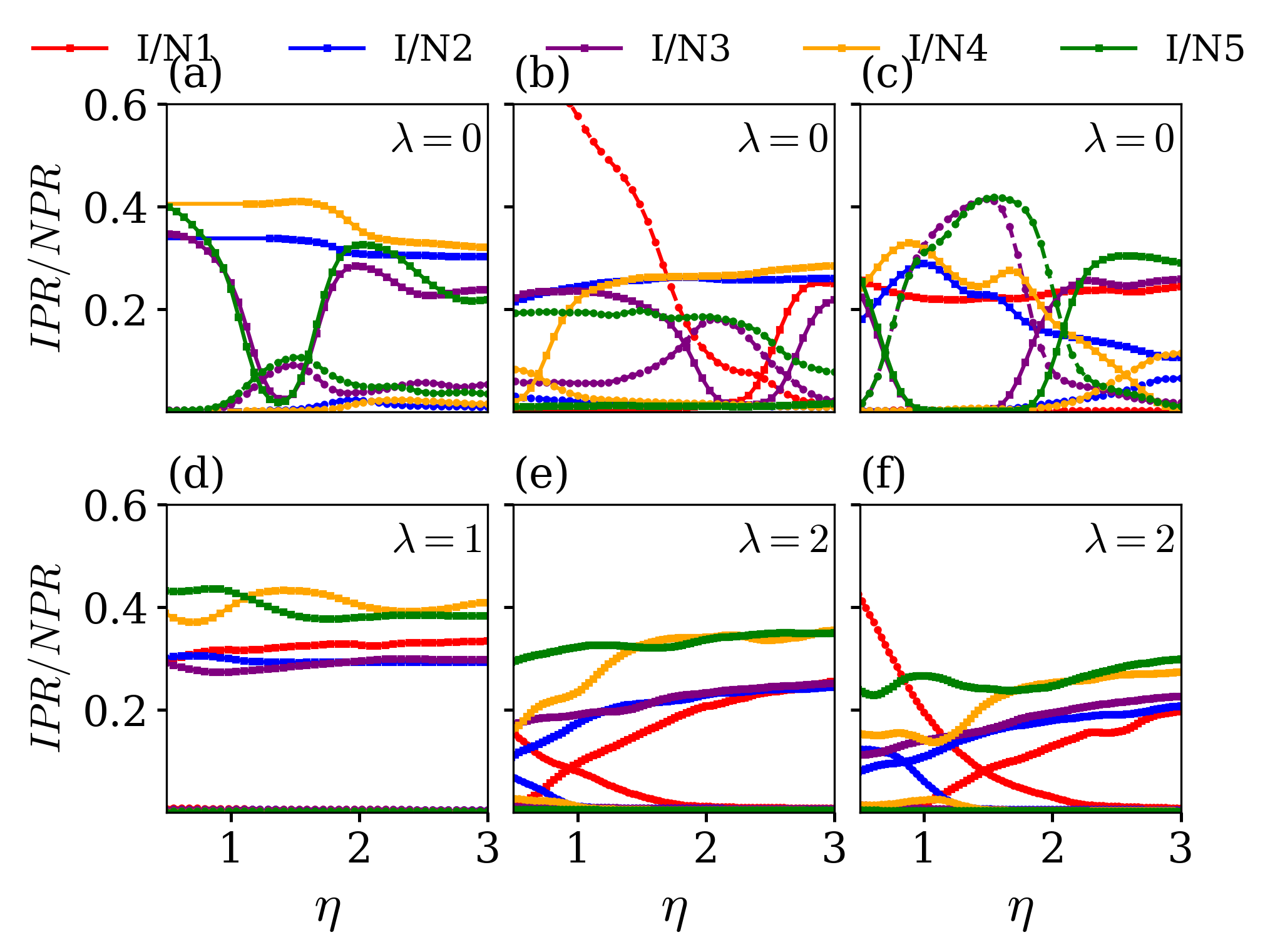}
	\caption{The IPR and NPR are plotted for various set of parameters. The red lines denote only finite $(v,w)$, with $(v,w)$=(1,0.5) for panel (a) and (1,3) in all other panel. All panel except (a) is with $(\alpha_{R_1},\alpha_{R_2},\gamma)=(2,1,0.5)$.  While blue and purple is with zero $(\alpha_{R_3},\alpha_{R_4})$, for orange and green it is (0.5,1).  $\gamma_1$ is zero (0.5) for blue, orange (purple, green). $\eta_1$ is also zero for blue, orange and made equal to $\eta$ for purple and green.  Panel (c) is with  DW. Panel (d) and (e) is with AA potential for different $\lambda$ as mentioned. Panel (f) shows combined effect of AA potential and DW. } 
	\label{ipr-domain-wall-AA-ap}
\end{figure}

\begin{figure}[h]
	\centering
	\includegraphics[width=1\linewidth]{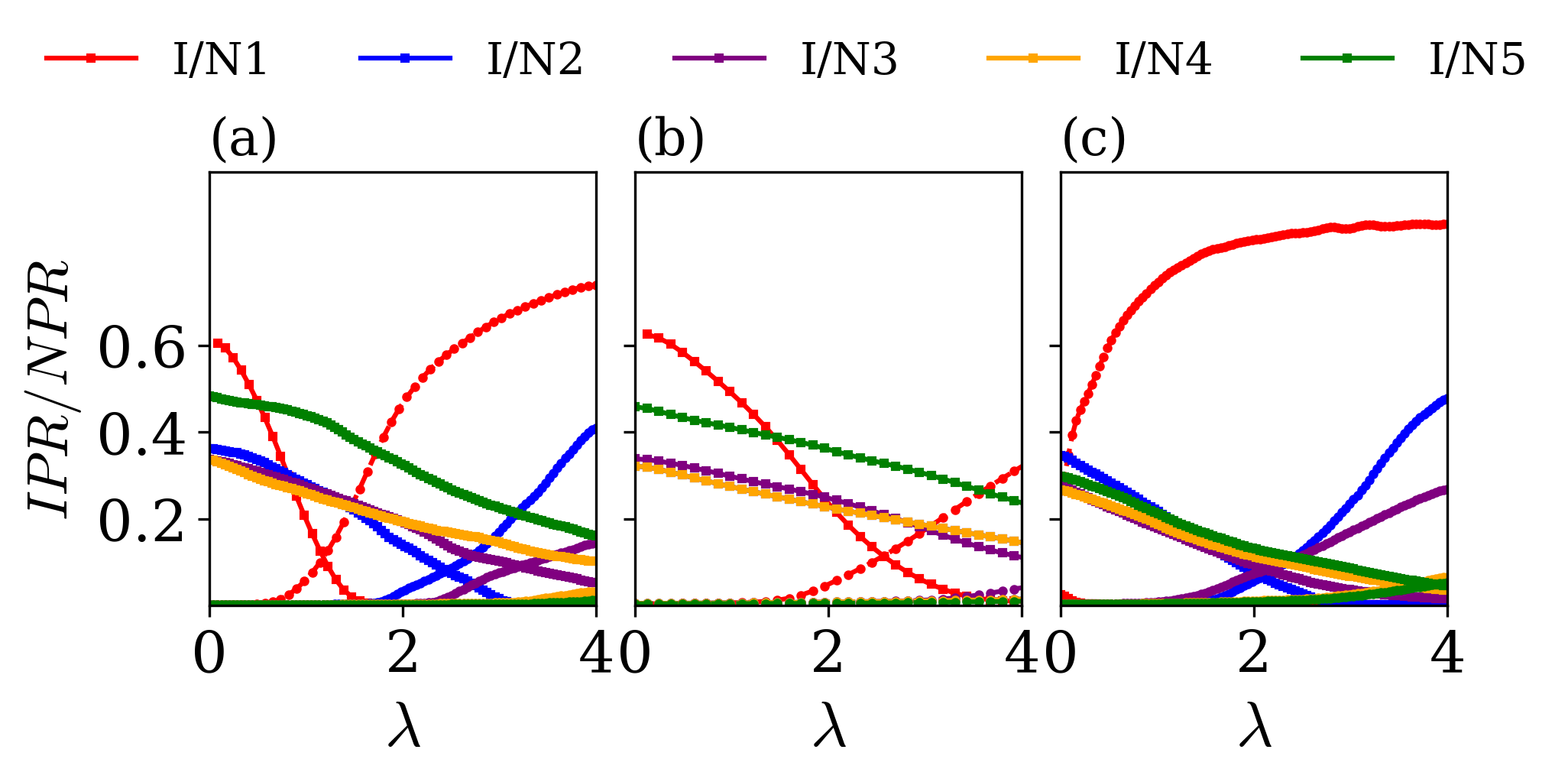}
	\caption{Various legends corresponds to that of Fig. \ref{ipr-domain-wall-AA}. For red plot,  panel (a) has $v>w$ with $(v,w)=1,0.5$ and  panel (b) and (c), $(v,w)=(1,3)$. We observe that AA potential influences the topological phase transitions \cite{tapan-2021-PhysRevLett.126.106803} in various capacity depending on the set of parameters. However panel (c), where DW potential is present AA potential effects minimaly as far as different parameters are concerned except the case of pure SSH model given by red plot.}
	\label{fig:ipr-npr-lambda}
\end{figure}

In Fig. \ref{fig:ipr-decay-ap}, we plot the IPR for  edge states for different sets of parameters for a comparison \cite{gabriel-PhysRevResearch.4.013185,Munoz2018,Bahmani-24}. For a better understanding we have considered the case of $v>w$ (in panel (a)-(f)) and $w>v$ (in panel (g) to (l)). Also to compare the effect of finite $\alpha_{R_3}$ and $\alpha_{R_4}$, we have kept them zero for odd rows and finite for even rows. For $v>w$, one observes that initially IPR is zero as there is no localized edge modes, however as $\alpha_{R_2}$ increases, IPR becomes finite due to transition to topological phase. For the red plot $\eta=0$ which makes the topological phase transition(TPT) for both sector at same value of $\alpha_{R_2}$. For finite $\eta$ (the dark and light green plots) show different values of $\alpha_{R_2}$ for TPT. A comparison between first two rows and last two rows show that  $v>w$ depicts a different trend in IPR than the $w>v$ case. In the later case, IPR first decreases with $\alpha_{R_2}$ and then it increases.\\
\indent
In Fig. \ref{ipr-domain-wall-one-chain-ap} we  plot IPR for the MGZE state in the presence of DW. The parameters of each panel in Fig. \ref{ipr-domain-wall-one-chain-ap} corresponds to panels in Fig. \ref{fig:ipr-decay-ap}. From the panel (a) to (f), we notice that  DW profile significantly increases the width of non-topological phase with $\alpha_{R_2}$.  Secondly in panels (g) to (l) (where $w>v$ originally), the DW induces large changes in IPR in comparison to corresponding panels in Fig. \ref{fig:ipr-decay-ap}.  Panels in first and third rows (similarly second and fourth rows) shows similarity in IPR showing a identical effect of DW in IPR on edge states. It shows that $(\alpha_{R_3}, \alpha_{R_4})$ plays a deciding role for IPR profile in the presence of DW.\\
\indent
In Fig. \ref{ipr-domain-wall-AA-ap}, we  compare effect AA and DW potential and plot IPR and NPR for various parameters. The NPR is defined as $ {\rm NPR}= \left( N\sum_i |a_i|^4 \right)^{-1}$ with $N$ being the total number of sites. A vanishing IPR and finite NPR signify a delocalized phase. For localized phase, they behave oppositely. The red curve denotes IPR and NPR for pure SSH model.   In panel (a), we note that green and purple (orange and blue) follow each other signifying their dependence on $\gamma_1, \eta_1$   which are same for this pairs. Panel (a) and (b) show differences in IPR and NPR in non-topological ($v>w$) and topological phase($w>v$) respectively.   Panel(c) shows that the effect of DW  (with the parameters of panel (b))). It shows re-entrant topological phase transition for green, purple and yellow lines. It exhibits how a finite $\gamma_1$ causes the re-entrant phase transition earlier for green and purple than orange for which $\gamma_1$ is zero.  Panel (d)  and (e) shows effect of AA potential only for $w>v$, it shows qualitatively different nature than in panel (c). Further, in panel (d), all IPR being zero (in comparison to panel (e)) indicates that one needs a critical $\lambda$ for the AA potential to have any topological phase conforming the earlier study \cite{tapan-2021-PhysRevLett.126.106803}(also see Fig.~\ref{fig:ipr-npr-lambda} for more comparison of competing effect of Aa and DW potential). Also AA potential induces a difference between green and purple (orange and blue) line, which is worth noting. Panel (f) shows that there is no qualitative changes when DW is present in addition to AA potential. It establishes stronger effect of DW potential in comparison to AA potential.

\section{Analytical understanding of localizaton of zero energy mid-gap states away from the edge in the presence of domain wall: Case of $\gamma\ne0$ \label{midcenter}.}
Here we briefly explain why Domain wall causes the localization  of the MGZE states to be maximum at the center of the chain. Then we  explain  explicitly the case of finite  $\gamma$ and show how it  causes the localization of the MGZE mode to be no longer at  the center. The procedure we describe here can be extended in a straight forward manner easily to explain all other cases. However the exact analytical understanding becomes increasingly complex, but can be numerically verified. To begin the stage we consider a SSH chain having $N_t$ number of unit cell with $N_t$ odd. This causes in total $N_t-1$ number of  intra-cell bonds associated with hopping amplitude $v$ and $N_t$ number of inter cell bonds with hopping amplitude $w$. In the presence of domain wall one can not use the notion of unit cell and translational invariant $v$ and $w$. Now the sites and bonds are labeled by the index `$m$' and `$j$' from the left. If we keep using the  previously used notion of unit cell index `$n$' one can easily map the index $m, j$ to $n$. The domain wall potential is described by the profile $u=u_0 \tanh (j-j_0) \bar{\xi}$ where $\bar{\xi}= \xi/a$ and $j_0$ is the middle bond. For our case the index $j_0=2 n_0-1$ where $n_0$ denotes the unit cell index at the middle with $n_0=\frac{N_t+1}{2}$. The Hamiltonian can be described as, 

\begin{eqnarray}
H &&= \sum_n  \left( v(n) c^{\dagger}_{n,A}c_{n,B} + w(n) c^{\dagger}_{n,B}c_{n+1,A} + h.c \right) \nonumber \\
&& ~~ +~~ \left( \gamma(n) c^{\dagger}_{n,A} c_{n+1,B} + h.c \right) .
\end{eqnarray} 

In the above $c_{n,A}$ and $c_{n,B}$ denote the annihilation operator defined on `$A$' and `$B$'  sub-lattices of $n$-th unit cell respectively. $v(n)$ denotes the intra-cell hopping amplitude, and $w(n)$ denotes the inter-cell hopping amplitude between $n$ and $n+1$-th unit cell. Similarly $\gamma(n)$ denotes the hopping amplitude between $A$ sub-lattice of $n$-th unit cell and $B$ sab-lattice of $(n+1)$-th unit cell. As the hopping is allowed between different sub-lattice only, any general eigenstate will be having a form $|\Psi \rangle= \sum \alpha_{m} c^{\dagger}_{m,A} + \beta_{m} c^{\dagger}_{m,B}$ where $\alpha_m$ and $\beta_m$ denote the amplitude of the wave-function at $A$ and $B$-sub-lattices in $m$-th unit cell respectively. The condition $H | \Psi \rangle=0$ yields the following recursion relations,
\begin{eqnarray}
&&w(n) \alpha_{n+1} + v(n) \alpha_n + \gamma(n-1)\alpha_{n-1} =0,  \\
&& w(n-1) \beta_{n-1} + v(n) \beta_n + \gamma(n) \beta_{n+1}=0.
\end{eqnarray}

For $\gamma=0$, the recurrence relation reduces to $\frac{\alpha_n}{\alpha_{n-1}}= - \frac{v(n-1)}{w(n-1)}$. In terms of $t_{j}$, the $v_m$ and $w_m$ is represented as $t_{2m-1}$ and $t_{2m}$ respectively. One finds that $\frac{v(m)}{w(m)}= \frac{1- u_0 \tanh(2m-1-j_0)\bar{\xi}}{1+ \tanh(2m-j_0) \bar{\xi}}$. The mid-side $j_0=2n_0-1$. This relation shows that when $m <n_0$, then $|v(m)| > |w(m)|$ implying $\alpha_{n}$ increases as $n$ increases and reaches maximum at $n=n_0$ and then again it decreases. This explain why the left edge state shifts its localization to the middle. The right edge state remains localized as it is due to similar reason. In the presence of $\gamma$ the recurrence relation is  obtained as,
\begin{eqnarray}
 \alpha_n &&= - \frac{v_{n-1}}{w_{n-1}} \alpha_{n-1} - \frac{\gamma}{w_{n-1}} \alpha_{n-2}.
 \end{eqnarray}

As the sign of $\alpha_n$ and $\alpha_{n-2}$ are opposite to that of $\alpha_{n-1}$ and as $v_{n-1}, w_{n-1}, \gamma$ are positive, the above equation says that the magnitude of $\alpha_{n}$ is decreased due to the presence of $\gamma$ by an amount given by the second term. However this is a qualitative statement as all the amplitudes are redistributed and it merely says that one can no longer  ascertain with certainty that the maximum amplitude happens  for $\alpha_{n_0}$. Now assuming that in the absence of very small $\gamma$, the values of $\alpha_n$ can be replaced by their value in the absence of $\gamma$, the decrease in magnitude of $\alpha_n$ is given by $-\frac{\gamma}{w_{n-1}} \alpha_{n-2}$. Now we can argue what is the amount by which the magnitude of $\alpha_{n}$ decreases in the presence of $\gamma$. It is useful to consider the following amplitudes $\alpha_{n_0}$ and $\alpha_{{n_0-1}}$  and their decreased amount $\delta_{n_0}$ and $\delta_{n_0-1}$. If $|\delta_{n_0}| > |\delta_{n_0-1}|$ and $|\delta_{n_0-1}| < |\delta_{n_0-2}|$, then we can quantitatively say that the maximum amplitude happens for $\alpha_{n_0-1}$. As the $\alpha_n$ and $\delta_n$ are defined in terms of $v(n), w(n)$, we need to map them to appropriate $t_{j}$ and the rules are easily found as $v_n=t_{2n-1}$ and $w_n=t_{2n}$. With this we find the following ratio,
\begin{eqnarray}
	\frac{\delta_{n_0}}{\delta_{n_0-1}}&&= \frac{1-u_0 \tanh 3\bar{\xi}}{1-u_0 \tanh \bar{\xi}} \frac{1+ u_0 \tanh 6\bar{\xi}}{1-u_0 \tanh 5\bar{\xi}} ,\\
		\frac{\delta_{{n_0-1}}}{\delta_{n_0-2}}&&= \frac{1-u_0 \tanh 5\bar{\xi}}{1-u_0 \tanh 3\bar{\xi}} \frac{1+ u_0 \tanh 8\bar{\xi}}{1-u_0 \tanh 7\bar{\xi}} .
\end{eqnarray} 

From the above relation we find that as $ | \delta_{n_0}| > |\delta_{n_0-1}|$, $\alpha_{n_0-1}$ may be greater in magnitude than $\alpha_{n_0}$. However there is a possibility that $\alpha_{n_0-2}$ is greater in magnitude than $\alpha_{n_0-1}$ as  $ | \delta_{n_0-1}| > |\delta_{n_0-2}|$. But the exact amplitude should be determined by the exact recursion relation and what it tells that the amplitude is redistributed in a non-trivial manner and the maximum amplitude no longer appears at $\alpha_{n_0}$. Similar arguments can also be made for the right edge mode which results in the recursion relation involving $\beta_{n}, \beta_{n-1}$.
\end{document}